\documentclass[pra,twocolumn,superscriptaddress]{revtex4-1}
\usepackage{amsmath,amssymb,graphicx,color,bm,epstopdf,float,caption,subcaption,tabularx,multirow}
\usepackage{scrextend}
\usepackage{hyperref}
\renewcommand{\vec}[1]{\mbox{\mathversion{bold}$#1$}}

\graphicspath{{fig2/}{fig4/}{fig5/}{fig6/}{fig7/}{fig8/}{fig9/}{fig10/}{fig11/}{fig12/}{fig13/}}

\begin{document}

\title{Electro-optical properties of phosphorene quantum dots}

\author{V. A. Saroka}
\email{v.saroka@exeter.ac.uk}
\affiliation{School of Physics, University of Exeter, Stocker Road, Exeter EX4 4QL, United Kingdom}
\affiliation{Institute for Nuclear Problems, Belarusian State University, Bobruiskaya 11, 220030 Minsk, Belarus}

\author{I. Lukyanchuk}
\affiliation{University of Picardie, Laboratory of Condensed Matter
Physics, Amiens, 80039, France}
\affiliation{L. D. Landau Institute for Theoretical Physics, Moscow, Russia}

\author{M. E. Portnoi}
\affiliation{School of Physics, University of Exeter, Stocker Road, Exeter EX4 4QL, United Kingdom}

\author{H. Abdelsalam}
\email{hazem.abdelsalam@etu.u-picardie.fr}
\affiliation{Department of Theoretical Physics, National Research Center, Cairo,12622, Egypt}

\date{\today}

\begin{abstract}
We study electronic and optical properties of single layer phosphorene quantum dots with various shapes, sizes, and edge types (including disordered edges) subjected to an external electric field normal to the structure plane. Compared to graphene quantum dots, in phosphorene clusters of similar shape and size there is a set of edge states with energies dispersed at around the Fermi level. These states make the majority of phosphorene quantum dots metallic and enrich the phosphorene absorption gap with low-energy absorption peaks tunable by the electric field. The presence of the edge states dispersed at around the Fermi level is a characteristic feature that is independent of the edge morphology and roughness.

\end{abstract}

\maketitle

\section{Introduction}

Phosphorene is a single layer of black phosphorous that has been recently isolated~\cite{CastellanosGomez2014}. Unlike its predecessor -- graphene~\cite{Novoselov2004} -- it has a significant band gap of about $2$~eV. Such a large band gap, in conjuntion with the carrier mobility up to $1000$~cm$^2$ V$^{-1}$ s$^{-1}$, is anticipated to be more practical, compared to graphene, for digital electronics~\cite{Li2014,Qiao2014}. However, any new material~\cite{Miro2014, Schwierz2015} which is put forward as a candidate to replace current silicon technology will have to catch up with it. In other words, it has to start from the end of Moore's law, to which current technology is rapidly approaching. The critical size limit is predicted to be $5$~nm; at this space scale quantum effects such as tunneling and carrier confinement affect device performance~\cite{SalmaniJelodar2014}. In this view, the effects due to the device's shape and size gain essential importance, thereby making their study in systems with edges such as ribbons and quantum dots a paramount priority. With respect to the optical properties spatial carrier confinement brings not only new challenges but also great advantages such as the decrease of the pumping threshold current in quantum dot lasers~\cite{Arakawa1982}.

Despite impressive recent achievements in the synthesis of nanostructures, such as the rise of self-assembling molecular engineering~\cite{Wu2003,Cai2010,Lu2011,Ruffieux2016} and nanolithography~\cite{Tapaszto2008,Jia2009}, the main problem with low dimensional structures is a precise control of their geometry and edge quality. Concurrently, techniques with the best outcomes are not easily transferable between materials. Therefore, properties and effects that are robust against disorder are of great importance for practical applications. Here we present the results of our search for universal features in the variety of phosphorene quantum dots.

The first attempts to synthesize phosphorene nanostructures have been undertaken. Along with the development of phosphorene synthesis by mechanical cleavage~\cite{CastellanosGomez2014,Li2014,Liu2014,Liu2014b}, liquid exfoliation~\cite{Brent2014,Yasaei2015,Woomer2015}, Ag$^{+}$ plasma thinning~\cite{Lu2014}, and pulse laser deposition~\cite{Yang2015}, few-layer nanoribbons~\cite{Masih2016} and phosphorene quantum dots~\cite{Sun2015,Zhang2015,Sofer2016} and have been produced. The theoretical results for regular shapes of phosphorene quantum dots have been reported. Besides the size effect in rectangular dots studied by first principles calculations~\cite{Niu2016}, the effects of the shape, size and external magnetic field were systematically analyzed within the effective tight-binding model~\cite{Zhang2015a}. Within this model the influence of the magnetic and in-plane electric fields in rectangular dots has also been considered~\cite{Li2016}. However, the effect of the electric field in somewhat more natural back gate geometry, which has been intensively studied for graphene clusters~\cite{Guclu2009,Guclu2010,Guclu2011,DaCosta2015,DaCosta2016,Guclu2016}, has not been investigated yet. The effects of edge disorder have not been investigated either. In the aforementioned studies the edges of the quantum dots were taken to be well-defined, whereas none of the currently available synthesis techniques ensure such a high edge quality, therefore revealing the effects of edge disorder is important. 

In this paper we present a tight-binding study of the electronic and optical properties of small phosphorene clusters with an applied electric field in back gate geometry. We also report on comparative analysis of so-called quasi-zero-energy states in phosphorene quantum dots with zero-energy states in graphene dots. 
We highlight differences between quasi-zero-energy states in phosphorene clusters and zero-energy states in graphene ones~\cite{Ezawa2007,Potasz2010,Zarenia2011,Espinosa2011,Guclu2013,Abdelsalam2015a,Ezawa2012,Abdelsalam2015b}.
In the present study phosphorene clusters with edge roughness are modeled and investigated by random fractals.
In particular, this model is employed to find the relation between the number of quasi-zero-energy states and the edge structure of the quantum dots and to outline a route to the design of dielectric clusters.

In what follows, we introduce structures in Sec.~\ref{sec:Structures}, provide theoretical details of calculations in Sec.~\ref{sec:TheoreticalModel}, discuss the results in Sec.~\ref{sec:Discussion} and summarize discussion in Sec.~\ref{sec:Conclusions}.

\section{\label{sec:Structures}Structures Classification}
\begin{figure}[htbp]
	\centering
  \includegraphics[width=0.4\textwidth]{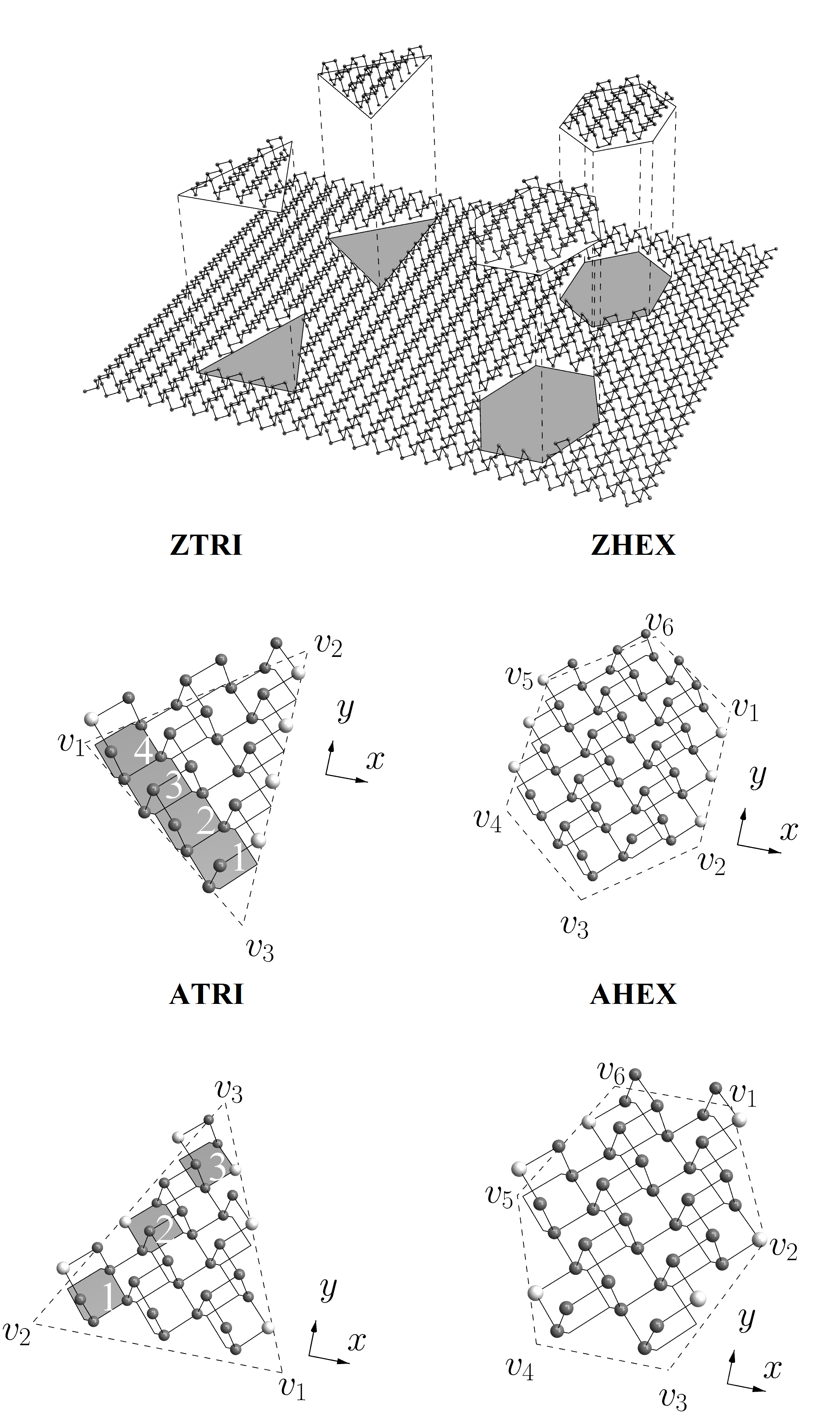}
\caption{Classification of phosphorene quantum dots. Shaded and numbered irregular hexagons are used for cluster size identification. The phosphorous atoms without ``a pair" in the opposite layer are highlighted in light gray.}
\label{fig:PhosphoreneQuantumDotsClassification}
\end{figure}

Phosphorene quantum dots (PQDs) are small monocrystal clusters of phosphorene. A single dot can be imagined as a piece of 2D phosphorene enclosed within a closed polygonal line without self-intersections. This general approach is useful for any 2D material, for instance, it can be used to define a graphene-based superlattice unit cell~\cite{Saroka2014,Saroka2015,Saroka2016}. Since boundaries in crystals tend to form along specific crystallographic directions, not every polygon is suitable for the role of the small cluster boundary. The  puckered honeycomb lattice structure of phosphorene restricts the variety of simple bounding polygons to triangles and hexagons. As shown in Fig.~\ref{fig:PhosphoreneQuantumDotsClassification} each of the bounding polygons used as a cutting mask admits the isolation of clusters with two different edge geometries. These two species correspond to graphene quantum dots with zigzag and armchair edges~\cite{Zarenia2011}, therefore, by analogy we refer to them as zigzag and armchair PQDs. Throughout this paper we use the following labelling convention: $\langle$edge type$\rangle$ $\langle$shape$\rangle$, where $\langle$edge type$\rangle$ is to be either ``Z'' or ``A'', meaning zigzag or armchair edge geometry, while $\langle$shape$\rangle$ defined as TRI or HEX means triangular or hexagonal shape.

Table~\ref{tab:BoundingPolygonVerteces} summarizes descriptions of the bounding polygons in terms of primitive translations: 
\begin{align}
\vec{a}_1 &= a (\cos \phi, \sin \phi) \, , & \vec{a}_2 &= a (\cos \phi, -\sin \phi) \, ,
\end{align}
where $a = |\vec{a}_1| = |\vec{a}_2| = 2.537$~{\AA} and $\phi = 40.11^{\circ}$ is the angle between either of the primitive translation and $Ox$-axis. The polygon vertex position $\vec{v}_i$ can be conveniently expressed as $\vec{v}_i = s \vec{\ell}_i$ in terms of the size factor $s$ and the elementary vectors $\vec{\ell}_i$ given in Table~\ref{tab:BoundingPolygonVerteces}. 
Note that to keep a precise correspondence with the graphene quantum dots~\cite{Potasz2010,Guclu2016} in the case of the ZTRI clusters the three phosphorene atoms closest to the vertexes must be removed from the structure.

\begin{table}[htbp]
\caption{ The vertex elementary vectors $\vec{\ell}_i = (n,m) = n \vec{a}_1 + m \vec{a}_2$ in the basis of the primitive translations $\vec{a}_1$ and $\vec{a}_2$ and size factors $s$ for various phosphorene quantum dots.}
\begin{tabular}{cccccc} \hline \hline
   & \multicolumn{4}{c}{Quantum dot type} \\ 
$\vec{\ell}_i$  & ZTRI & ZHEX & ATRI & AHEX \\ \hline
   1 & $(0,0)$ & $(1,0)$ & $(0,1)$ & $(2,-1)$  \\ 
   2 & $(1,0)$ & $(0,1)$ & $(-1,0)$ & $(1,1)$ \\
   3 & $(0,1)$ & $(-1,1)$ & $(1,-1)$ & $(-1,2)$ \\
   4 & -- & $(-1,0)$ & -- & $(-2,1)$ \\
   5 & -- & $(0,-1)$ & -- & $(-1,-1)$ \\
   6 & -- & $(1,-1)$ & -- & $(1,-2)$ \\ \hline
   $s$ & $N + 1$ & $N$ & $N +1/2$ & $N - 1/2$ \\ \hline \hline
\end{tabular}
\label{tab:BoundingPolygonVerteces}
\end{table}

\begin{figure}[htbp]
	\centering
	\begin{subfigure}{.33\textwidth}
        \centering
         \includegraphics[width=\textwidth]{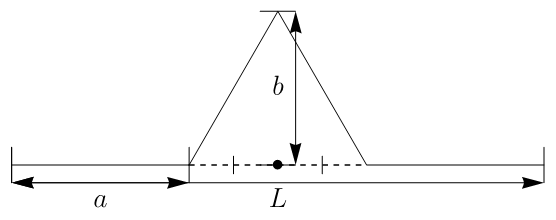}
    \caption{}
\end{subfigure}%

	\begin{subfigure}{.33\textwidth}
        \centering
         \includegraphics[width=\textwidth]{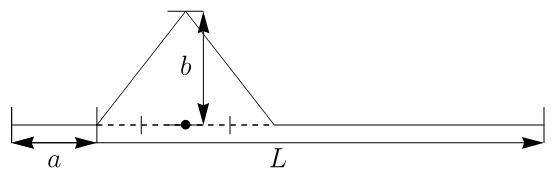}
    \caption{}
\end{subfigure}%

	\begin{subfigure}{.33\textwidth}
        \centering
         \includegraphics[width=\textwidth]{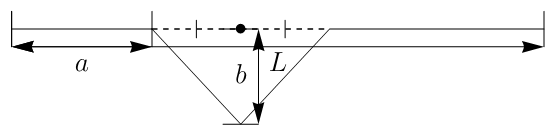}
    \caption{}
\end{subfigure}%

	\begin{subfigure}{.4\textwidth}
        \centering
         \includegraphics[width=\textwidth]{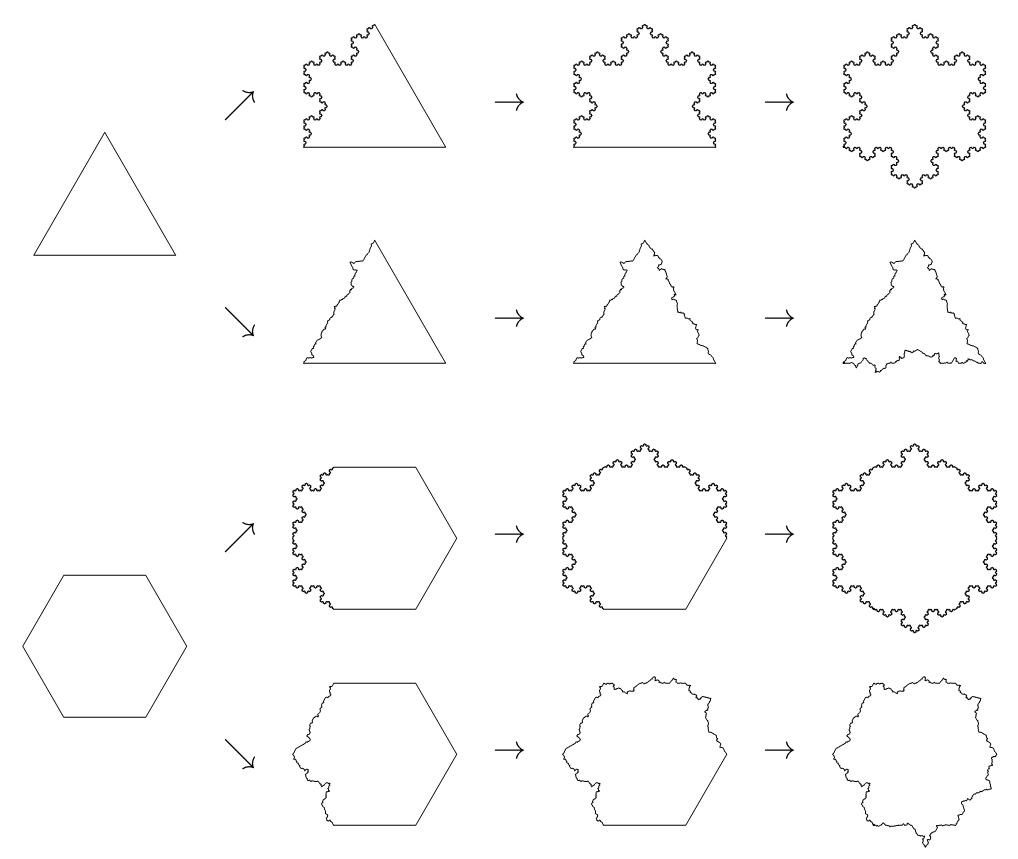}
    \caption{}
\end{subfigure}%
\caption{The bounding polygon randomization. (a) The triangular notch used in the ideal Koch curve generation together with the main geomertical parameters. (b) and (c) The outward and the inward notches with $a$ and $b$ parameters randomized. (d) The replacement of a polygon edge with the ideal and randomized Koch curve for the regular triangle and hexagon.}
\label{fig:BoudingFractal}
\end{figure}

As has been mentioned above, the synthesis of small phosphorene clusters by liquid exfoliation does not result in well-defined edges~\cite{Sun2015,Zhang2015}. On one hand, it is evident that none of the top-down methods (nanolithography, plasma etching, etc.) can ensure atomically smooth edges. On the other hand, the bottom-up approaches based on organic precursors rapidly developing for graphene nanostructures~\cite{Cai2010,Ruffieux2016} cannot be adapted to the phosphorene structures. Some methods that can be classified as intermediate ones, for instance, graphene quantum dot production by the decomposition of C$_{60}$ fullerenes~\cite{Lu2011}, cannot be straightforwardly adapted for phosphorene dots either. In other words, dealing with phosphorene quantum dots one inevitably faces a problem of edge roughness. In this view, the polygon edges mentioned above should be irregular. To model this edge roughness we adopt a random fractal approach that in comparison to some other approaches~\cite{Espinosa2011,EspinosaOrtega2013} preserves the initial triangular or hexagonal morphology of the dot. Our approach is somewhat similar to that implemented in Ref.~\cite{Kosior2017} but it avoids vacancies in the interior of the dots. Instead of keeping the fractional Hausdorff dimension fixed~\cite{Kosior2017}, in modeling the edge disorder it is more important to make sure the distribution of the vacancies and additional atoms at the structure edge avoids disorder in the interior. This problem is similar to that of modeling coastlines in geophysics~\cite{BookMandelbrot1983}. Here we take up a practical rather than strictly mathematical approach. The following algorithm does not exclude completely self-intersections for triangular clusters, but it is quite robust in this sense for hexagonal ones. The occasional self-intersections resulting in cluster decay into smaller pieces can be interpreted as phosphorene  debris being a by-product of the dot synthesis. To model the edge roughness each bounding polygon edge is replaced by a Koch curve~\cite{Koch1904,BookMandelbrot1983} generated after $5$ iterations with random parameters~\cite{Saroka2017a}. The Koch curve is a fractal structure that can be obtained by replacing the central one third of the initial edge, $L$, with a triangular notch as shown in Fig.~\ref{fig:BoudingFractal} (a) and repeating this operation with each newly produced edge. As shown in Figs.~\ref{fig:BoudingFractal} (b) and (c) the position of the notch, $a$, and its height, $b$, can be sown evenly in the interval $(0,1)$. Also, the direction of the notch can be randomly chosen between the inward and outward as presented in Figs.~\ref{fig:BoudingFractal} (b) and (c). The result of such a randomization is given in Fig.~\ref{fig:BoudingFractal} (d). Thus, the edge roughness can be modeled by replacing the bounding polygon with the randomized fractal line -- teragon~\cite{BookMandelbrot1983}. The above-described procedure is intended to imitate the result of various uncontrolled fluctuations in the conditions of quantum dot synthesis. Ideally, it should be supplemented with the edge relaxation via geometry optimization procedure, as has been done for graphene quantum dots~\cite{Voznyy2011,Bugajny2017}. However, to reveal the pure effect of the edge roughness, we neglect such a relaxation and assume that all the atoms has the same coordinates as they would have within the 2D phosphorene. The described procedure results in an inhomogeneous distribution of the hopping integrals at the edges of the structure. We note that this type of disorder is different from previously studied, for instance, for graphene quantum dots, where the on-site energies were varied either throughout the whole structure or its edges~\cite{Jaworowski2013}, or in 2D phosphorene~\cite{Yuan2015}, where vacancies and impurity atoms were distributed randomly throughout the whole structure.

\section{\label{sec:TheoreticalModel} Theoretical Model}
\begin{figure}[htbp]
	\centering
  \includegraphics[width=0.4\textwidth]{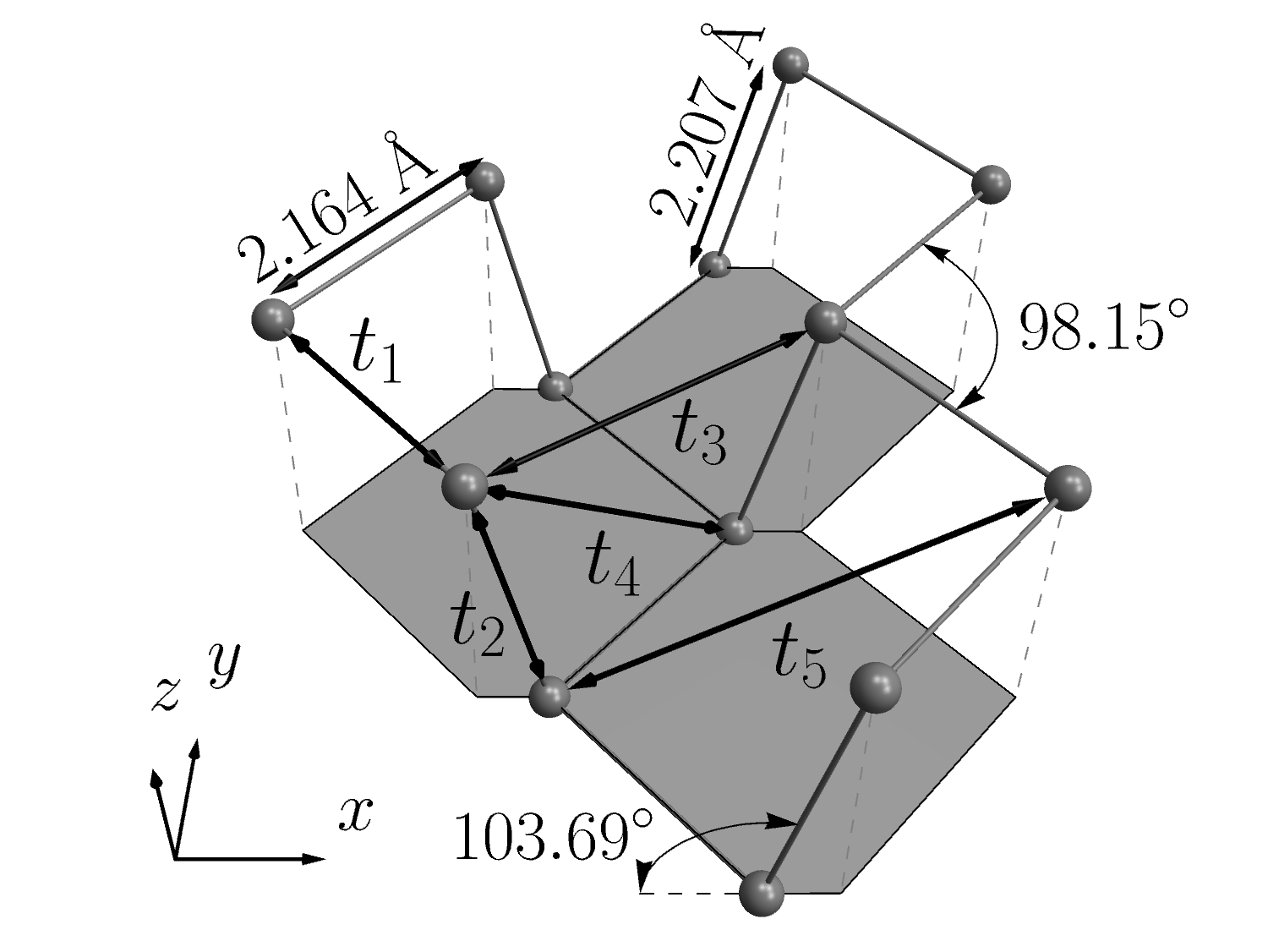}
\caption{The tight-binding and structural parameters of the phosphorene lattice.}
\label{fig:TightBindingParemeters}
\end{figure}

In general the matrix form of the Hamiltonian is obtained by expanding electron wave functions in an orthogonal basis set and calculating the matrix elements of the Hamiltonian operator between the basis functions. The Hamiltonian, then, is an  $n \times n$ matrix
\begin{equation}
\label{eq:TightBindingHamiltonian}
H =  \left(
\begin{array}{cccccc}
h_1 & h_{12} & h_{13} & h_{14} & \cdots & h_{1n} \\
h_{21} &  h_2 & h_{23} & h_{24} & \cdots & h_{2n} \\
h_{31} & h_{32} & h_3 & h_{43} & \cdots & h_{3n} \\
h_{41} & h_{42} & h_{43} & h_4 & \cdots & h_{4n} \\
\vdots &   &   &  &  & \vdots \\
h_{n1} & \cdots   & \cdots & \cdots & \cdots & h_n
\end{array}
\right) \, ,
\end{equation}
where $n$ is the number of functions in the basis set. Within the single orbital tight-binding model the basis functions are atomic orbitals. Thus, the matrix element $h_{ij} = \int_{V} \psi_i \hat{H} \psi_j d^3 \vec{r}$, with basis functions $\psi_{i}$, is referred to as the hopping integral between the $i$-th and $j$-th atomic sites. The hoping integral $h_{ii} = h_i$ is usually referred to as the on-site energy. Within the single orbital model the dimension of the matrix Hamiltonian is naturally equal to the number of atoms in the structure. The $p_z$-orbital tight-binding model has been widely used for the investigation of graphene structures~\cite{Chung2016} including monolayer~\cite{Ezawa2007,Guclu2013} and bilayer~\cite{DaCosta2015,DaCosta2016} graphene QDs. A similar model has been deployed for group IV 2D materials~\cite{Wu2016} and GaAs monolayers~\cite{Chung2017}. Unlike carbon orbitals in graphene, in phospherene phoshorous atomic orbitals are $sp^3$ hybridized. Therefore phosphorene band structure is described by a mixture of the $s$ and $p$ orbitals. However, the largest contribution to the wave function close to the Fermi level comes from $p_z$-orbitals~\cite{Takao1981,Takao1981a,Rudenko2015}. As a result
the low energy spectrum of single layer phosphorene can be described by an effective tight-binding model accounting for only $p_z$-orbitals~\cite{Rudenko2014}. It has been shown by Rudenko and Katsnelson that within this tight-binding model it is sufficient to consider only a few nearest-neighbour hopping integrals for correct description of the low-energy electronic properties of single and double layer phosphorene ~\cite{Rudenko2014}. If the distance between the $i$-th and $j$-th atoms is one of $d_i$, presented in Table~\ref{tab:TightBindingAndStructuralParameters}, then $h_{ij}$ in matrix~\eqref{eq:TightBindingHamiltonian} is equal to the coupling parameter $t_i$, presented in Table~\ref{tab:TightBindingAndStructuralParameters} and depicted in Fig.~\ref{fig:TightBindingParemeters}. If the distance between the atoms does not match any of $d_i$ then we set $h_{ij}=0$. The on-site energies are taken to be zero; $h_i = 0$. This effective tight-binding  model  has  been  widely  adopted to carry out systematic studies on monolayer phosphorene nanoribbons~\cite{Ezawa2014,TaghizadehSisakht2015,Grujiс2016} and phosphorene quantum dots~\cite{Zhang2015a,Li2016}.

\begin{table}[htbp]
\caption{The tight-binding, $t_i$, and structural, $d_i$, parameters used for phosphorene based quantum dots.}
\begin{tabular}{c>{\centering}p{2.5cm}>{\centering\arraybackslash}p{2.5cm}} \hline \hline
 No.  & $t_i$~\footnote{\label{footnote1} Ref.~\cite{Rudenko2014}}, eV & $d_i$~\footnote{\label{footnote2} Obtained from Ref.~\cite{CastellanosGomez2014}}, \AA \\ \hline
   1  &  $-1.220$ & $2.164 $  \\
   2  &  $3.665$ & $2.207 $  \\
   3  &  $- 0.205$ & $2.956 $  \\
   4  &  $- 0.105$ & $3.322 $  \\
   5  &  $- 0.055$ & $3.985 $  \\ \hline \hline
\end{tabular}
\label{tab:TightBindingAndStructuralParameters}
\end{table}

Applying a static electric field to the considered system adds the following potential to the on-site energy:
\begin{equation}
    U = e \vec{E} \cdot \vec{r} \, ,
\end{equation}
where $\vec{E}$ is the electric field strength and $\vec{r}$ is the radius-vector of the given atomic site. For the electric field applied perpendicular to the flat structure parallel to the $xOy$ plane the on-site energy is defined as 
\begin{equation}
    h_j = \int_{V} \psi_i \hat{U} \psi_i d^3 \vec{r} = e E z_j \, .
\end{equation}
For the phosphorene quantum dots in question $ z_j = d_2 \cos \left( \varphi - \pi/2 \right)$, where $\varphi = 103.69^{\circ}$, for atoms in the upper plane and it is zero for atoms in the lower plane.

The study of optical properties of a finite system requires evaluating of the matrix elements of dipole moment or position operator. These matrix elements are conventionally referred to as optical matrix elements. To calculate these matrix elements we present the electric field of the incident electromagnetic wave as $\vec{E} = E \, \vec{e}_p$, where $E$ is the magnitude of the electric field and $\vec{e}_p$ is a unit vector specifying the polarization of the incident wave. In what follows we consider a linearly polarized optical excitation propagating normally to the $xOy$ plane, but our results can be easily generalized for an arbitrary incident angle and polarization. Then $\vec{e}_p$ is a constant vector and without losing generality it can be chosen to be along the $Ox$-axis, i.e. $\vec{e}_p = (1,0)$. In this way the position operator $\vec{r}$ is reduced to its projection onto the plane wave polarization vector, which for the given case is just the $x$ coordinate.

Next we have to convert the $x$ coordinate matrix element,
\begin{equation}
\label{eq:PositionOperatorXProjectionMatrixElement}
x_{lm} = \int_V \Psi_{l}^{\ast} \, x\, \Psi_{m}\, d^3\vec{r} \, ,
\end{equation}
to that of the tight-binding model. For this purpose we expand the electron wave function $\Psi_i$ over a set of functions $\{\psi_j\}_{j=1 \ldots n}$ forming a complete orthonormal basis: 
\begin{equation}
\label{eq:WaveFunctionAsLinearSuperpositionOfBasisFunctions}
\Psi_i = \sum_{j=1}^{n} c_{ij} \, \psi_j \, .
\end{equation}
Substituting this expansion into Eq.~\eqref{eq:PositionOperatorXProjectionMatrixElement} yields
\begin{equation}
\label{eq:PositionOperatorXProjectionMatrixElement1}
x_{lm} = \sum_{i,j} c^{\ast}_{lj} c_{mi} \int_V \psi_{j}^{\ast} \, x\, \psi_{i}\, d^3\vec{r} \, .
\end{equation}
Within the orthogonal nearest-neighbor tight-binding approximation
\begin{equation}
\int_V \psi_{j}^{\ast} \, x\, \psi_{i}\, d^3 \vec{r} = x_i \delta_{ij} = X_{ij} \, 
\end{equation}
or equivalently the matrix form of the $x$ coordinate operator in the tight-binding model is
\begin{equation}
\label{eq:AtomicRingProblemPositionOperatorXProjection}
X =  \left(
\begin{array}{cccccc}
x_1 & 0 & 0 & 0 & \cdots & 0 \\
0 &  x_2 & 0 & 0 & \cdots & 0 \\
0 & 0 & x_3 & 0 & \cdots & 0 \\
0 & 0 & 0 & x_4 & \cdots & 0 \\
\vdots &   &   &  &  & \vdots \\
0 & \cdots   & \cdots & \cdots & \cdots & x_n
\end{array}
\right) \, ,
\end{equation}
where $x_i$ are the $x$-coordinates of the atomic positions in the structure. In fact, the coefficients, $c_{mi}$, introduced in Eq.~\eqref{eq:WaveFunctionAsLinearSuperpositionOfBasisFunctions} are the components of the eigenvectors  $\widetilde{C}_m$ of the matrix Hamiltonian given by Eq.~\eqref{eq:TightBindingHamiltonian}. Thus, the matrix form of Eq.~\eqref{eq:PositionOperatorXProjectionMatrixElement1} is
\begin{equation}
\label{eq:PositionOperatorXProjectionMatrixElement2}
x_{lm} =\widetilde{C}^{\dagger}_l X \widetilde{C}_m = \sum_{j=1}^{n} c_{lj}^{\ast} c_{mj} x_j\, ,
\end{equation}
where ``$\dagger$'' denotes the Hermitian conjugate.

Utilizing the matrix elements given by Eq.~\eqref{eq:PositionOperatorXProjectionMatrixElement2}, we calculate the oscillator strength of a dipole oscillator~\cite{BookHaug2003} as
\begin{equation}
S_x(\varepsilon_{i,f}) = \dfrac{2 m}{\hbar^2} \left\vert x_{if} \right\vert^{2} \varepsilon_{i,f} \, ,
\label{eq:OscillatorStrength}
\end{equation}%
where $m$ is the free electron mass, $\varepsilon_{i,f}= \varepsilon_{f}- \varepsilon_{i}$ is the energy of a single-electron transition between the initial and final states with energies $\varepsilon_{i}$ and $\varepsilon_{f}$, respectively. The knowledge of the oscillator strength allows one to calculate the optical absorption cross-section~\cite{Lee2002,Yamamoto2006}:
\begin{equation}
\label{eq:AbsorptionCrossSection}
  \sigma_x (\varepsilon) \sim \sum_{i,f} S_x(\varepsilon_{i,f}) \delta (\varepsilon-\varepsilon_{i,f}) \, ,
\end{equation}
where summation is carried out over all possible transitions between the valence and conduction states; $\delta (\varepsilon-\varepsilon_{i,f})$ is the Dirac delta-function. The losses due to scattering on phonons, inhomogenuities etc. can be taken into account phenomenologically by replacing the Dirac delta-function by a Gaussian with a broadening parameter $\alpha$: 
\begin{equation}
\label{eq:AbsorptionCrossSectionFinal}
\sigma_x (\varepsilon)\sim \sum_{i,f}S(\varepsilon_{i,f})\exp \left[ -\frac{(\varepsilon - \varepsilon_{i,f})^{2}}{\alpha ^{2}}\right] \,.
\end{equation}

\section{\label{sec:Discussion} Results and discussion}
The calculations are carried out for zigzag triangular (ZTRI), zigzag hexagonal (ZHEX), armchair triangular (ATRI), and armchair hexagonal (AHEX) quantum dots (QDs). For the model study we choose ZTRI, ZHEX, ATRI and AHEX PQDs with $n=222$, $384$, $216$ and $366$, characterized by the edge length $L= |\vec{v}_2 - \vec{v}_1| \approx 35.5$~{\AA}, $26.2$~{\AA}, $33$~{\AA} and $23.7$~{\AA}, respectively. 

\subsection{\label{sec:EnergySpectra}Energy spectra}
\begin{figure}[htbp]
\centering
\begin{subfigure}{.25\textwidth}
  \centering
  \includegraphics[width=\textwidth]{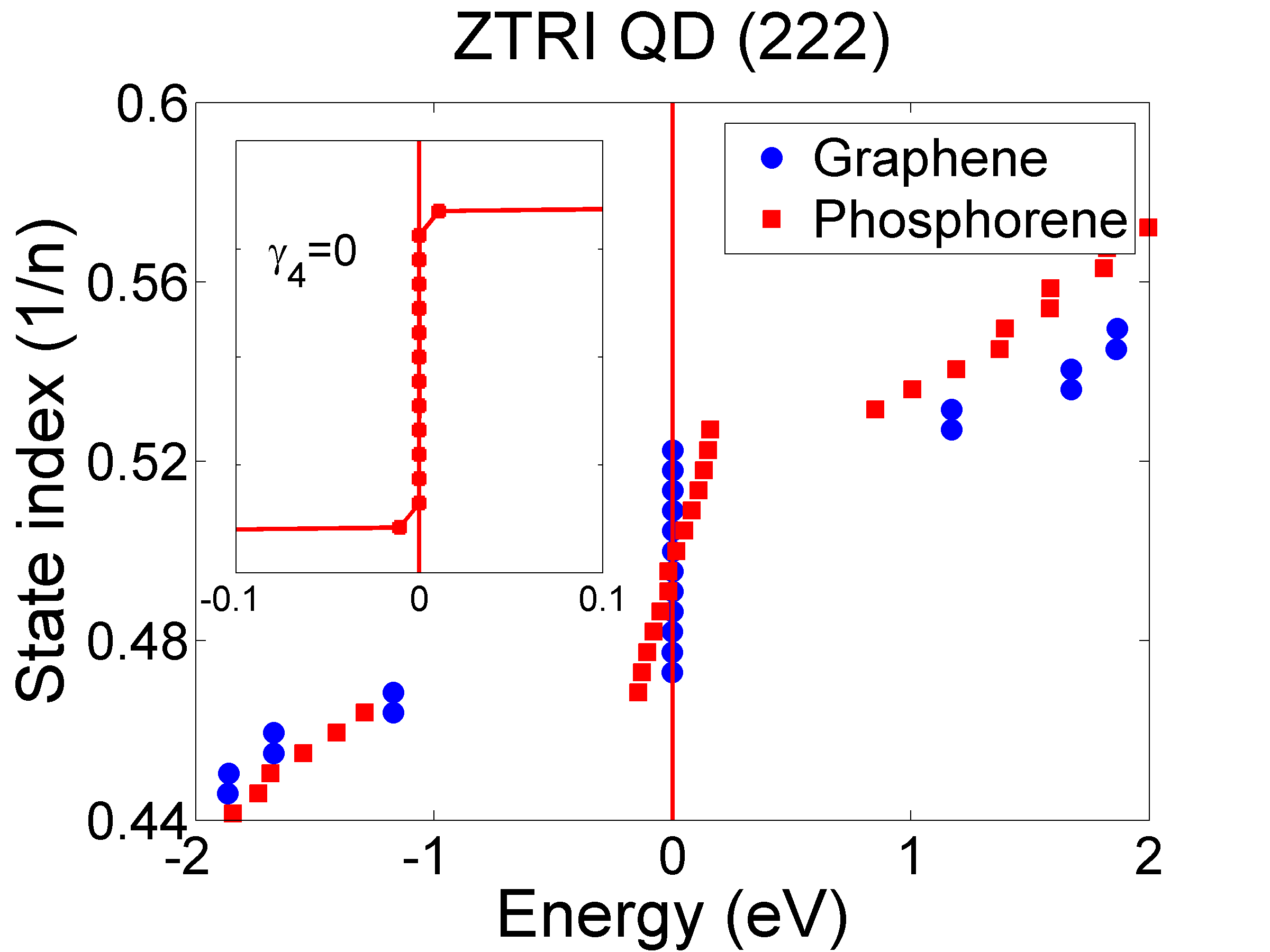}
  \caption{}
\end{subfigure}%
\begin{subfigure}{.25\textwidth}
  \centering
 \centerline{\includegraphics[width=\textwidth]{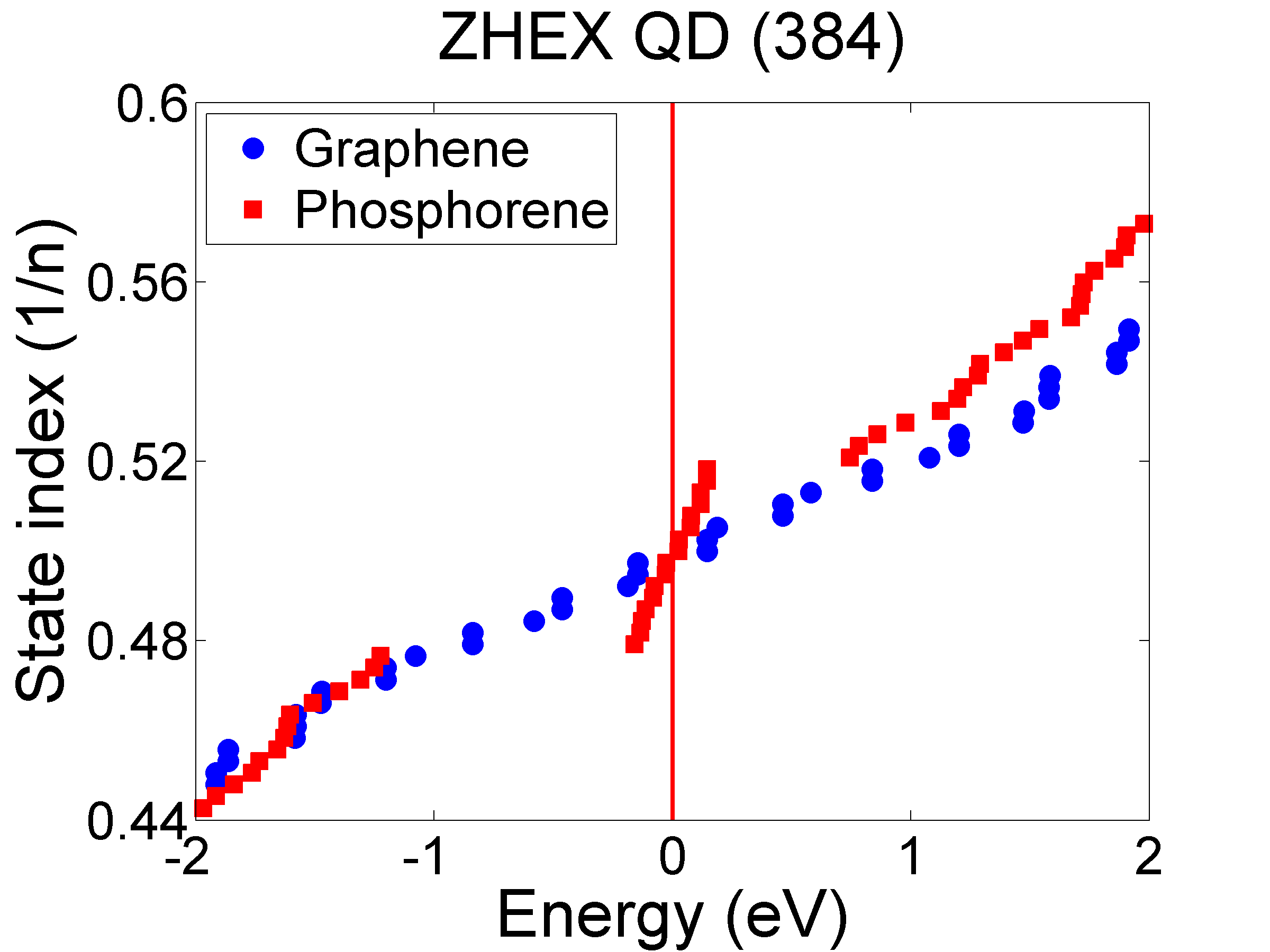}}
  \caption{}
\end{subfigure}

\begin{subfigure}{.25\textwidth}
  \centering
  \includegraphics[width=\textwidth]{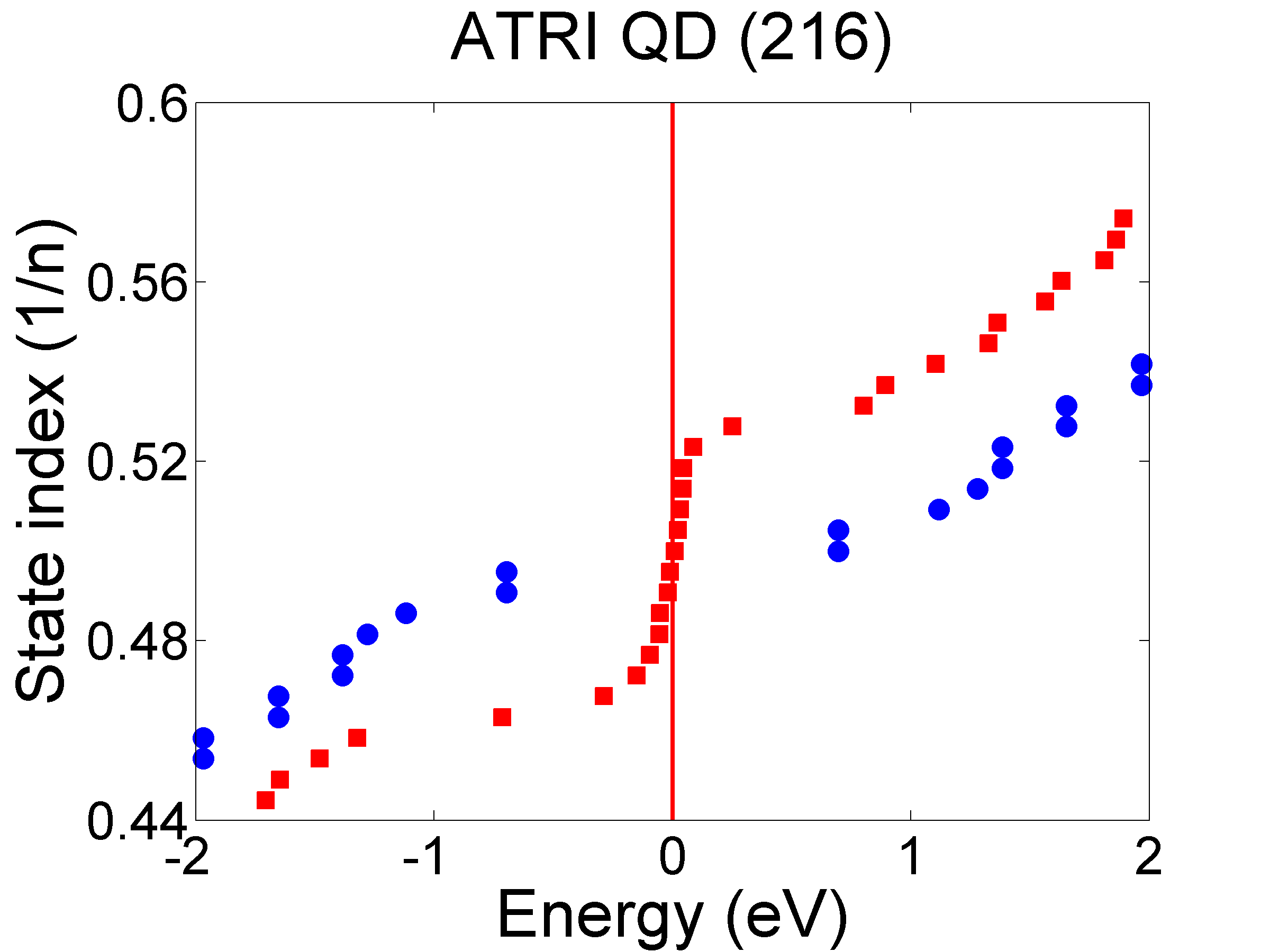}
  \caption{}
\end{subfigure}%
 \begin{subfigure}{.25\textwidth}
  \centering
 \centerline{\includegraphics[width=\textwidth]{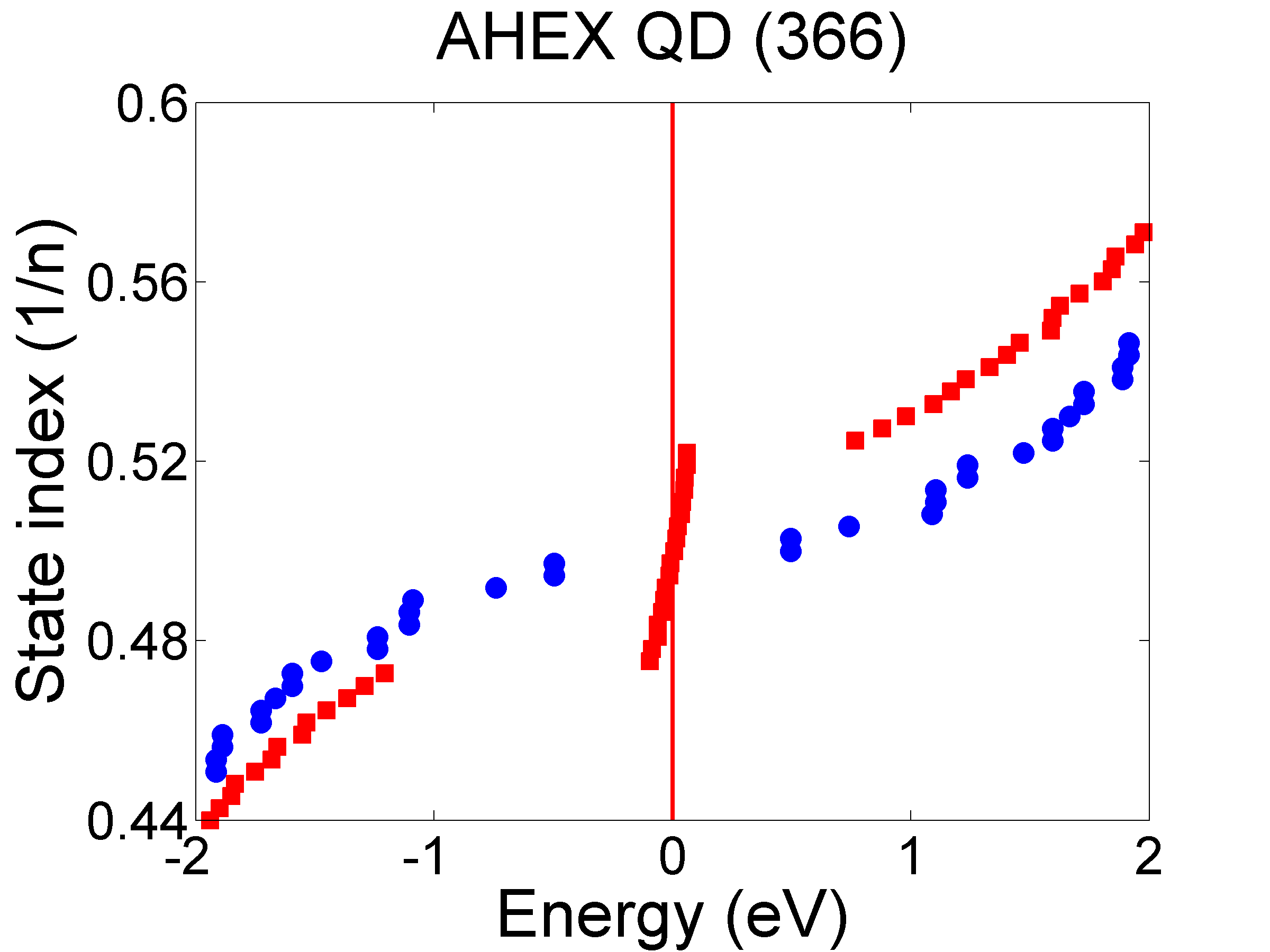}}
  \caption{}
\end{subfigure}%
\caption{Energy levels of phosphorene QDs versus graphene QDs for triangular and hexagonal shapes with both zigzag and armchair termination.}
\label{fig:EnergyLevelsPQDsVsGQDss}
\end{figure}

We start with the comparison of the phosphorene quantum dots' (PQDs) energy levels  with  those of graphene quantum dots (GQDs) as presented in Fig.~\ref{fig:EnergyLevelsPQDsVsGQDss}. A peculiar group of energy states is observed in the low-energy (close to $E=0$~eV) part of the spectrum in all the selected cluster shapes. These states do not exist in most of their counterparts -- graphene quantum dots. As one can see, they completely modify the electronic properties of PQDs compared to GQDs. For instance, the group of states in ATRI phosphorene QDs totally fills the energy gap, providing conducting armchair phosphorene QDs (Fig.~\ref{fig:EnergyLevelsPQDsVsGQDss} (c)) in contrast to  ATRI  graphene QDs,  where the energy gap ensures the semiconducting behaviour. The states dispersed near the Fermi level of an undoped dot, i.e. $\varepsilon_{F} =0$~eV, are localized at the structure edges. In what follows we refer to these edge states in PQDs as quasi-zero energy states (QZES) and denote the number of such states by $N_{\text{QZES}}$. 

By setting the coupling parameter $t_4 = 0$ we reveal the origin of the dispersion asymmetry of the QZES. As seen in the inset of Fig.~\ref{fig:EnergyLevelsPQDsVsGQDss} (a), when $t_4 = 0$, the asymmetry disappears and $N_{\text{QZES}}$ splits into two sets. The first set contains $12$ ZES positioned exactly at $\varepsilon_{F}$ as in graphene, whereas the second set contains only $2$ states that are symmetrically arranged with respect to $\varepsilon_{F}$: one from the conduction band and another from the valance band. From Figure~\ref{fig:EnergyLevelsPQDsVsGQDss}(a) we conclude that there are $12$ ZES in GQD with $n = 222$ and there are $14$ edge states smeared asymmetrically around the Fermi level in the PQD when all the coupling parameters are included. Thus, unlike in triangular graphene quantum dots where $N_{\text{ZES}}=N-1$~\cite{Guclu2016} in corresponding PQDs we have $N_{\text{QZES}}=N+1$.

The origin of the QZES in ZTRI PQDs can be found by setting the coupling parameters $t_3 = t_4 = t_5 = 0$  and varying $t_{2}$ with respect to $t{_1}$~\cite{Montambaux2009,Pereira2009,Ezawa2014}.
Obviously, when $t_1=t_2$, the total number of edge states (ZES and QZES) is equal for graphene and phosphorene ZTRI QDs. At $t_1 = -1.22$~eV and $t_2 = 2$~eV, the  number of ZES is still the same as in graphene but the conduction and valance states in PQD  move towards the ZES. With the continuous increase of $t_2$, the two energy states, one from the valence band and the other from conduction band, become well separated from their bands and move towards ZES which decreases the energy gaps between ZES and valence and conduction bands. As $t_2$ increases to $3.665$~eV~\cite{Rudenko2014} the number of edge states increases to $14$ states symmetrically separated from  conduction and valence by $\varepsilon_g \simeq 1.5$~eV.  We  found that the  number of states (two states) split-off from the bulk bands does not depend on the cluster size.

Adding $t_3$ and $t_5$ decreases the energy gap between the edge states and bulk states from $1.5$~eV to $ 1.2$~eV with no change to the distribution of the QZES.

In Figure~\ref{fig:EnergyLevelsPQDsVsGQDss}(b), the energy levels of the hexagonal graphene QD with zigzag termination and $n=384$ are compared with those of the corresponding phosphorene QD.
We note that for this small size ($n=384$) the ZHEX graphene QD has no edge states, whereas for the same size ZHEX phosphorene QD there are $16$ edge states smeared around the Fermi energy. To investigate the origin of QZES in hexagonal zigzag phosphorene we apply the same strategy as used in triangular phosphorene clusters. At $t_1=t_2$ it has the same energy spectrum as for hexagonal graphene clusters. However, at $t_2=2$~eV a new set of energy states ($16$ energy levels for $n=384$ atoms) fills the energy gap. Increasing $t_2$ to $3.665$~eV leads to gathering of the $16$ states with a very small dispersion forming edge states isolated from the bulk bands by $\varepsilon \simeq 1.4$~eV. The effect of $t_3$ and $t_5$ is the same as in triangular clusters, i.e. the decreasing of the energy gap between edge states and bulk states. Introducing $t_4 = -0.105$~eV generates the antisymmetric displacement of the edge states with respect to the bulk states and a small increase in their dispersion. 

The number of new edge states, $N_{\text{QZES}}$, in ZHEX-phosphorene dots increases by increasing the size of the cluster. It is given by $N_{\text{QZES}} =2N$, where $N$ is the number of hexagons at the edge.

Figure~\ref{fig:EnergyLevelsPQDsVsGQDss}(c) shows a comparison between the energy levels of armchair triangular quantum dots of graphene and phosphorene with $n=216$. In the ATRI graphene QD  there is a noticeable energy gap $\varepsilon_g \simeq 1.3$~eV due to the size effect,  while in the ATRI phosphorene QD the energy gap disappears. QZES in triangular armchair phosphorene QDs are dispersed inside the energy gap (Fig.~\ref{fig:EnergyLevelsPQDsVsGQDss}(c) red squares) giving rise to a  cluster with zero energy gap. The total number of edge states is $2N$, similar to the case of ZHEX-phosphorene QDs.  Figure~\ref{fig:EnergyLevelsPQDsVsGQDss}(d) compares the QZES in hexagonal armchair phosphorene QDs to hexagonal graphene QDs with $n=366$. The total number of QZES in an AHEX phosphorene quantum dot is $N_{\text{QZES}} = 2(2N-1)$.

\begin{table}[htbp]
\caption{The number of quasi-zero energy states as a function of the quantum dot size for various dot shapes. $N_{\text{QZES}} = N_{\mathrm{top}} + N_{\mathrm{bottom}} = |N_2-N_1|$, where $N_{\mathrm{top, bottom}}$ are the number of edge atoms and $N_{1,2}$ are the total number of atoms in the top and bottom layers of the phosphorene dot, respectively. $N$ is the number of hexagonal elements at the edge as shown in Fig.~\ref{fig:PhosphoreneQuantumDotsClassification}.}
\begin{tabular}{cccccc} \hline \hline
   & \multicolumn{4}{c}{Quantum dot type} \\ 
   & ZTRI & ZHEX & ATRI & AHEX \\ \hline
   $N_{\text{QZES}}$ & $ N + 1 $ & $2N $ & $2N$ & $2(2N-1) $  \\ 
   
   $N$ & $\sqrt{n+3} - 2$ & $\sqrt{\dfrac{n}{6}}$ & $\dfrac{\sqrt{12n + 9} - 3}{6}$ & $\dfrac{\sqrt{2n - 3} + 3}{6}$ \\ \hline \hline
\end{tabular}
\label{tab:QZESTable}
\end{table}

Thus, we conclude that the origin of QZES is the distribution of the phosphorene atoms in two layers and  $t_2 > t_1$. Table~\ref{tab:QZESTable} summarizes the relations between the number of QZES, $N_{\text{QZES}}$, and the structure size for various types of phosphorene QDs. A general rule valid for all PQD types can be formulated as follows: the number of QZES is equal to the total number of atoms, which are not connected to nearest neighbour atoms by $t_{2}$. The number of QZES in all types of phosphorene QDs can also be expressed as the sum of QZES localized at the top ($N_{\text{top}}$) and the bottom ($N_{\text{bottom}}$) layer, i.e. $N_{\text{QZES}} = N_{\text{top}}+N_{\text{bottom}}$.

\begin{figure}[htbp]
\centering
\begin{subfigure}{.25\textwidth}
  \centering
  \includegraphics[width=\textwidth]{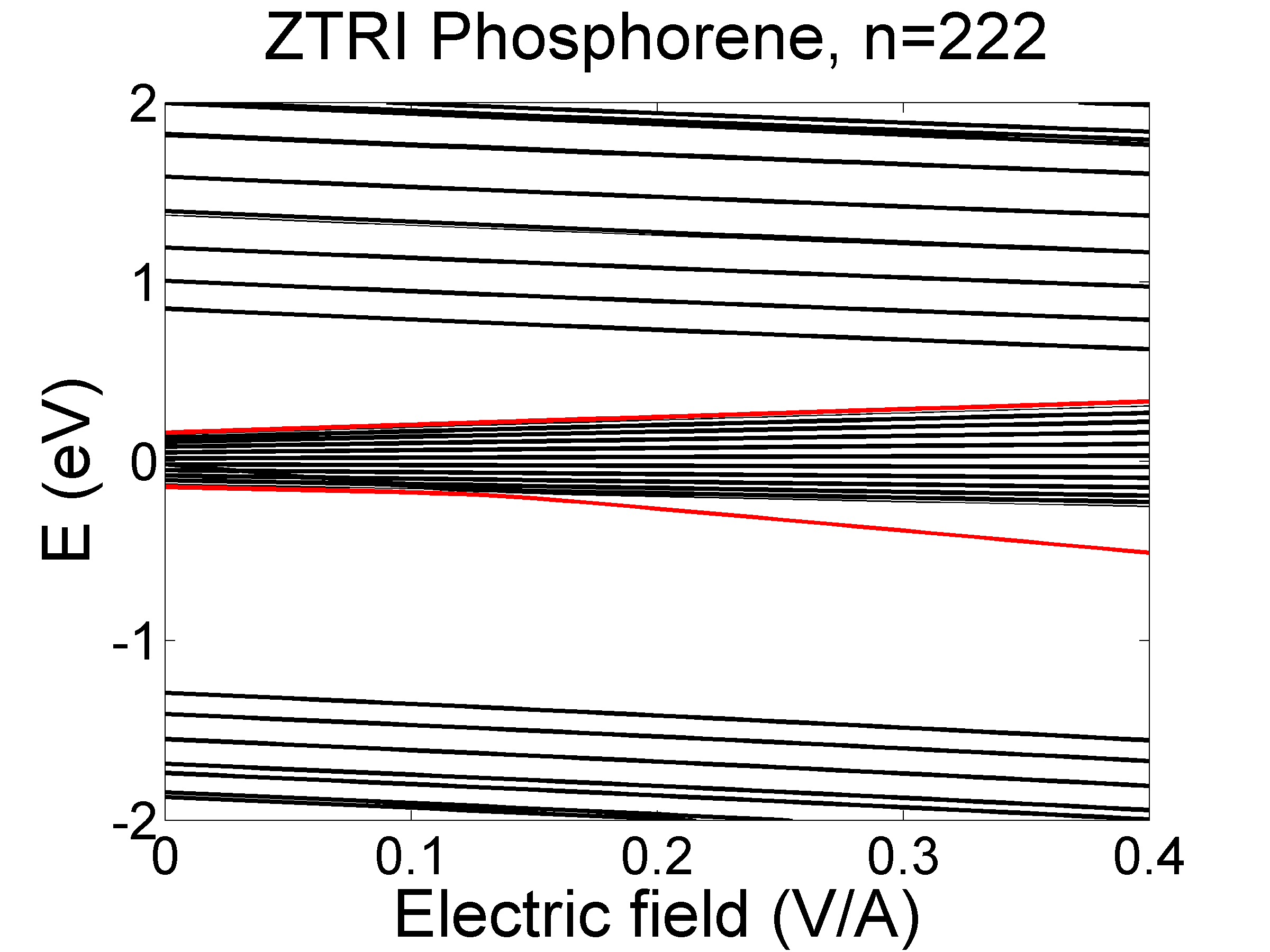}
  \caption{}
\end{subfigure}%
\begin{subfigure}{.25\textwidth}
  \centering
 \centerline{\includegraphics[width=\textwidth]{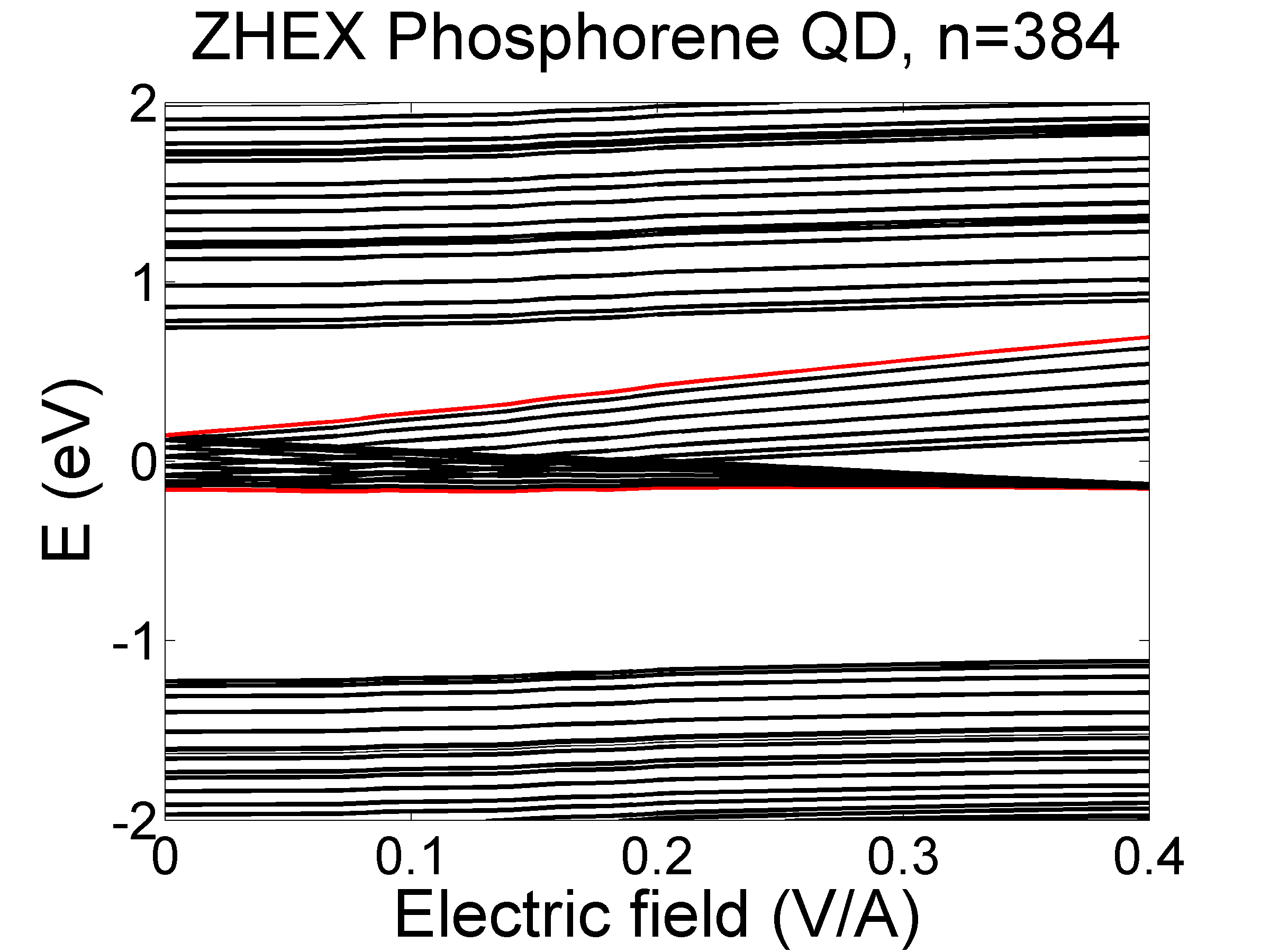}}
  \caption{}
\end{subfigure}

\begin{subfigure}{.25\textwidth}
  \centering
  \includegraphics[width=\textwidth]{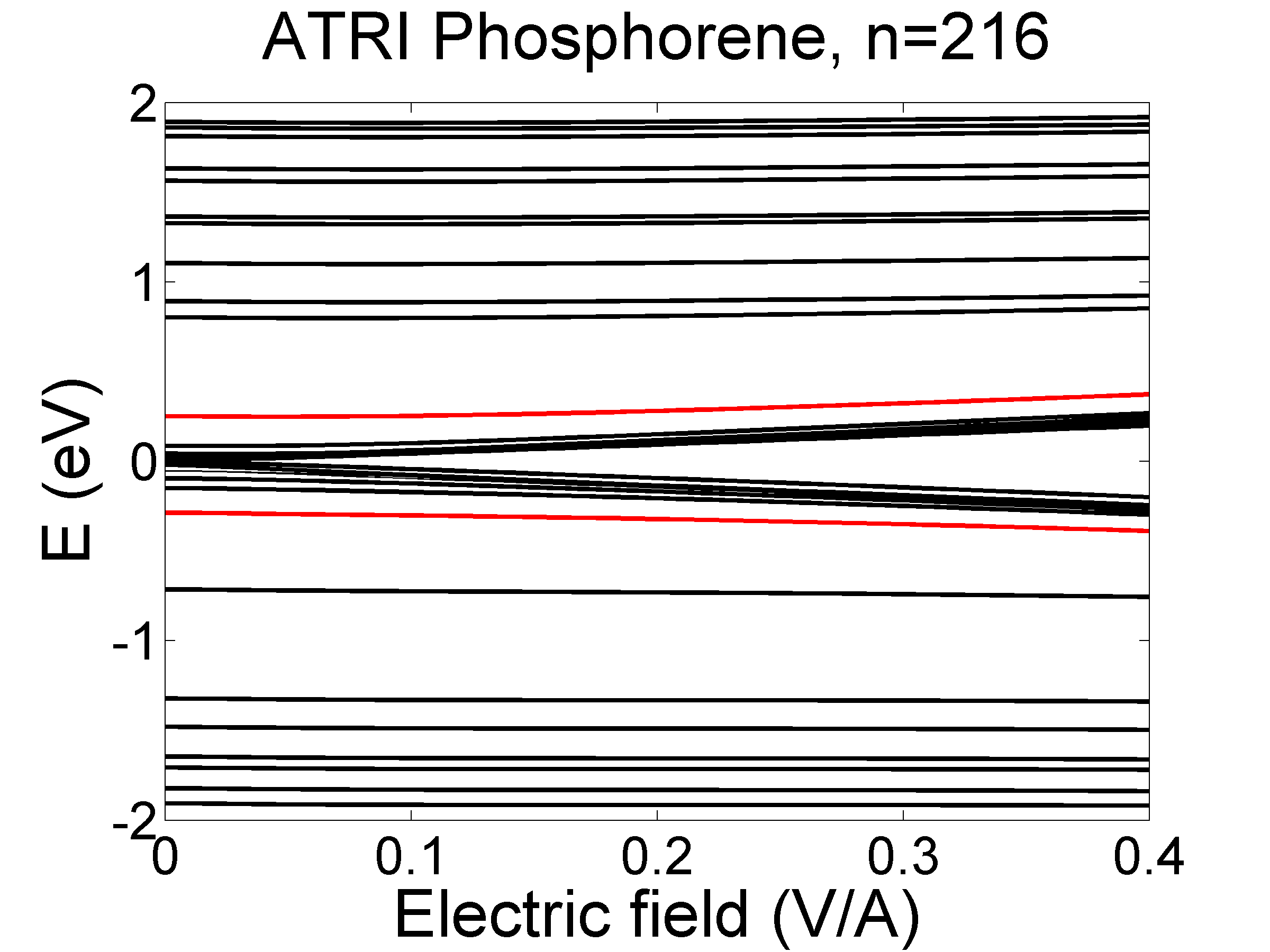}
  \caption{}
\end{subfigure}%
 \begin{subfigure}{.25\textwidth}
  \centering
 \centerline{\includegraphics[width=\textwidth]{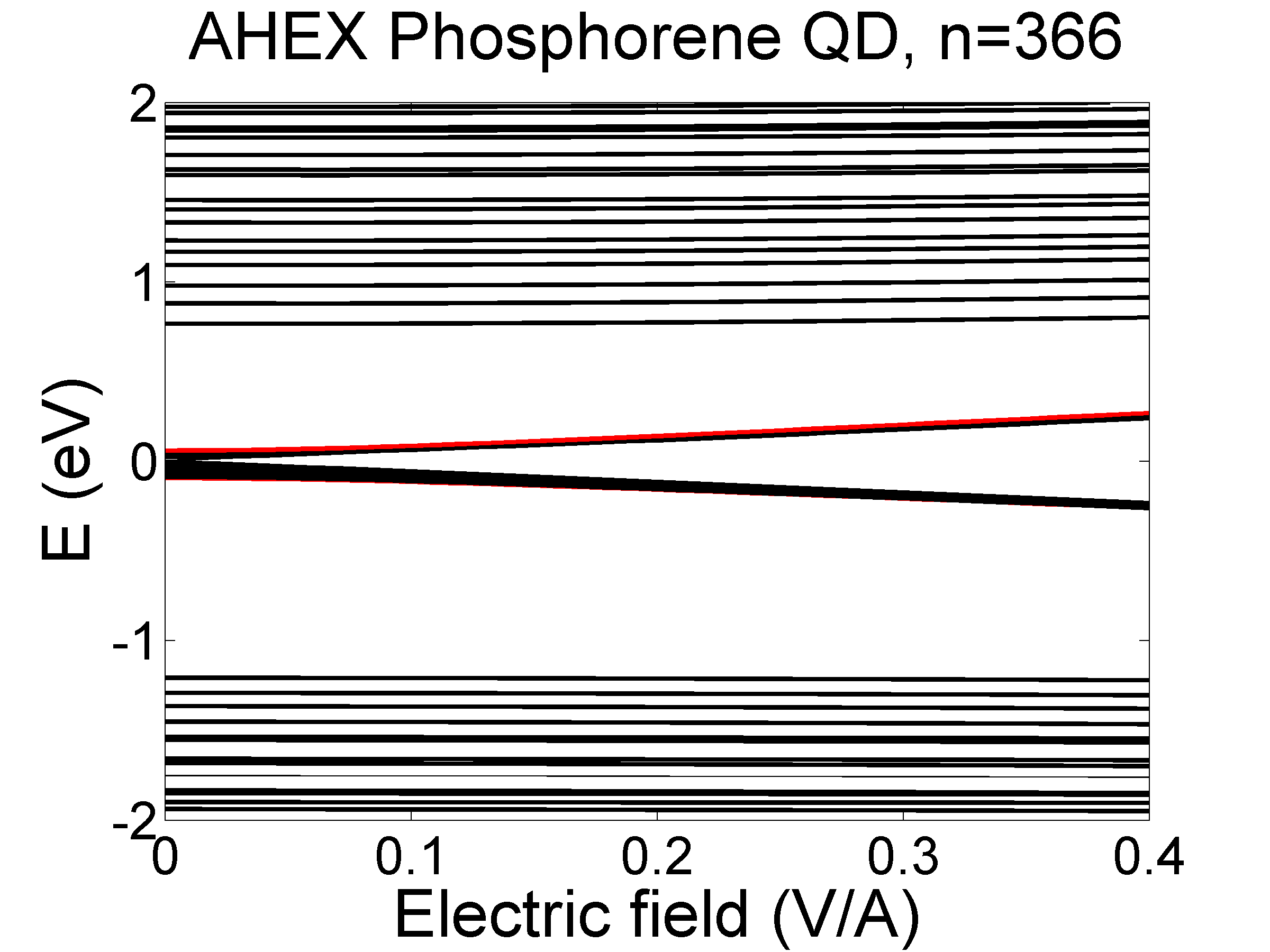}}
  \caption{}
\end{subfigure}%
\caption{The energy levels as a continuous function of the electric field strength for (a) ZTRI; (b) ZHEX; (c) ATRI and (d) AHEX phosphorene quantum dots. }
\label{E-field}
\end{figure}

To study the effect of the electric field on the energy levels of the four types of QDs, we plotted in Fig.~\ref{E-field} their energy levels dependence on the electric fields up to $E=0.4$~V/{\AA}.  The upper limit of the field applied perpendicular to the dot plane was chosen such that it facilitates comparison with the same field geometry for phosphorene nanoribbons~\cite{Grujiс2016}. Similar electric field strength has been also used in the studies of the bilayer phosphorene~\cite{Dai2014,Dolui2015}. Notably, in all the dots except the ZTRI ones the conduction and valence band states are almost insensitive to the field, thereby the primary effect comes form the QZES. An important difference between quantum dots with zigzag and armchair edges can be seen. 
The latter exhibit splitting of the QZES in two dense groups of states [Fig.~\ref{E-field}(c, d)], whereas in the former QZES split into dense and dispersed groups[Fig.~\ref{E-field}(a, b)]. In the case of ZHEX quantum dots, presented in Fig.~\ref{E-field}(b), the dense group contains the same number of energy states as the dispersed one. For ZTRI quantum dots only one state split from the main dispersed group as shown in Fig.~\ref{E-field}(b). In all the cases the states from different groups belong to different layers of the phosphorene structure.

\subsection{Optical properties}
In order to study the effect of puckering on the optical properties of PQDs we compare in this section the absorption spectra of different phosphorene QDs and the spectra of similarly shaped graphene QDs. The optical absorption cross-section, $\sigma(\varepsilon)$, was calculated for $x$-, $y$-, and $z$-polarizations of the incident electromagnetic wave. Throughout this paper the quantum dot orientation with respect to the coordinate system is fixed as presented in Fig.~\ref{fig:PhosphoreneQuantumDotsClassification}, Gaussian broadening $\alpha = 0.02$~eV, and temperature $T = 0$~K. In the case of graphene it was found that $\sigma_{x}$ and $\sigma_{y}$ are almost the same, whereas $\sigma_{z} = 0$~\cite{Yamamoto2006}, therefore for graphene QDs we consider only the $\sigma_{x}$ cross-section. In contrast to this, the absorption cross-sections due to $x$- and $y$-polarizations are considerably different in phosphorene QDs. The absorption of a $z$-polarized incident wave is tiny compared to that of  $x$- or $y$-polarizations therefore it is not discussed hereafter. 
\begin{figure}[htbp]
\centering
\begin{subfigure}{.25\textwidth}
  \centering
  \includegraphics[width=\textwidth]{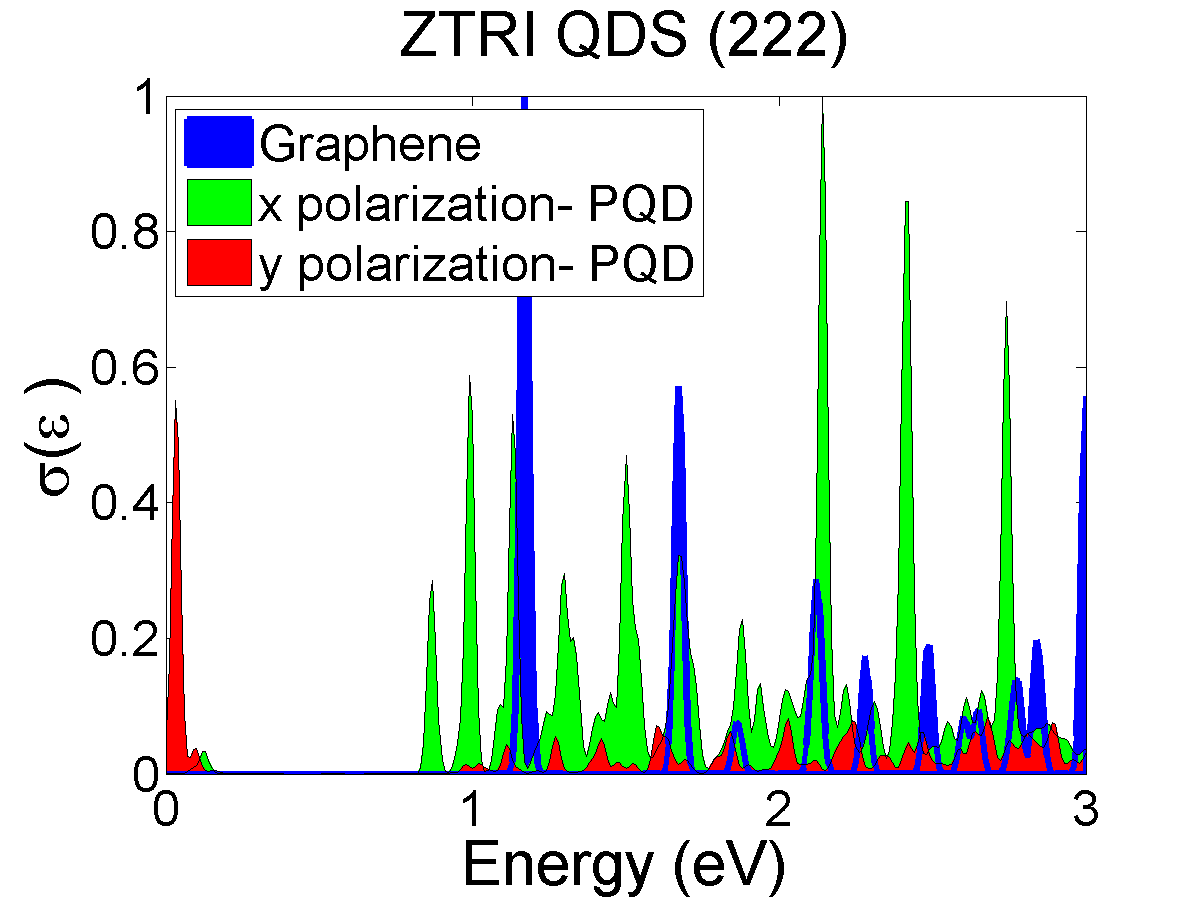}
  \caption{}
\end{subfigure}%
\begin{subfigure}{.25\textwidth}
  \centering
 \centerline{\includegraphics[width=\textwidth]{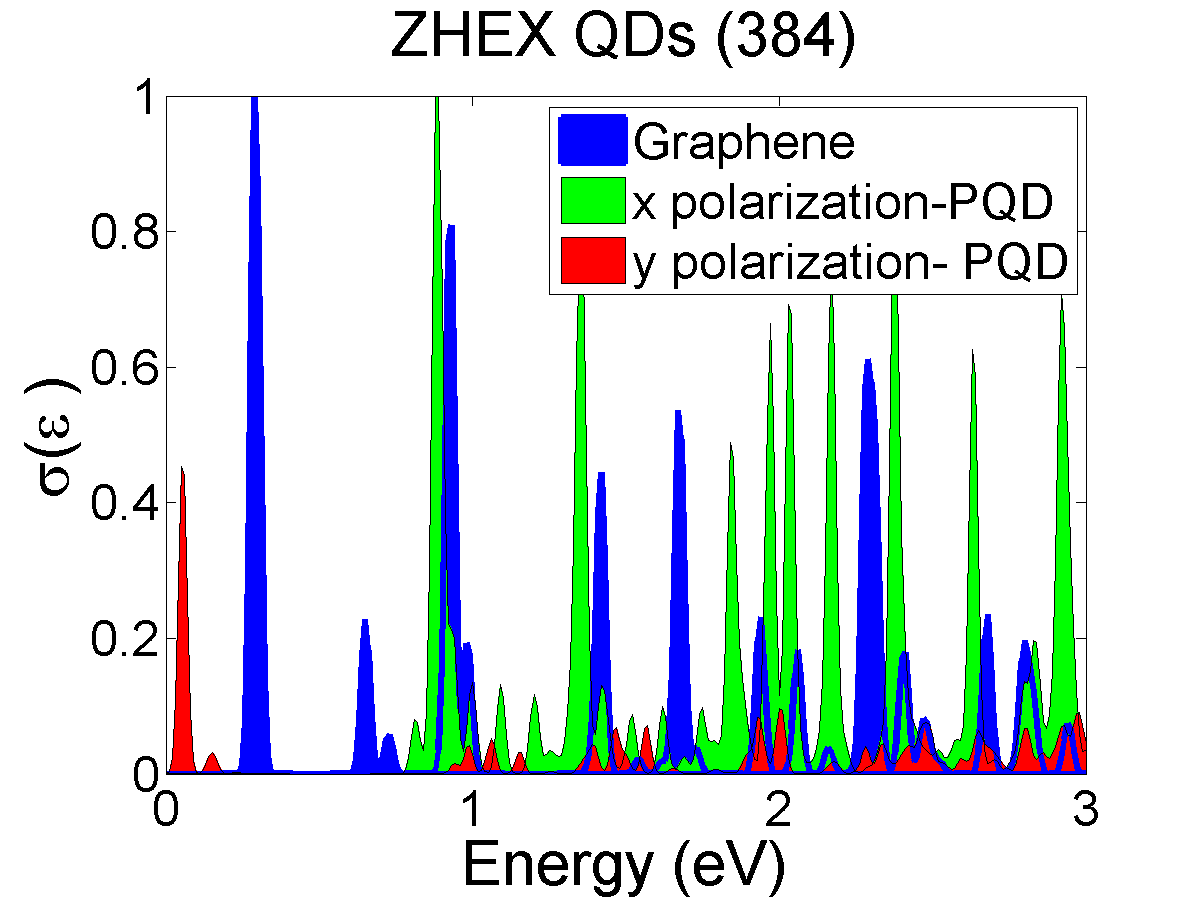}}
  \caption{}
\end{subfigure}

\begin{subfigure}{.25\textwidth}
  \centering
  \includegraphics[width=\textwidth]{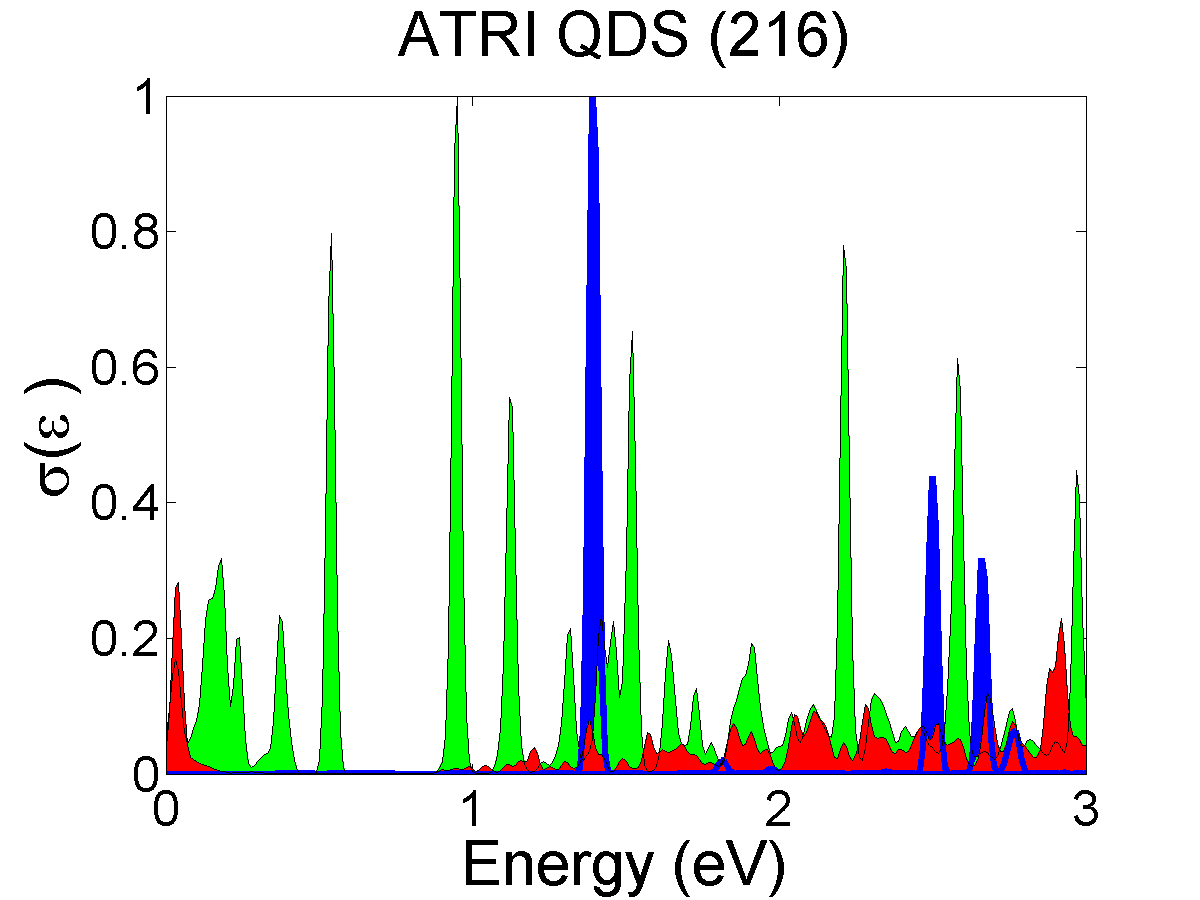}
  \caption{}
\end{subfigure}%
 \begin{subfigure}{.25\textwidth}
  \centering
 \centerline{\includegraphics[width=\textwidth]{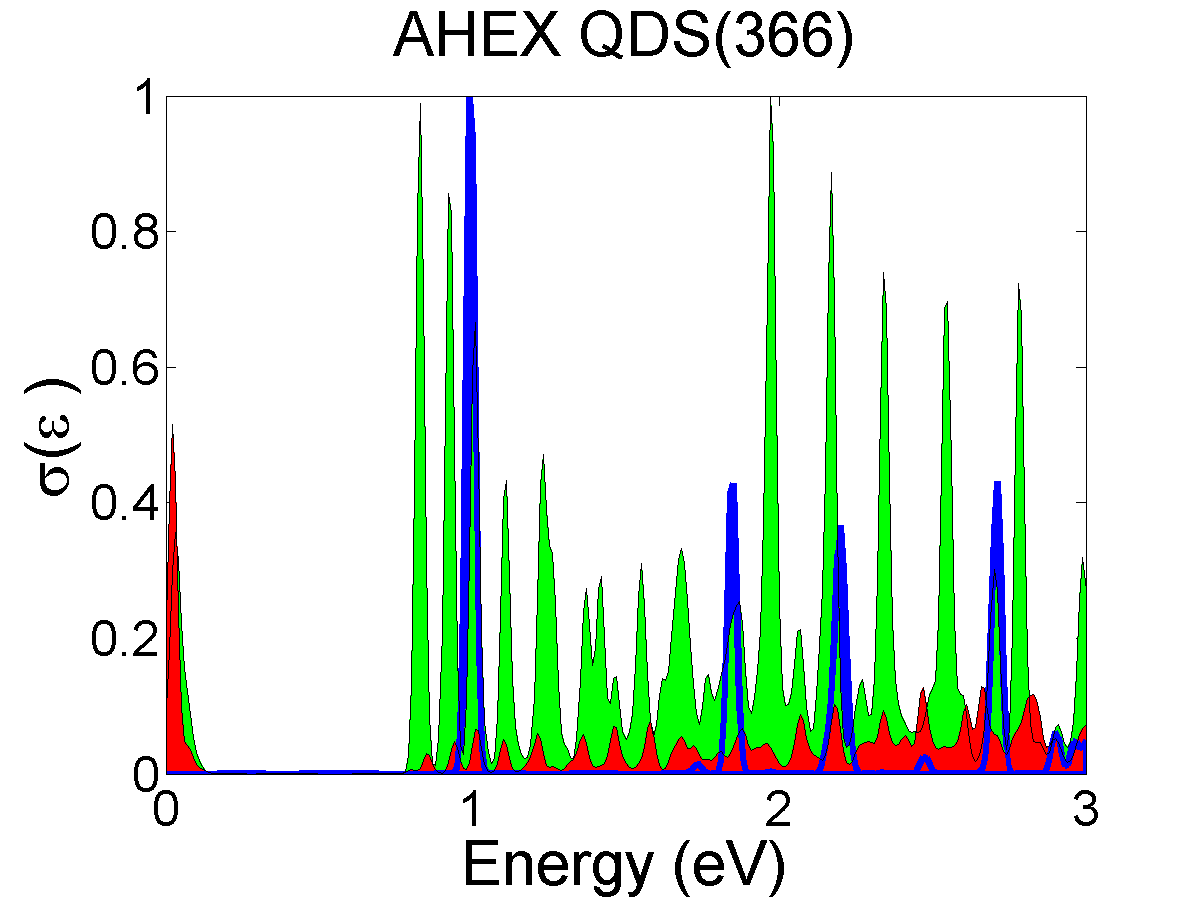}}
  \caption{}
\end{subfigure}%
\caption{Optical absorption cross-sections of different graphene QDs versus the corresponding phosphorene QDs.}
\label{fig:OpticalAbsorptionGQDsvsPQDs}
\end{figure}

Figure~\ref{fig:OpticalAbsorptionGQDsvsPQDs} shows the absorption spectra of various graphene and phosphorene QDs: (a) ZTRI, (b) ZHEX, (c) ATRI, and (d) AHEX QDs. The blue absorption peak at $\varepsilon \approx 1.2$~eV in Fig.~\ref{fig:OpticalAbsorptionGQDsvsPQDs} (a) represents the optical transition from  the highest occupied energy level (HOEL) in valance band to ZES and from ZES to the lowest unoccupied energy level (LUEL) in the conduction band~\cite{Abdelsalam2016}. This single blue peak in graphene  QDs is replaced by  four green  peaks representing the same transitions in phosphorene QDs as can be seen from Fig.~\ref{fig:OpticalAbsorptionGQDsvsPQDs} (a). The three peaks  between $\varepsilon \approx 0.8$~eV and  $\varepsilon \approx 1.2$~eV originate from transitions between the HOEL and the smeared group of QZES and between these states and the LUEL. The $y$-absorption  peak around $\varepsilon \approx 0.05$~eV represents transition between the QZES. Another important difference between phosphorene and graphene QDs is that the number of absorption peaks in the low energy region is higher in phosphorene QDs than graphene QDs due to the wider spread of QZES. This dense absorption spectrum may be useful for the detection of a wider spectrum of incident wavelengths.

In the case of ZHEX flakes, presented in Fig.~\ref{fig:OpticalAbsorptionGQDsvsPQDs} (b), one can see a blue shift of the absorption edge in the phosphorene QD with $n=384$ compared to the position of the absorption edge in the corresponding graphene QD. This shift is due to the fact that the energy gap in the PQD is larger than that in the GQD. As can be seen from Fig.~\ref{fig:EnergyLevelsPQDsVsGQDss} (b), in the ZHEX PQD the gap between QZES and bulk conduction and valance bands is $\varepsilon_g \approx 1$~eV, whereas for the ZHEX graphene QD with $n=384$ the energy gap between the conduction and valence bands is $\varepsilon_g \approx 0.2$~eV.

The optical absorption cross-section of ATRI QDs is shown in Fig.~\ref{fig:OpticalAbsorptionGQDsvsPQDs} (c). The absorption gap which was a characteristic feature of the armchair triangular graphene QDs, totally disappears in the puckered phosphorene QDs and an almost continuous absorption spectrum in the region $\varepsilon < 1$~eV  is observed for triangular phosphorene clusters with armchair terminations. 

The absorption cross-sections of AHEX QDs presented in Fig.~\ref{fig:OpticalAbsorptionGQDsvsPQDs} (d) also has an increased number of absorption peaks and low-energy $y$-polarized peaks in the absorption gap for the AHEX PQD  which is absent in the AHEX GQD. The plots of different PQDs show a considerable difference between the absorption spectra for incident wave polarized in $x$-and $y$-direction. The  linear dichroism observed in selected PQDs is due to the anisotropic nature of phosphorene~\cite{Tran2014,Yuan2015}. It makes phosphorene QDs an ideal medium for optical polarizers required in various optoelectronics applications~\cite{Nan2009,Qiao2014}.
\begin{figure}[htbp]
\centering
\begin{subfigure}{.25\textwidth}
  \centering
  \includegraphics[width=\textwidth]{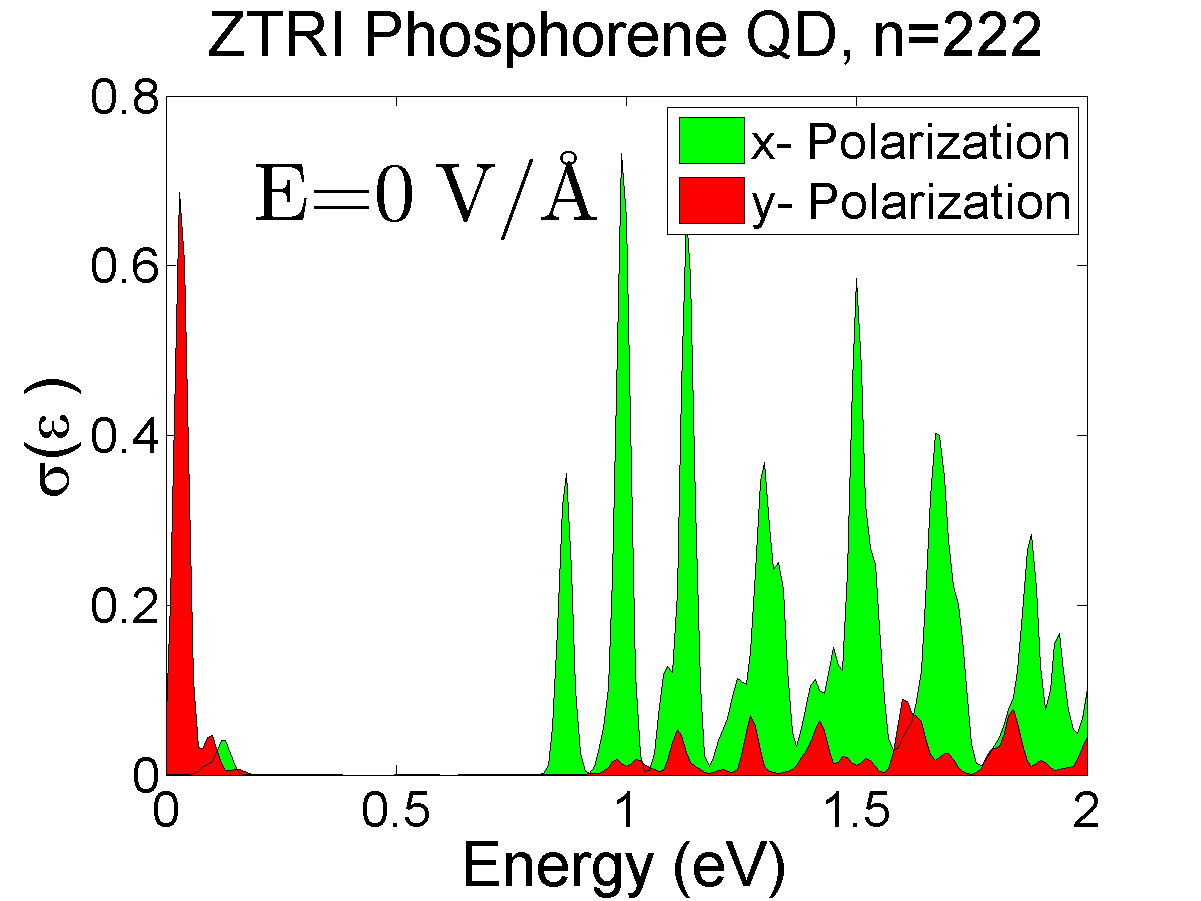}
  \caption{}
\end{subfigure}%
\begin{subfigure}{.25\textwidth}
  \centering
 \centerline{\includegraphics[width=\textwidth]{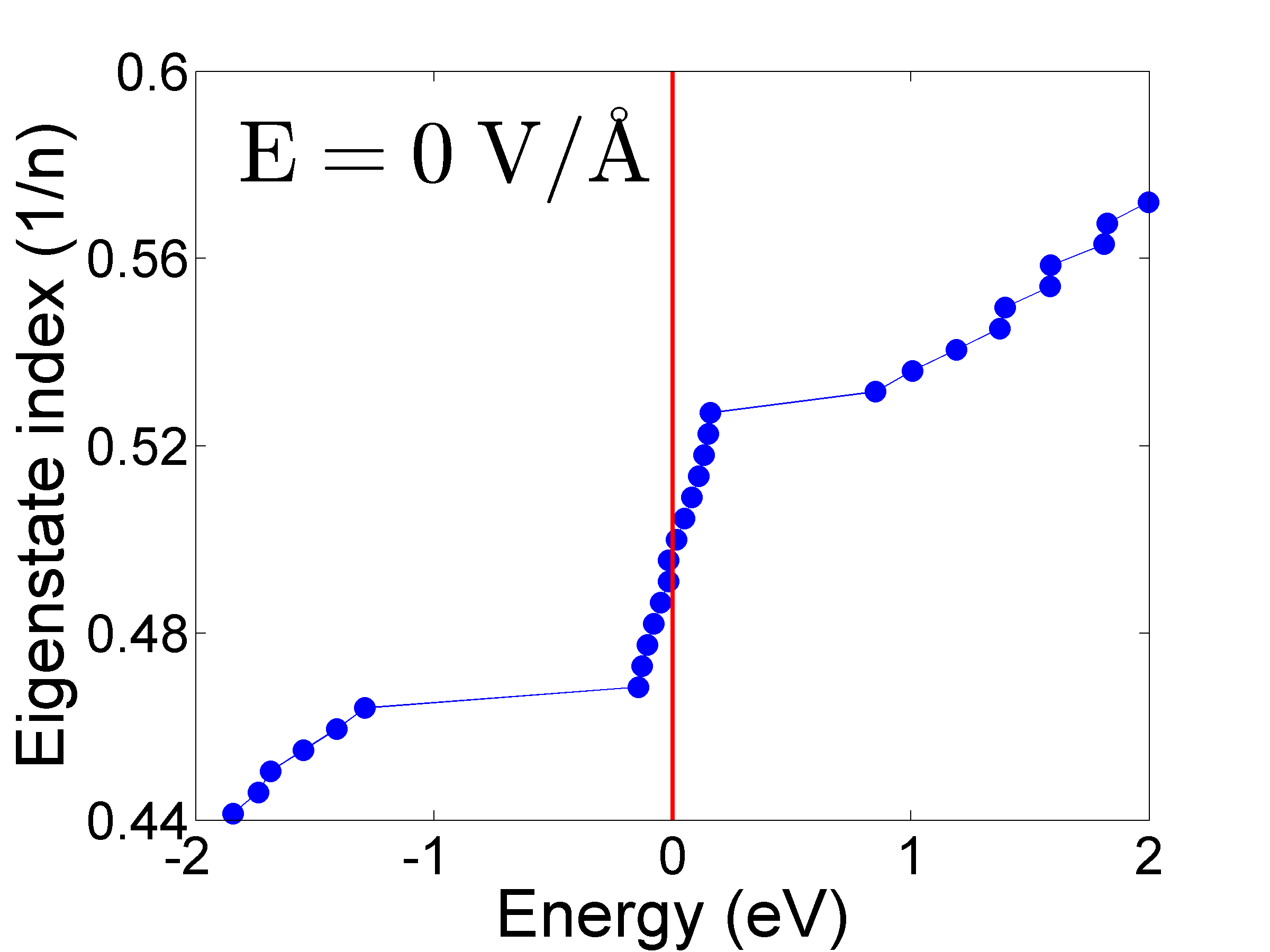}}
  \caption{}
\end{subfigure}

\begin{subfigure}{.25\textwidth}
  \centering
  \includegraphics[width=\textwidth]{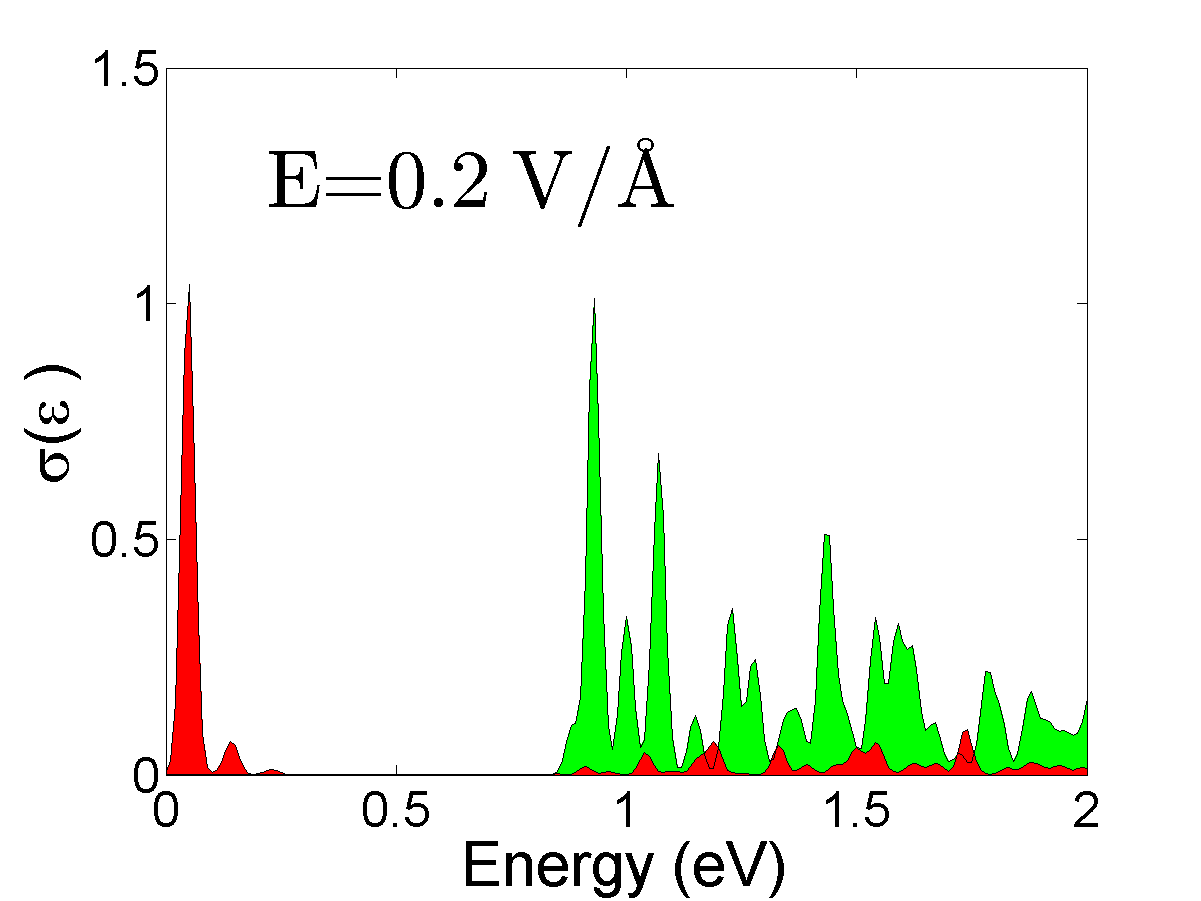}
  \caption{}
\end{subfigure}%
 \begin{subfigure}{.25\textwidth}
  \centering
 \centerline{\includegraphics[width=\textwidth]{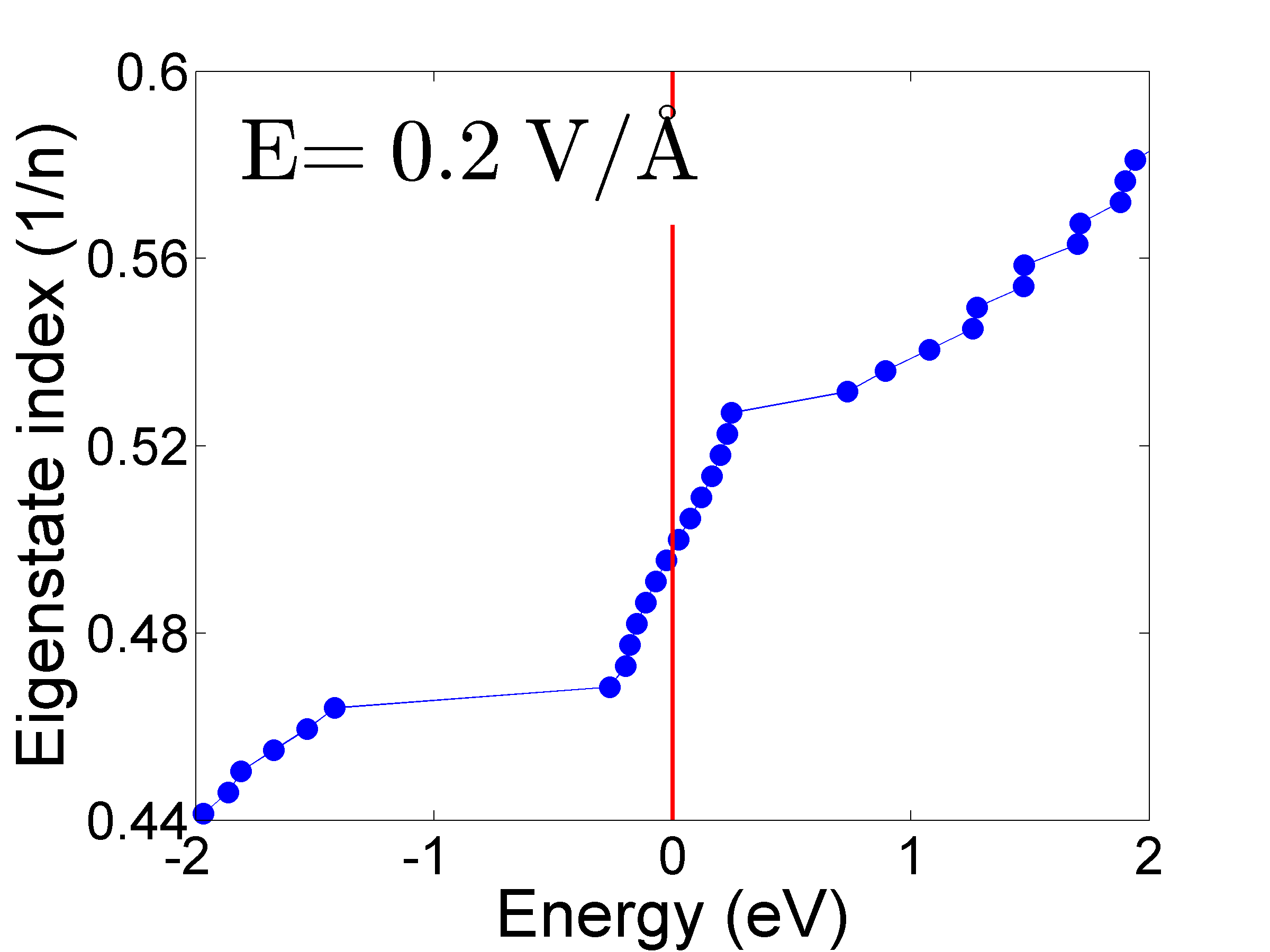}}
  \caption{}
\end{subfigure}%

\begin{subfigure}{.25\textwidth}
 \centering
 \includegraphics[width=\textwidth]{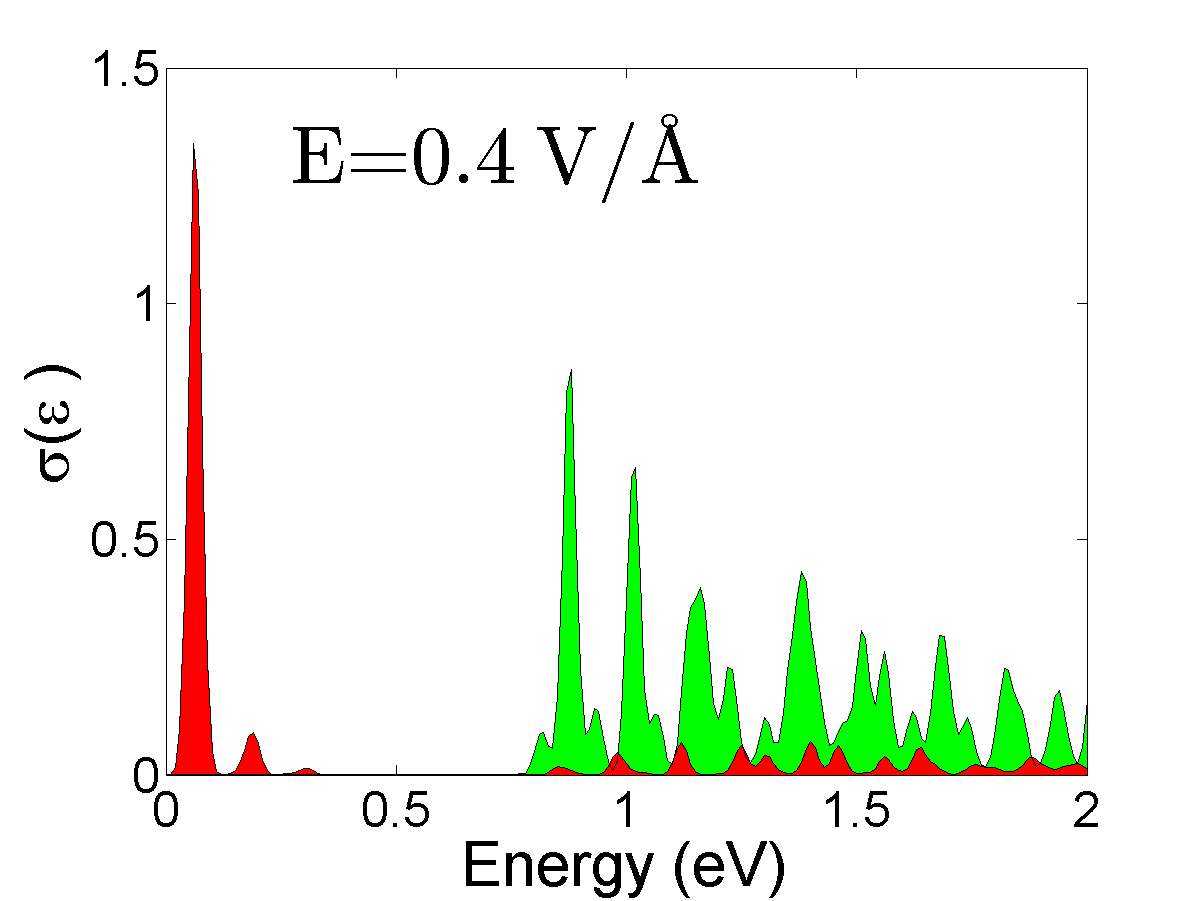}
  \caption{}
\end{subfigure}%
\begin{subfigure}{.25\textwidth}
 \centering
 \centerline{\includegraphics[width=\textwidth]{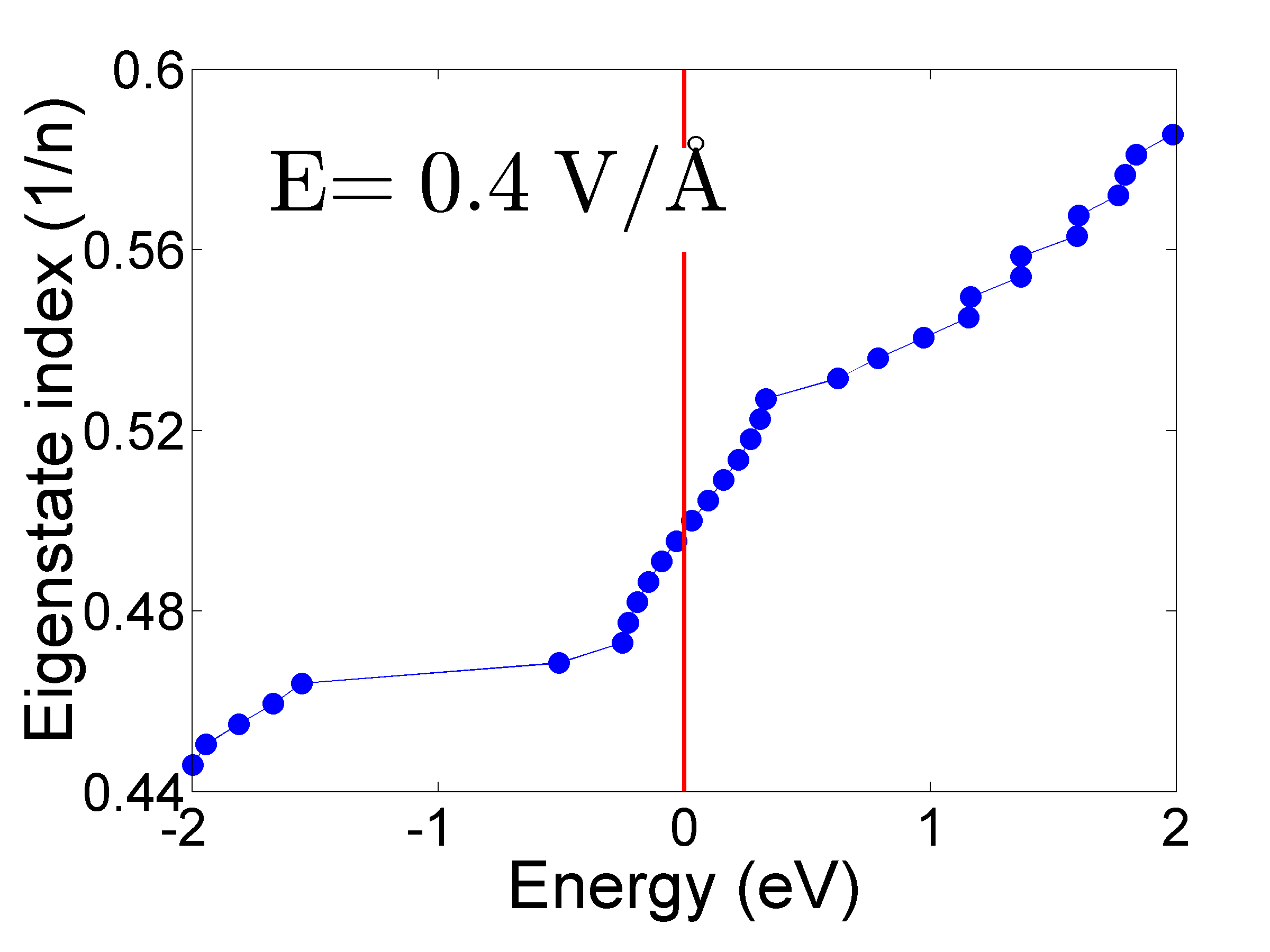}}
  \caption{}
  \label{fig:sub8}
\end{subfigure}
\caption{The effect of a perpendicular electric field on the optical absorption cross-section (a, c, e) and the corresponding energy levels (b, d, f) of zigzag triangular phosphorene QDs.}
\label{fig:fig:AbsorptionZTRIPQD}
\end{figure}

\subsection{\label{sec:ElectricFieldEffect}Electric field effect}
Let us discuss the effect of an electric field applied normally to the structure plane on optical absorption of triangular and hexagonal phosphorene QDs with zigzag and armchair terminations. As in Section~\ref{sec:EnergySpectra}, we choose the electric field strength used in Ref.~\cite{,TaghizadehSisakht2015,Grujiс2016}, $E=0.4$~V/{\AA}, as the upper limit and supplement the obtained results with calculations for $E=0.2$~V/{\AA}. The chosen upper limit is close to the electric field strength used in other studies~\cite{Dai2014,Dolui2015}.

\subsubsection{Zigzag edges}
Optical absorption cross-sections of ZTRI phosphorene QDs for different values of the electric field are shown in Fig.~\ref{fig:fig:AbsorptionZTRIPQD}. It is seen from Fig.~\ref{fig:fig:AbsorptionZTRIPQD} (b, d) that the electric field  increases the QZES dispersion and shifts QZES and conduction band states towards each other.
As can be seen from Fig.~\ref{E-field} (a), at $E>0.2$~V/{\AA} only one QZES moves towards the valance band. The same can also be seen in Fig.~\ref{fig:fig:AbsorptionZTRIPQD} (f). In order to discuss the effect of shifting QZES and conduction band states towards each other on the optical transitions, let us consider the three intense peaks around $\varepsilon = 1$~eV, at $E=0$ V/{\AA} in Fig.~\ref{fig:fig:AbsorptionZTRIPQD} (a). These peaks are due to  transitions from the HOEL to the group of QZES above the Fermi level (peak at $\varepsilon \approx 1.2$~eV) and from the QZES below the Fermi level to the LUEL (peaks at $\varepsilon \approx 0.8$~eV and $\varepsilon \approx 1$~eV ). At $E=0.4$~V/{\AA} the optical transition at  $\varepsilon \approx 1.2$~eV disappears from the low energy absorption and shifts towards higher energies, the positions of the other two peaks at $\varepsilon \approx 0.8$~eV and $1$~eV stay almost the same as shown in Fig~\ref{fig:fig:AbsorptionZTRIPQD} (e). Such a behaviour suggests that the transitions occur from the bottom of the QZES band so that the increase of the QZES dispersion eliminates the effect of the approaching of the conduction bands states towards the QZES as a group. 

\begin{figure}[htbp]
\centering
\begin{subfigure}{.25\textwidth}
  \centering
  \includegraphics[width=\textwidth]{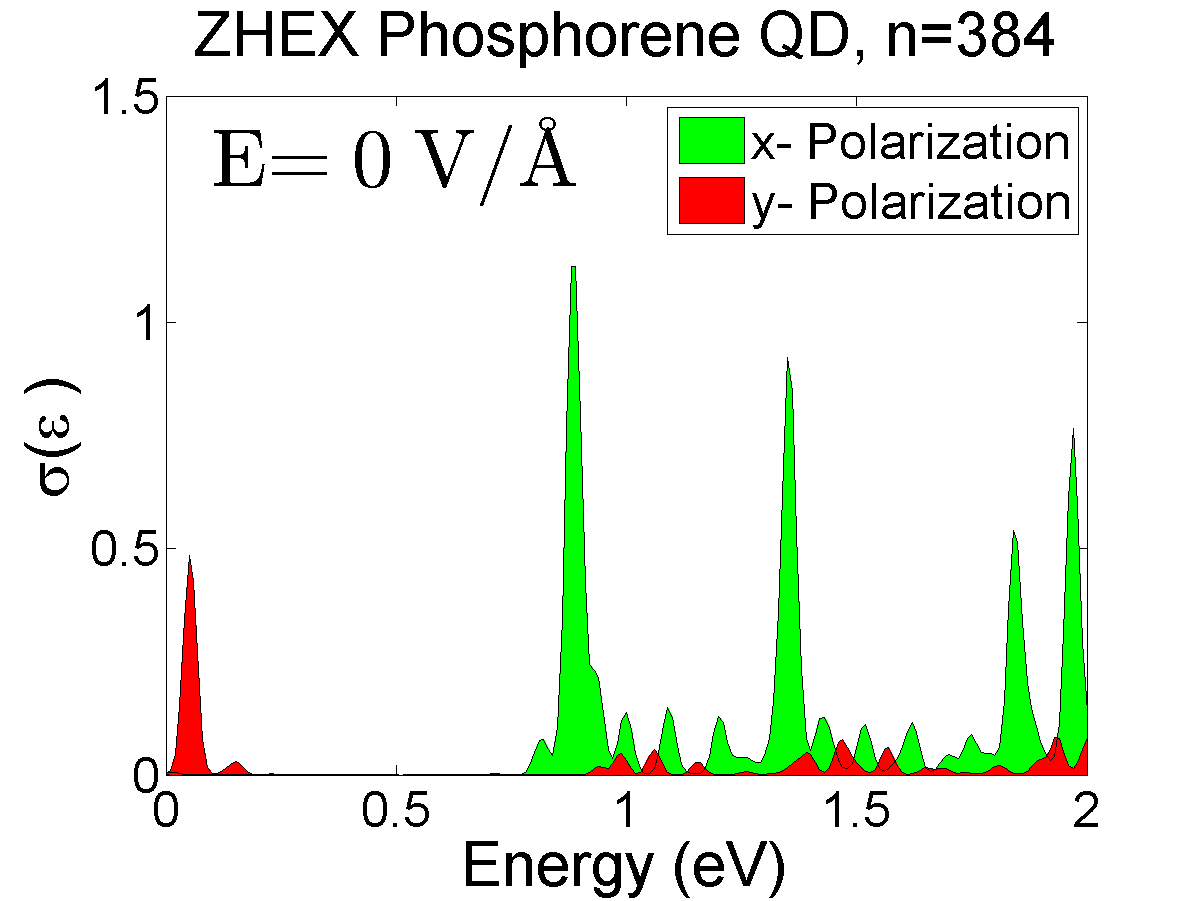}
  \caption{}
\end{subfigure}%
\begin{subfigure}{.25\textwidth}
  \centering
 \centerline{\includegraphics[width=\textwidth]{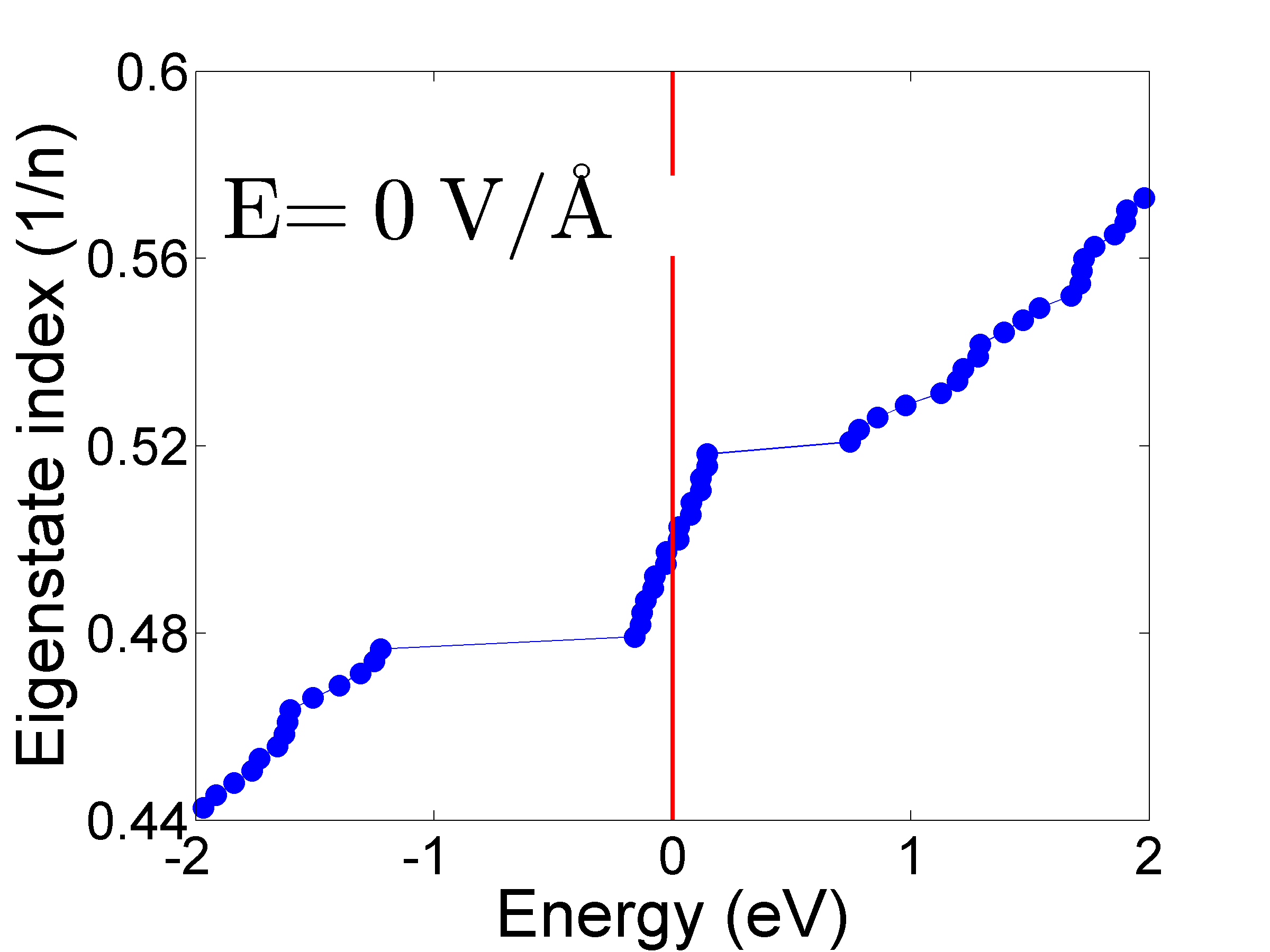}}
  \caption{}
\end{subfigure}

\begin{subfigure}{.25\textwidth}
  \centering
  \includegraphics[width=\textwidth]{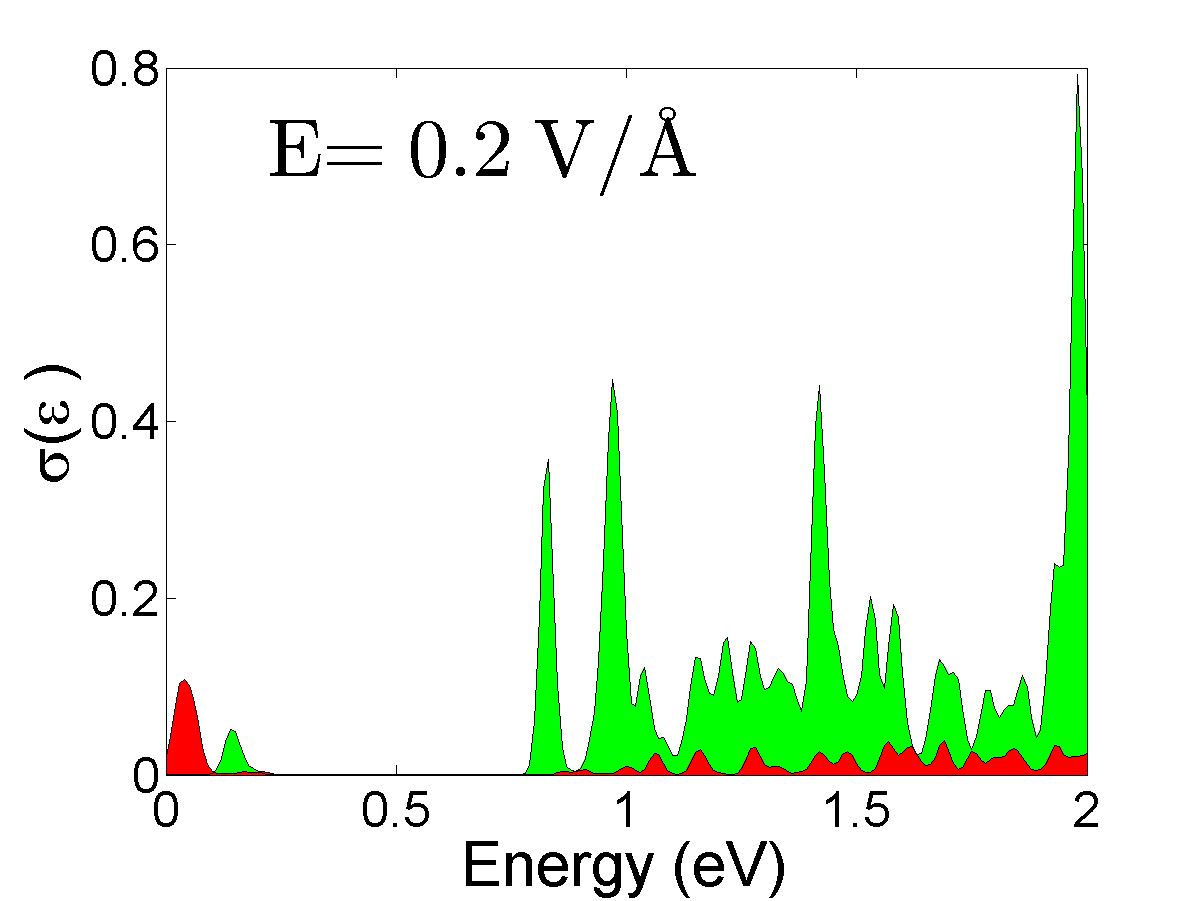}
  \caption{}
\end{subfigure}%
 \begin{subfigure}{.25\textwidth}
  \centering
 \centerline{\includegraphics[width=\textwidth]{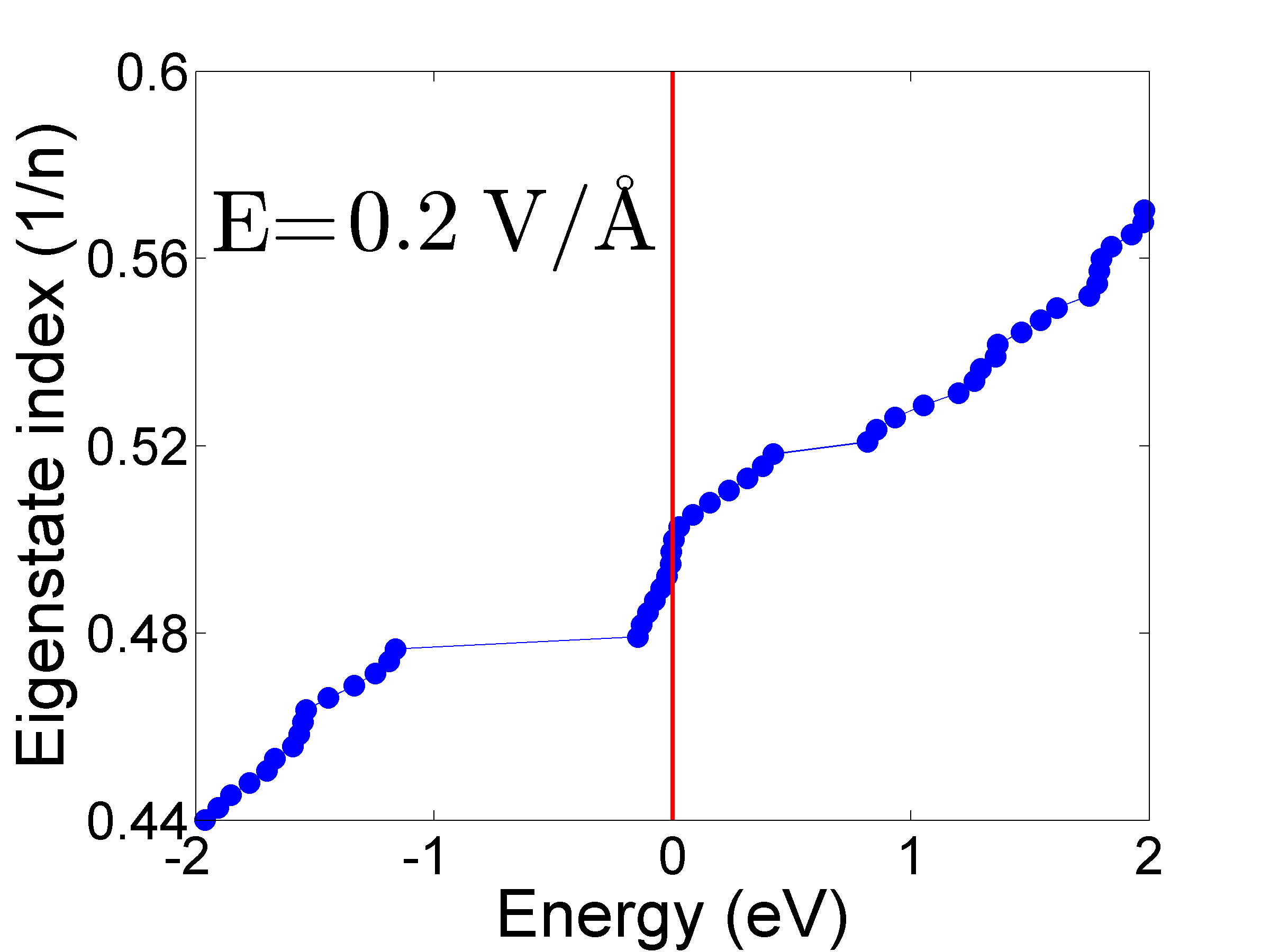}}
  \caption{}
\end{subfigure}%

\begin{subfigure}{.25\textwidth}
  \centering
  \includegraphics[width=\textwidth]{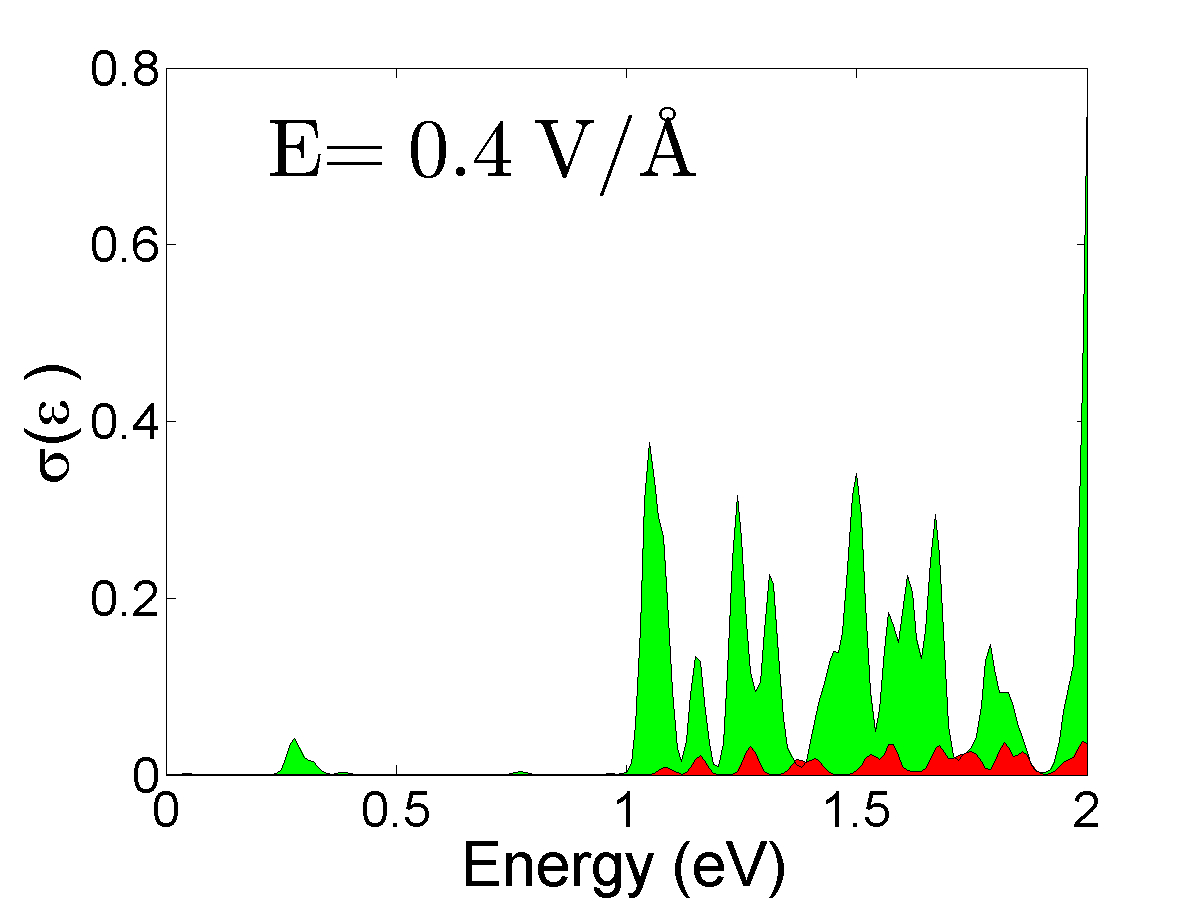}
  \caption{}
\end{subfigure}%
\begin{subfigure}{.25\textwidth}
  \centering
 \centerline{\includegraphics[width=\textwidth]{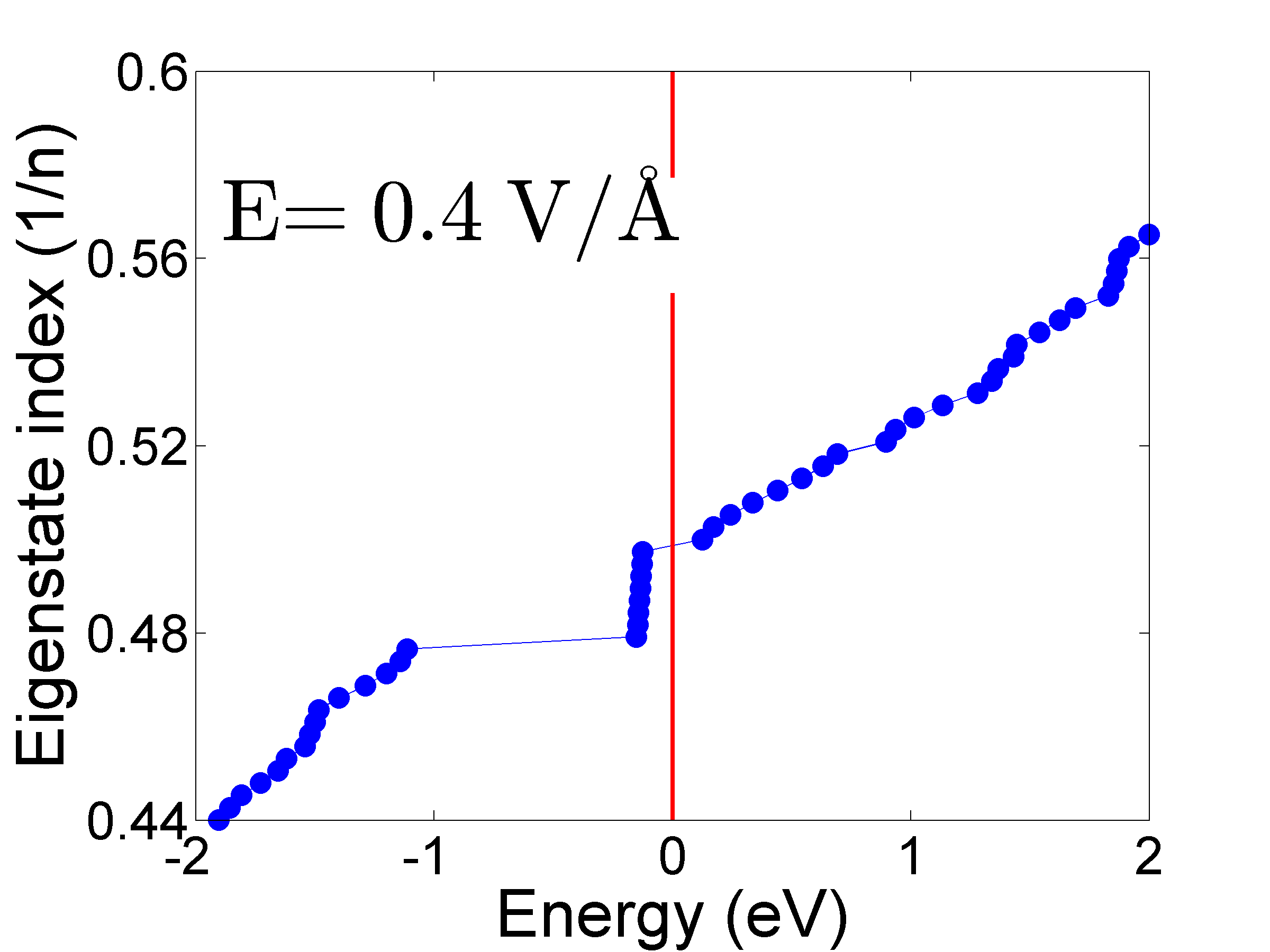}}
  \caption{}
  \label{fig:sub9}
\end{subfigure}
\caption{The optical absorption cross-section (a, c, e) and  energy levels (b, d, f) of zigzag hexagonal phosphorene QDs at different values of electric field.}
\label{fig:AbsorptionZHEXPQD}
\end{figure}

It can be seen from Fig.~\ref{fig:fig:AbsorptionZTRIPQD} (a, c, e) that with increasing field there is a noticeable decrease in the intensities of absorption peaks at $\varepsilon > 0.8$~eV compared to the prominent low-energy $y$-polarized peak. This can be attributed to the decrease of transition matrix elements because the positions of the peaks stay nearly the same. It is aslo worth noting that the red absorption peak for an incident $y$-polarized electromagnetic wave at $\varepsilon \simeq 0.05$ eV, (Fig.~\ref{fig:AbsorptionZHEXPQD} (a)), experiences a blue shift with increasing electric field which can be attributed to the increased smearing of the QZES as a function of the applied field.

The absorption cross-section of hexagonal zigzag phosphorene QDs in Fig.~\ref{fig:AbsorptionZHEXPQD} shows a totally different  behavior under the influence of electric field. The strong absorption peak, see Fig~\ref{fig:AbsorptionZHEXPQD} (a), around  $\varepsilon = 0.9$~eV occurs due to transitions from the group of edge states to the LUEL. At $E = 0.2$~V/{\AA} the intensity of this peak decreases  (see Fig.~\ref{fig:AbsorptionZHEXPQD} (b)) and vanishes at $E = 0.4$ V/{\AA}. 

The group of edge states at $E=0.4$~V/{\AA} splits into two  groups. As seen from Fig.~\ref{fig:AbsorptionZHEXPQD} (f) the group above $\varepsilon_F$ spreads in the  energy gap between edge states and the conduction band states.  The red absorption peak at $\varepsilon \approx 0.05$~eV, corresponding to an upper edge of the highly topical terahertz frequency range, shown in Fig.~\ref{fig:AbsorptionZHEXPQD} (a) experiences a decrease in intensity at $E = 0.2$ V/{\AA} and disappears at $E = 0.4$ V/{\AA}. This effect results from the energy gap opening between QZES shown in Fig.~\ref{fig:AbsorptionZHEXPQD} (f).

\subsubsection{Armchair edges}
\begin{figure}[htbp]
\centering
\begin{subfigure}{.25\textwidth}
  \centering
  \includegraphics[width=\textwidth]{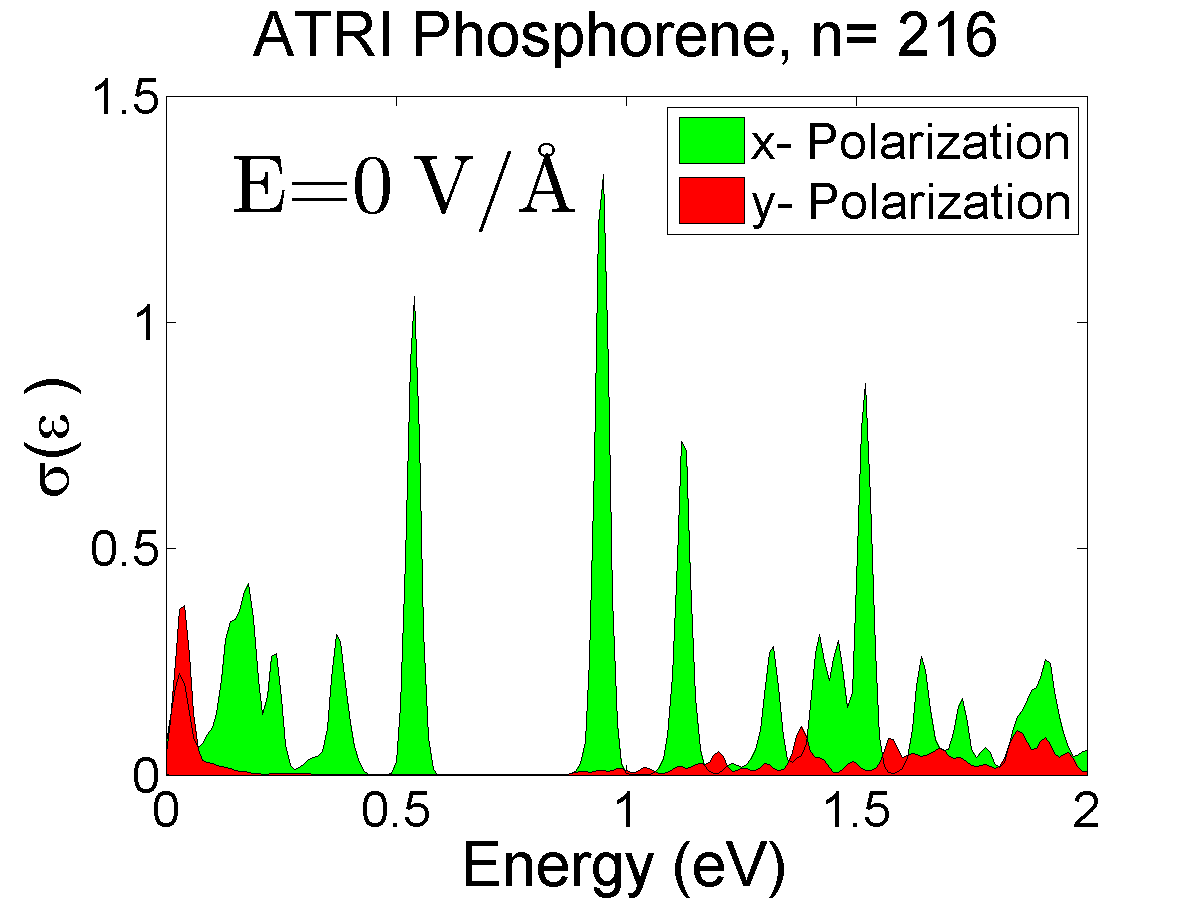}
  \caption{}
\end{subfigure}%
\begin{subfigure}{.25\textwidth}
  \centering
 \centerline{\includegraphics[width=\textwidth]{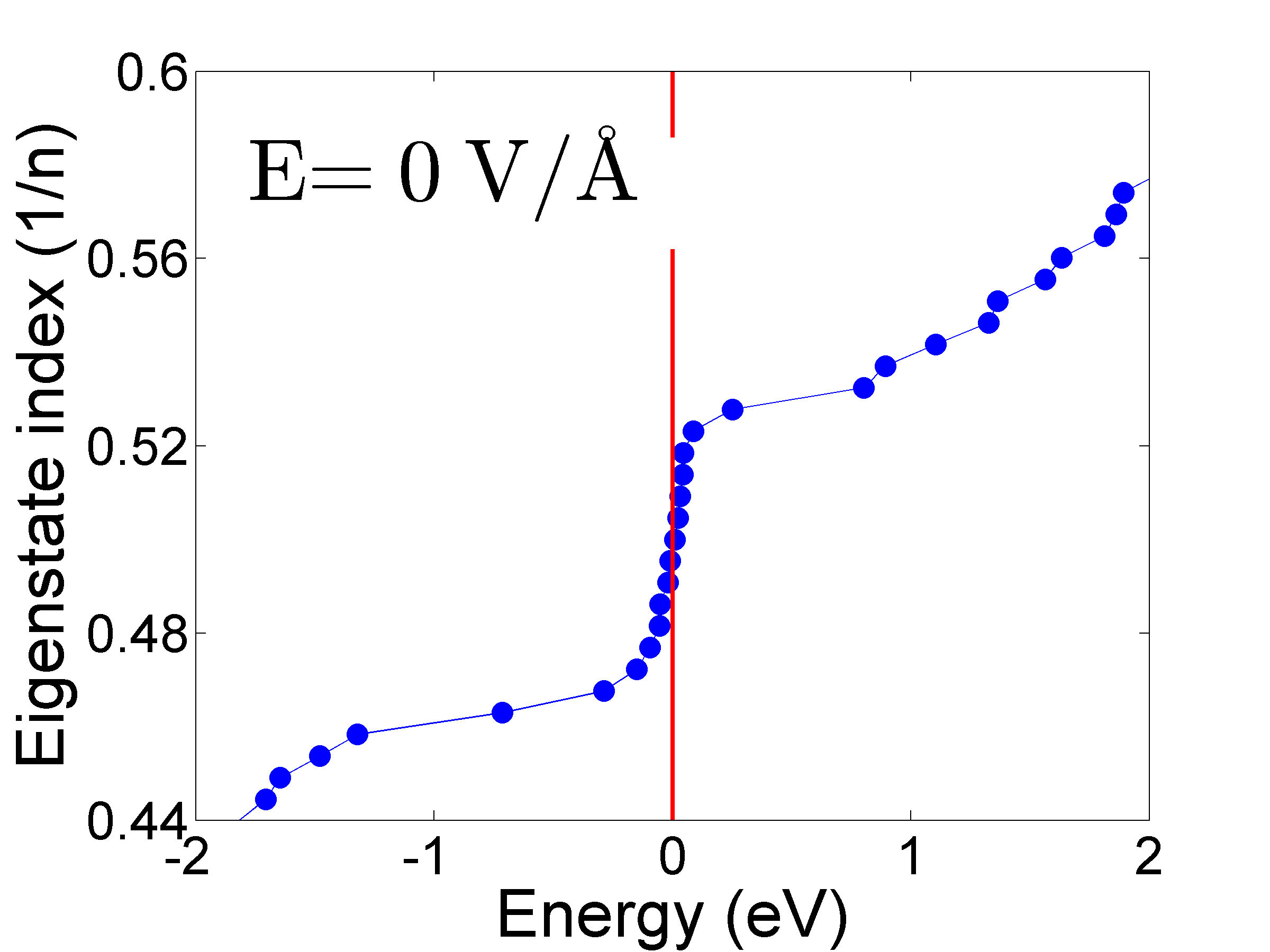}}
  \caption{}
\end{subfigure}

\begin{subfigure}{.25\textwidth}
  \centering
  \includegraphics[width=\textwidth]{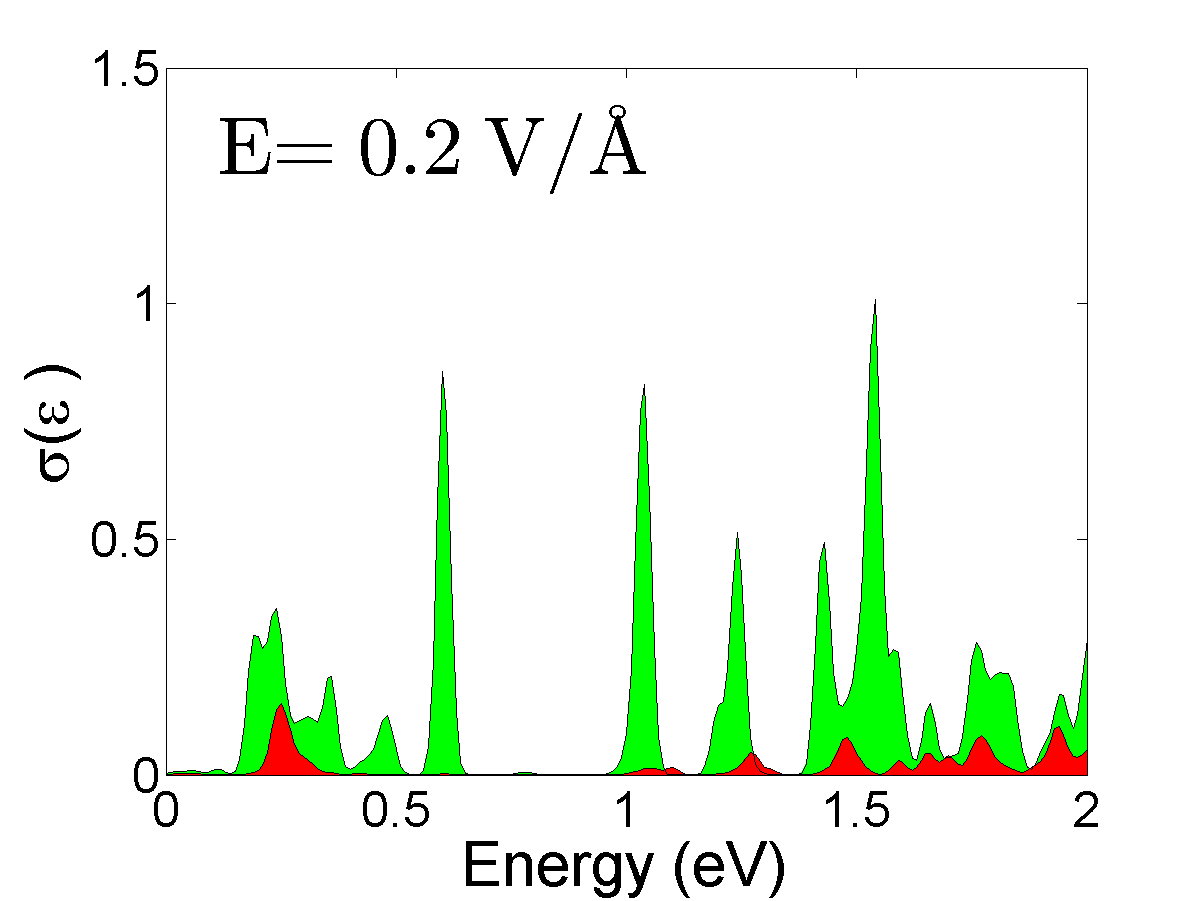}
  \caption{}
\end{subfigure}%
 \begin{subfigure}{.25\textwidth}
  \centering
 \centerline{\includegraphics[width=\textwidth]{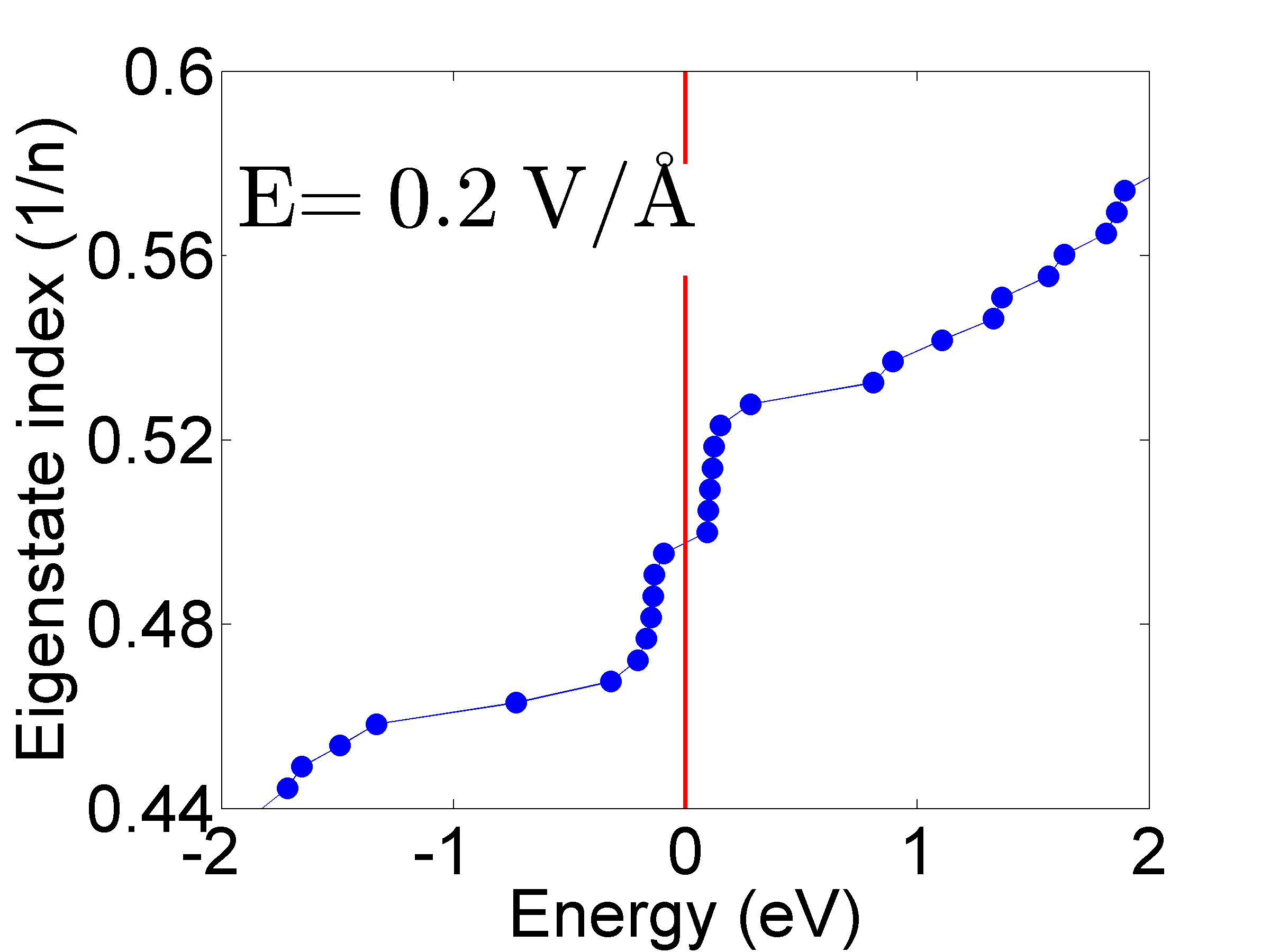}}
  \caption{}
\end{subfigure}%

\begin{subfigure}{.25\textwidth}
  \centering
  \includegraphics[width=\textwidth]{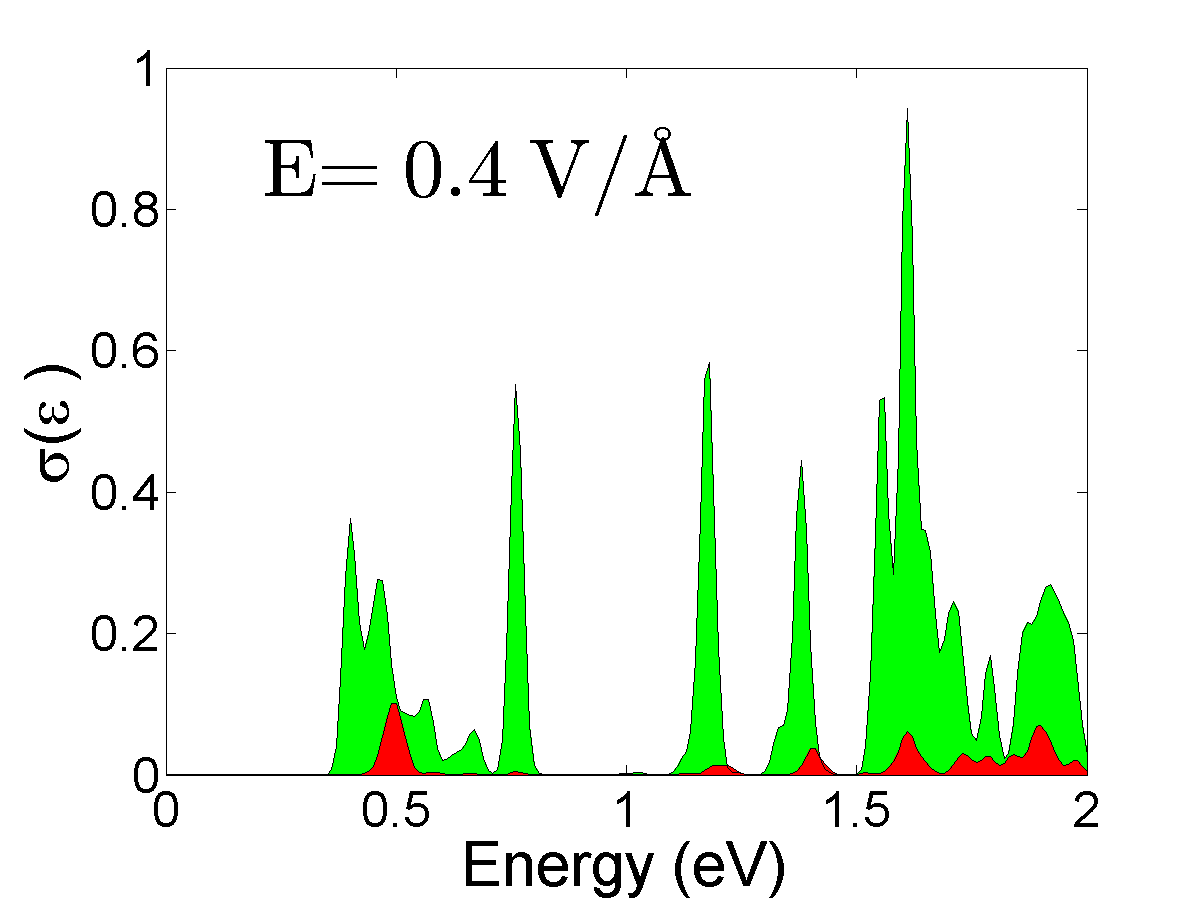}
  \caption{}
\end{subfigure}%
\begin{subfigure}{.25\textwidth}
  \centering
 \centerline{\includegraphics[width=\textwidth]{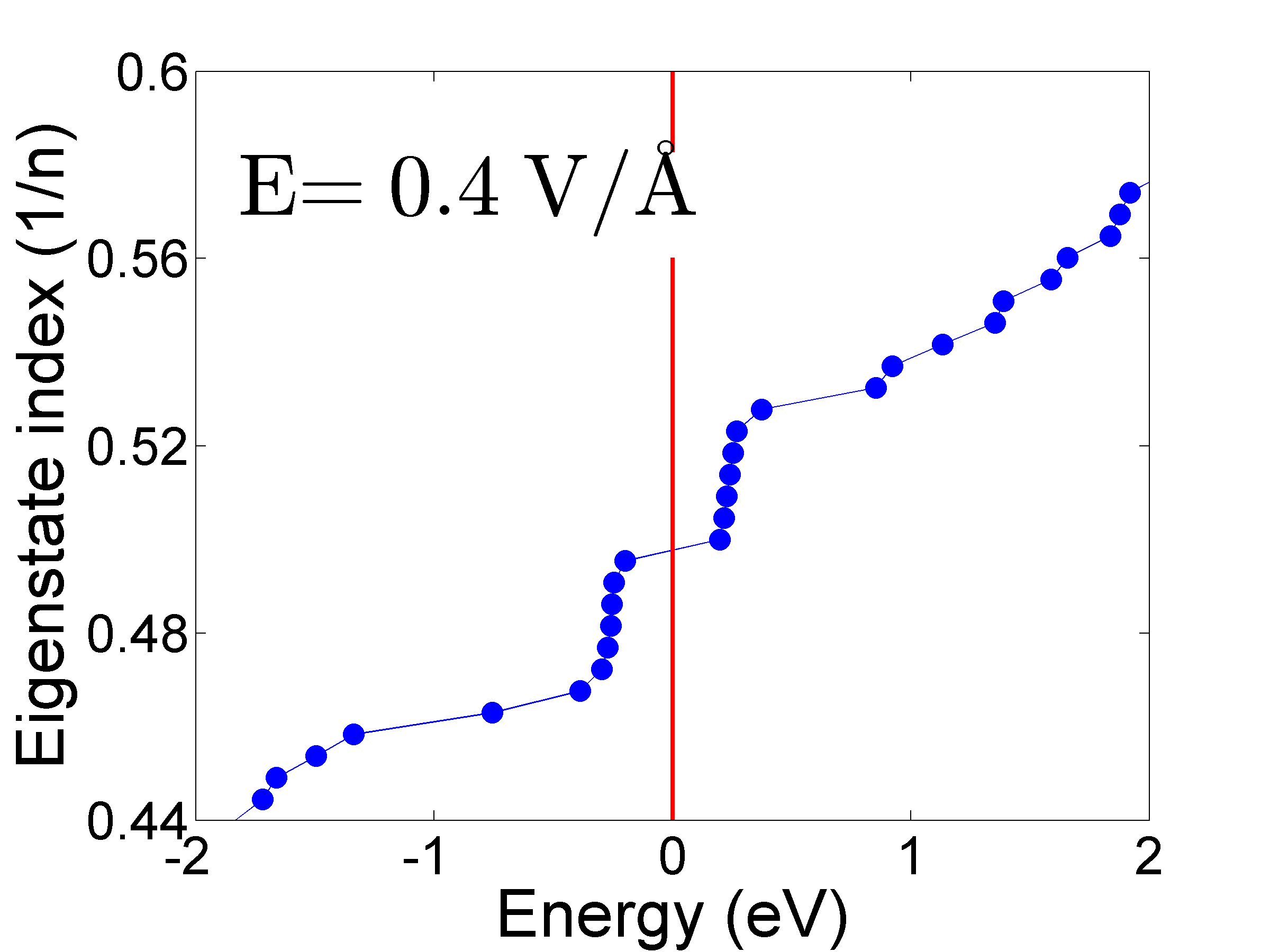}}
  \caption{}
  \label{fig:sub10}
\end{subfigure}
\caption{ The optical absorption cross-section (a, c, e) and the corresponding energy levels (b, d, f) of triangular armchair phosphorene QDs under the effect of an electric field.}
\label{fig:AbsorptionATRIPQD}
\end{figure}

\begin{figure}[htbp]
\centering
\begin{subfigure}{.25\textwidth}
  \centering
  \includegraphics[width=\textwidth]{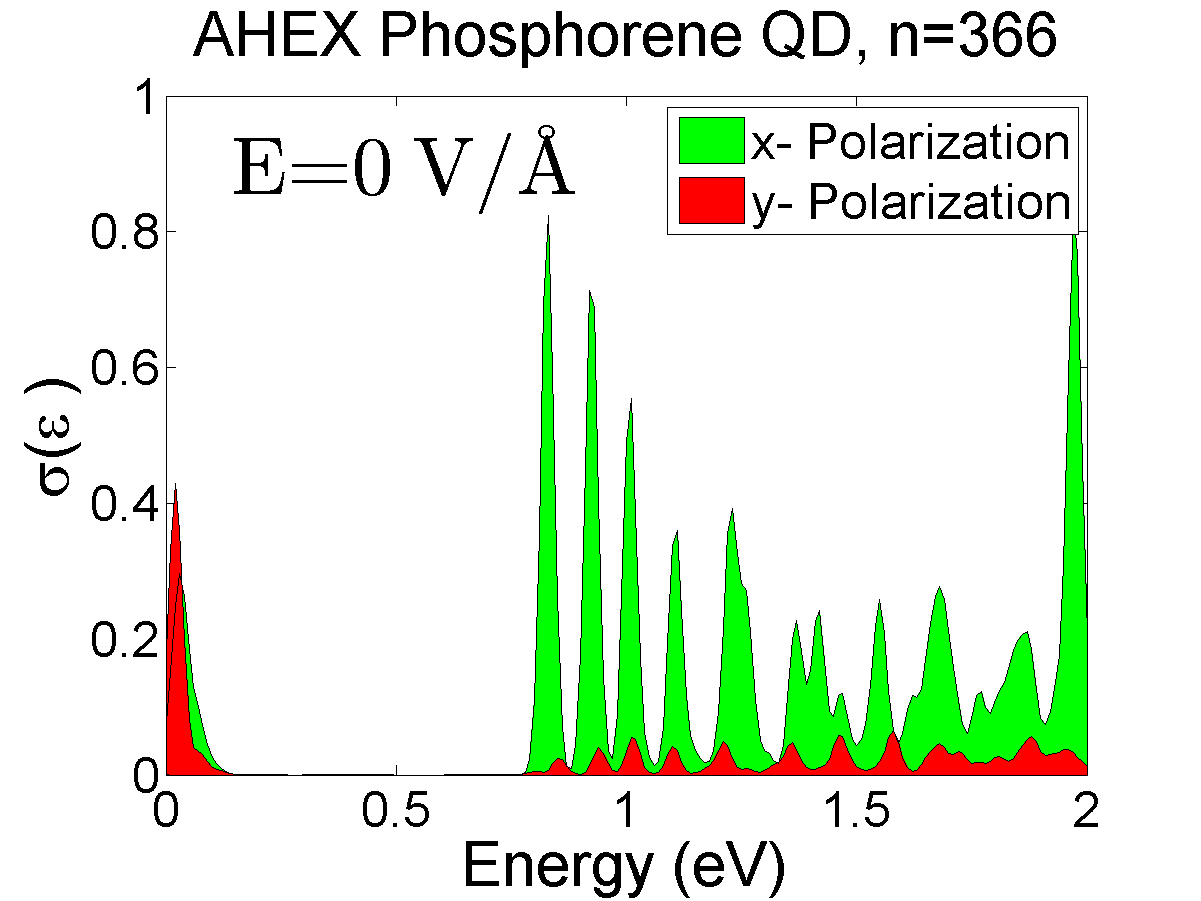}
  \caption{}
\end{subfigure}%
\begin{subfigure}{.25\textwidth}
  \centering
 \centerline{\includegraphics[width=\textwidth]{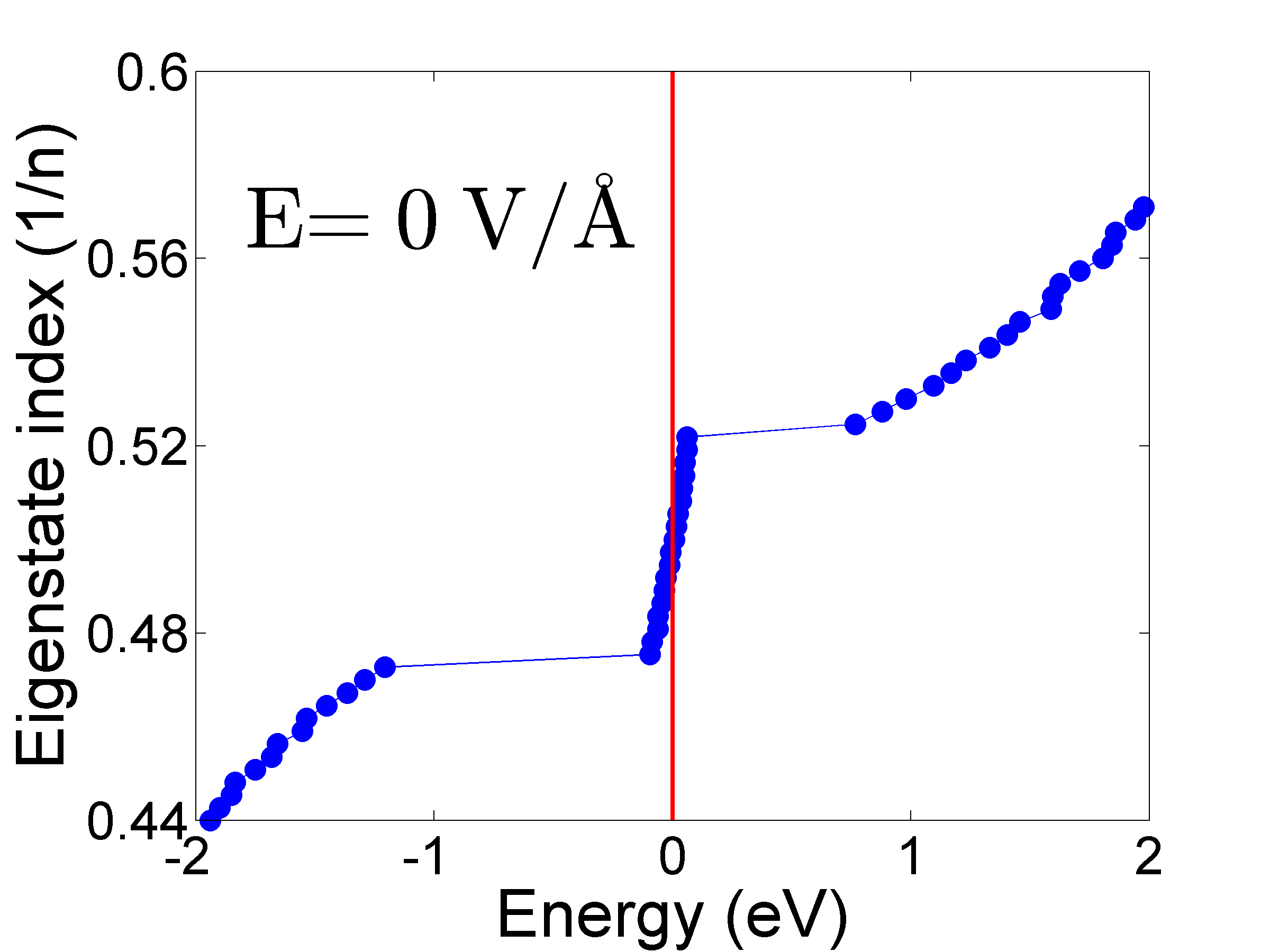}}
  \caption{}
\end{subfigure}

\begin{subfigure}{.25\textwidth}
  \centering
  \includegraphics[width=\textwidth]{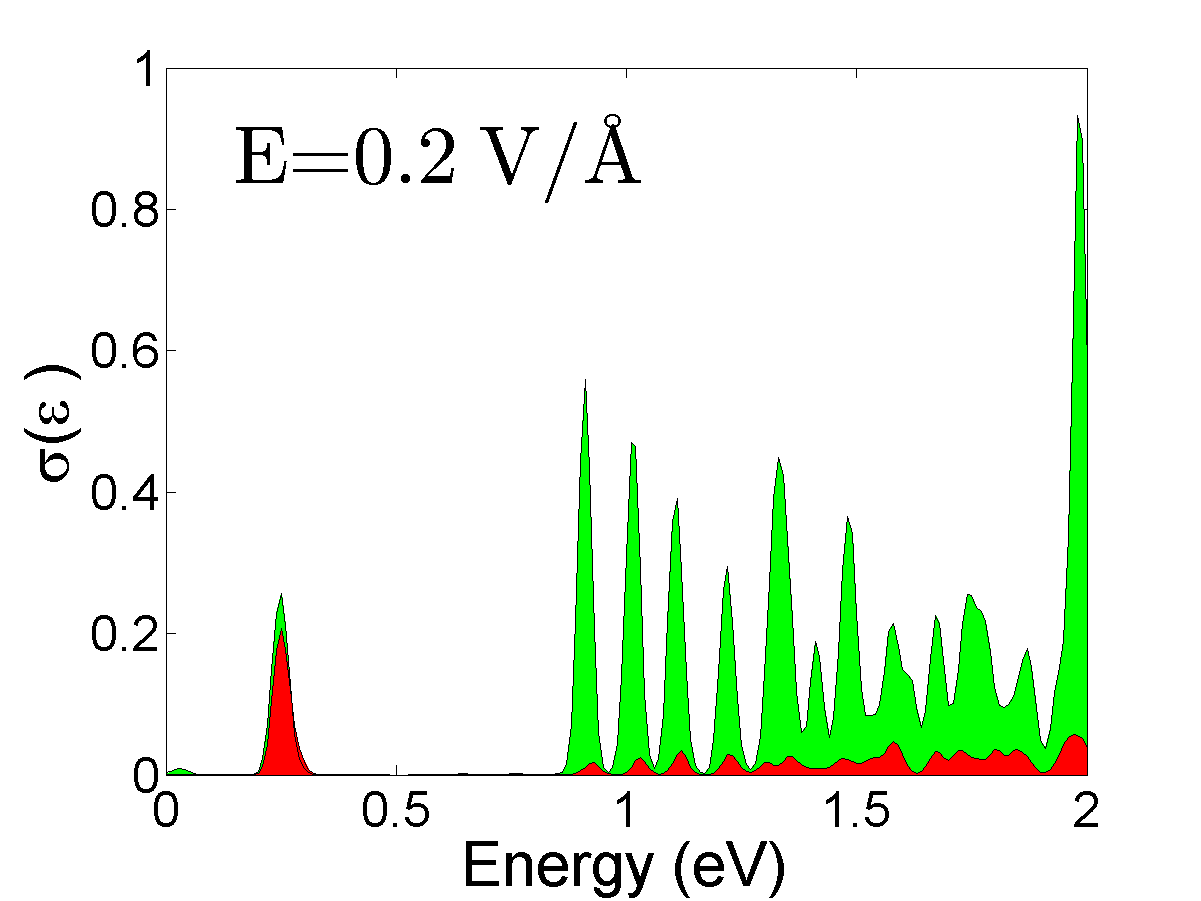}
  \caption{}
\end{subfigure}%
 \begin{subfigure}{.25\textwidth}
  \centering
 \centerline{\includegraphics[width=\textwidth]{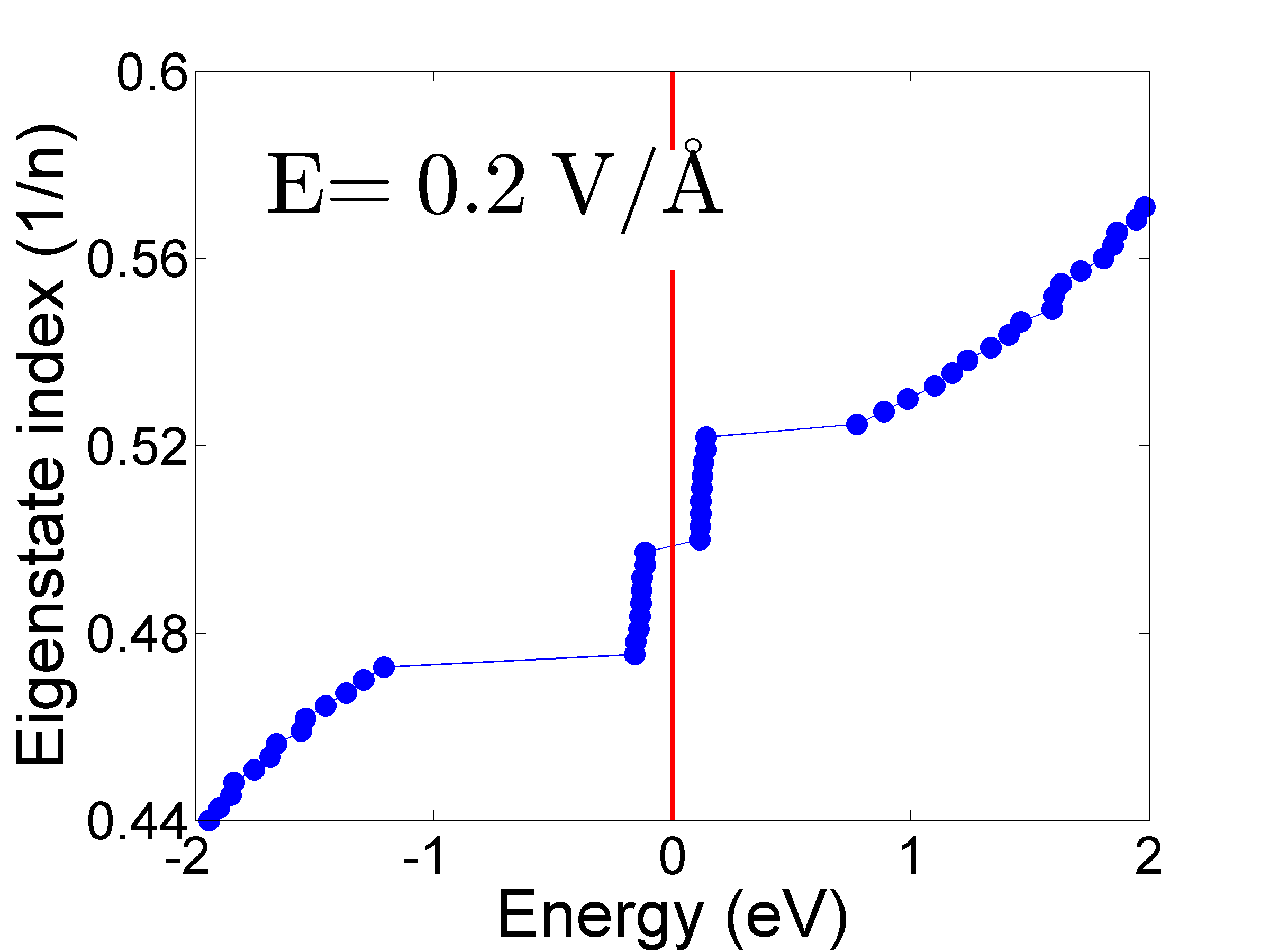}}
  \caption{}
\end{subfigure}%

\begin{subfigure}{.25\textwidth}
  \centering
  \includegraphics[width=\textwidth]{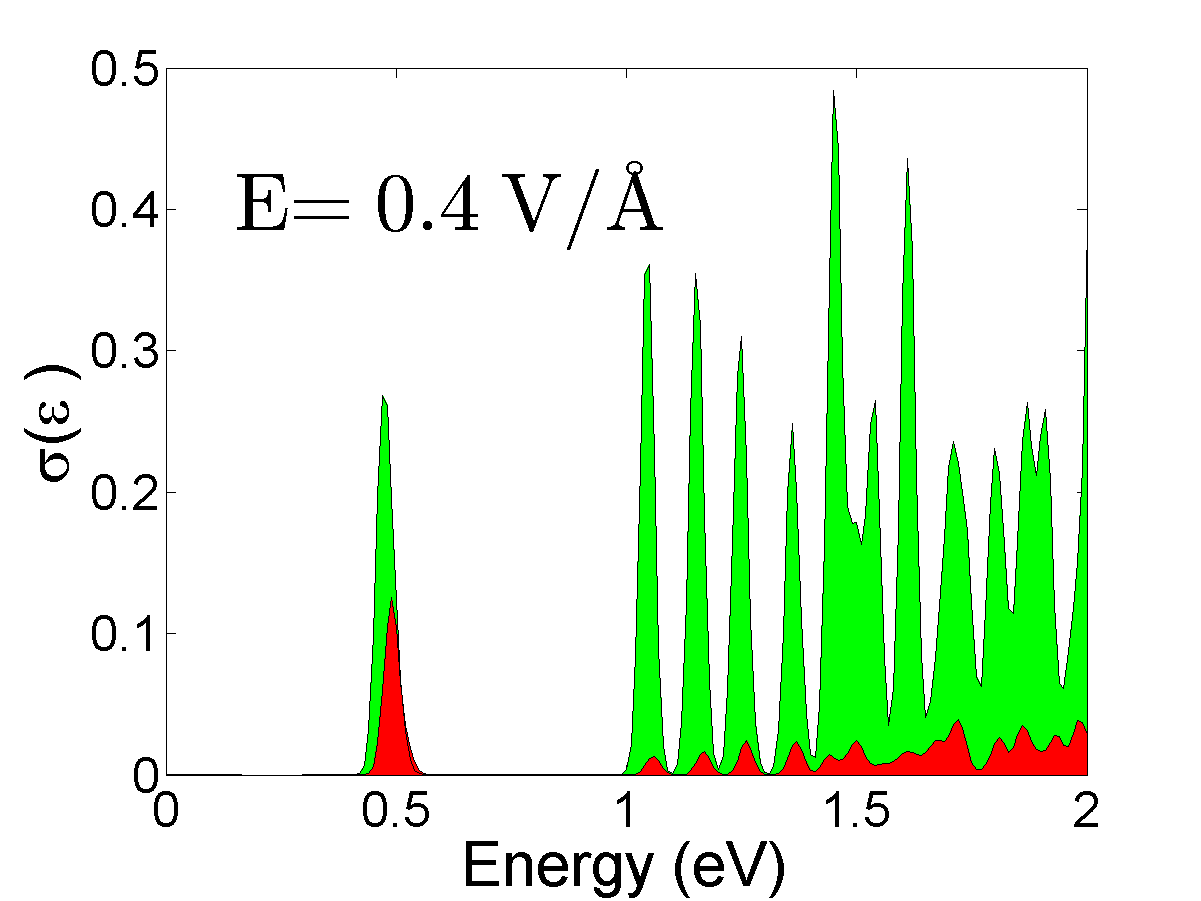}
  \caption{}
\end{subfigure}%
\begin{subfigure}{.25\textwidth}
  \centering
 \centerline{\includegraphics[width=\textwidth]{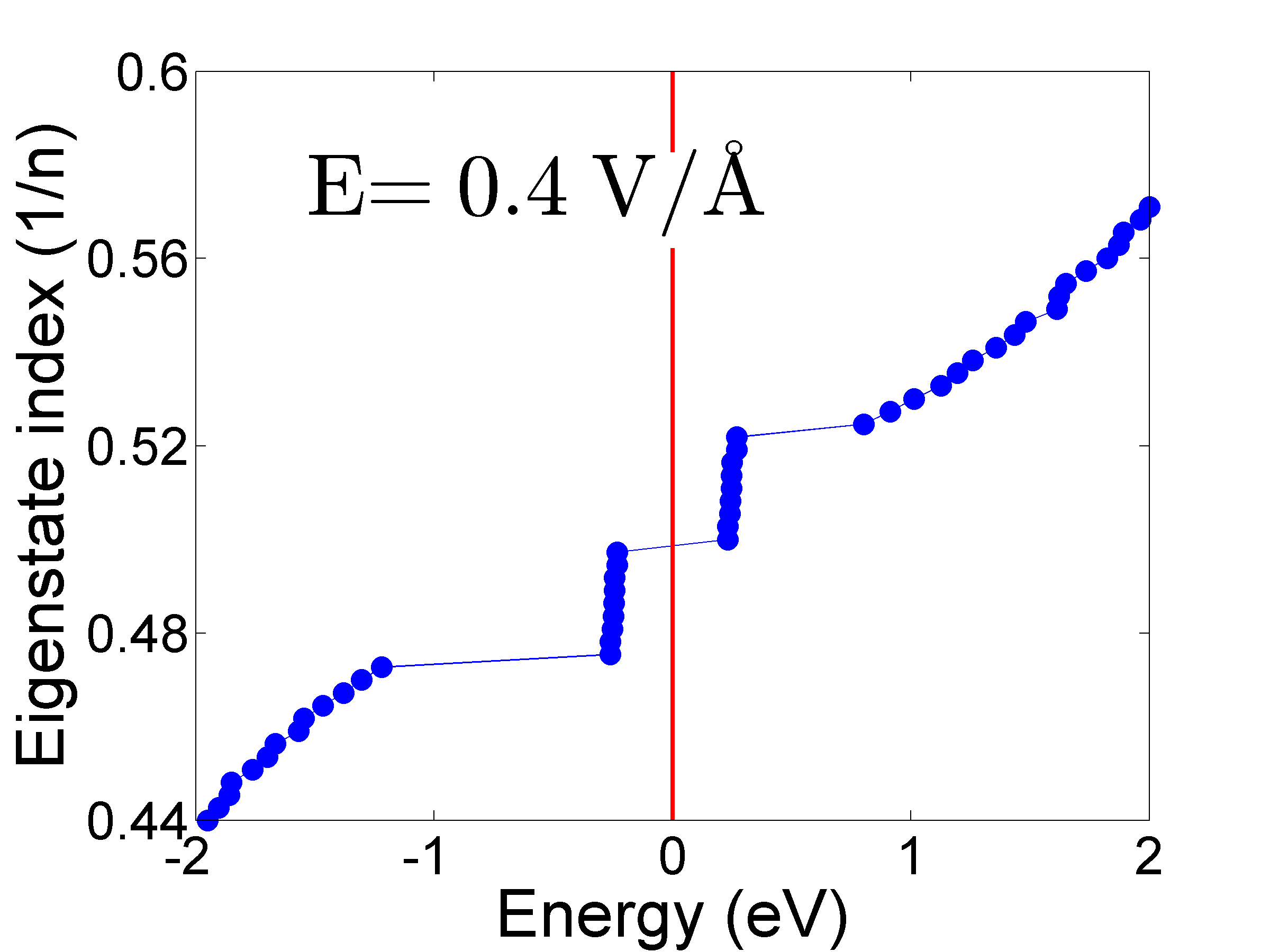}}
 \caption{}
  \label{fig:sub11}
\end{subfigure}
\caption{ The effect of a perpendicular electric field on the optical absorption cross-section (a, c, e) and corresponding energy levels (b, d, f) of hexagonal armchair phosphorene QDs.}
\label{fig:AbsorptionAHEXPQD}
\end{figure}

The energy levels and absorption cross-sections of  triangular (Fig.~\ref{fig:AbsorptionATRIPQD}) and hexagonal (Fig.~\ref{fig:AbsorptionAHEXPQD}) phosphorene quantum dots with armchair edges  are studied under the effect of an electric field applied perpendicular to the structure plane: $n=216$ for the triangular case and $n=366$ for the hexagonal case. 

Figure~\ref{fig:AbsorptionATRIPQD} (b, d, f) shows that the ATRI edge states split into two groups, like in ZHEX QDs, with an energy gap between them that increases with increasing the applied electric field. The result of this new energy gap is a blue shift in the edge of the optical absorption cross-section, as shown in Fig.~\ref{fig:AbsorptionATRIPQD} (a, c, e). Moreover, the absorption peaks due to transitions between edge states with a photon energies $\varepsilon$ ranging from $0$ to  $0.5$~eV shown in Fig.~\ref{fig:AbsorptionATRIPQD} (a), have a comparable intensity with the peaks corresponding to the transitions from edge states to conduction (valance) band states. 
This behaviour is opposite to that in triangular bilayer graphene QDs with zigzag edges where transitions between edge states are weak~\cite{Abdelsalam2016}. It also contrasts with selection rules in zigzag graphene nanoribbons where transitions between the edges states are strictly forbidden~\cite{Chang2006,Hsu2007,Chung2011,Sasaki2011,Saroka2017}. However, it is somewhat similar to the edge state transitions in triangular graphene QDs with the excitonic effects taken into account~\cite{Guclu2010}. By increasing the electric field transitions between the edges states in the ZHEX PQD shift to a higher energy due to the opening of energy gap, Fig.~\ref{fig:AbsorptionATRIPQD} (d, f), and the number of transition peaks decreases as a result of the reduction in smearing of the edge states (see Fig.~\ref{fig:AbsorptionATRIPQD} (f)).

Figure~\ref{fig:AbsorptionAHEXPQD} shows the electronic states and optical absorption cross-sections of 
hexagonal phosphorene QDs with armchair termination. In this case, the smearing of the edge 
states is small, Fig.~\ref{fig:AbsorptionAHEXPQD} (b, d, f), and the transitions between them are given by two strong absorption peaks (shown in red and green) at $\varepsilon\simeq 0.05$~eV. By varying the electric field we can control the position of these two peaks in the absorption spectrum. For instance, at $E= 0.2$~V/{\AA} they are situated at $\varepsilon \simeq 0.25$~eV, therefore it is possible to generate controllable optical transitions within the energy gap of the hexagonal phosphorene QDs with armchair terminations. Moreover, it can be seen that at $E=0$~V/{\AA}, Fig.~\ref{fig:AbsorptionAHEXPQD} (a), the red absorption peak (for incident $y$-polarized electromagnetic wave) has a higher intensity than the green peak ($x$-polarized incident wave). However, by increasing the electric field the situation is inverted: the green peak becomes more intense than the red peak. Again, as in ZHEX QDs, the QZES optical transitions depend strongly on the opening of the energy gap between QZES which can be controlled by the applied electric field. At zero field the energy gap is almost zero, see Fig. \ref{fig:AbsorptionAHEXPQD} (b), which promotes a strong $y$-absorption peak. At high values of the electric field the energy gap increases leading to decrease in the intensity of the $y$-absorption peak (red peak in Fig. \ref{fig:AbsorptionAHEXPQD} (e)) and increase in the $x$-absorption (green peak at $\varepsilon \approx 0.5$~eV in Fig. \ref{fig:AbsorptionAHEXPQD} (e)). Therefore, we conclude that the intensity of the $x$-absorption peak is directly proportional to the opening of the energy gap between QZES and $y$-absorption peak intensity is inversely proportional to the energy gap.

\subsection{Edge roughness}
In this section we study the effect of edge disorder on the electronic spectra and optical properties of PQDs. The edge disorder was modeled as described at the end of Section~\ref{sec:Structures} for all types of quantum dots considered in Section~\ref{sec:ElectricFieldEffect}; this means that the edges of the initial bounding polygons were replaced with random Koch curves. We consider replacement for AHEX, ATRI, ZHEX and ZTRI types of phosphorene dots with the number of atoms $n=366,216,384$, and $222$, respectively. For random structures we keep the same notations as for the original regular structure with `(r)' appended at the end, e.g. AHEX is changed to AHEX(r). We also clearly indicate for each structure the new number of atoms $n$. The random structures have edges of neither armchair nor zigzag type but their initial shape and crystallographic orientation are preserved to some extent.
\begin{figure}[htbp]
\centering
\begin{subfigure}{.223\textwidth}
    \centering
    \includegraphics[width=\textwidth]{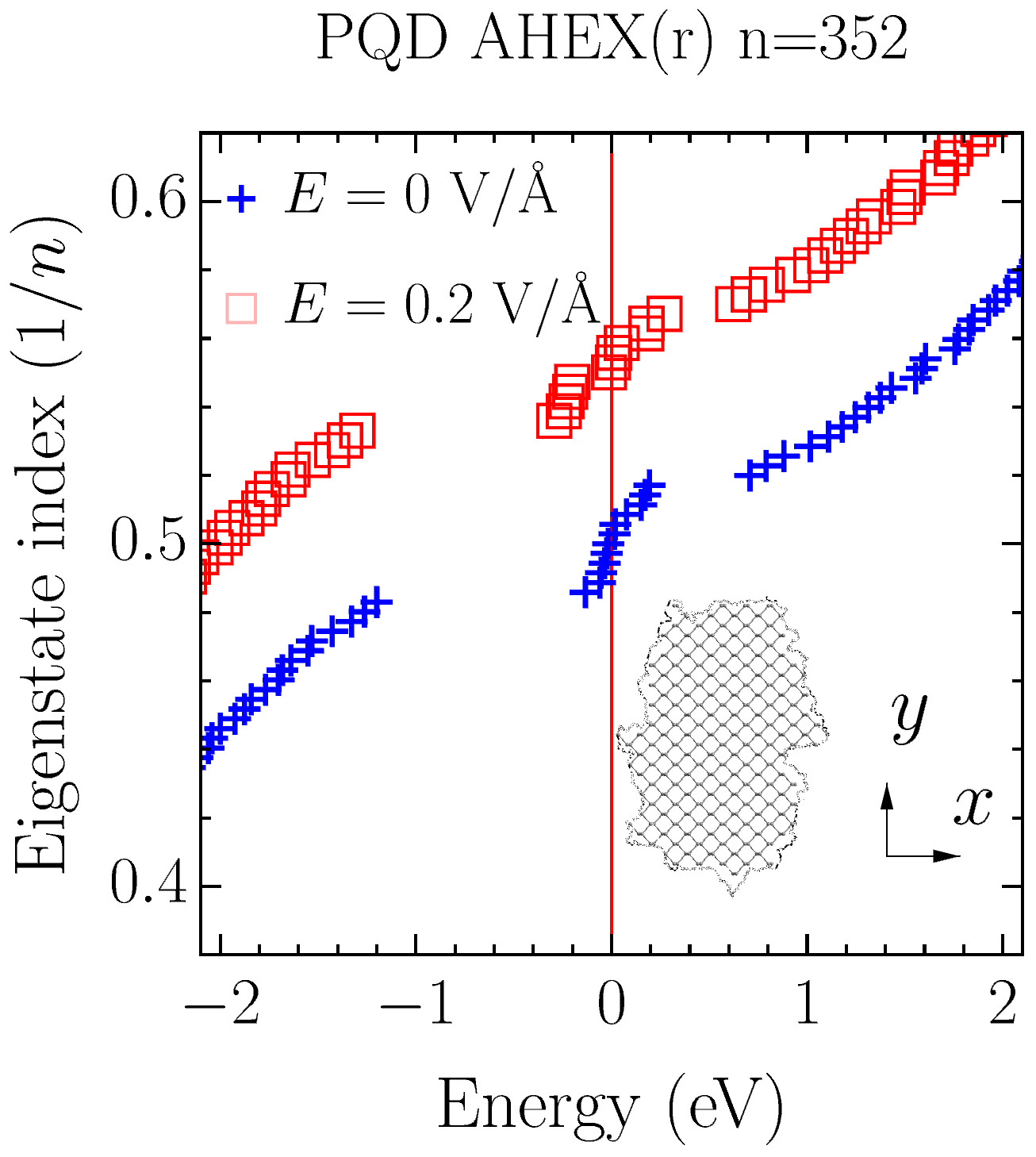}
    \caption{}
\end{subfigure}%
\begin{subfigure}{.277\textwidth}
    \centering
    \includegraphics[width=\textwidth]{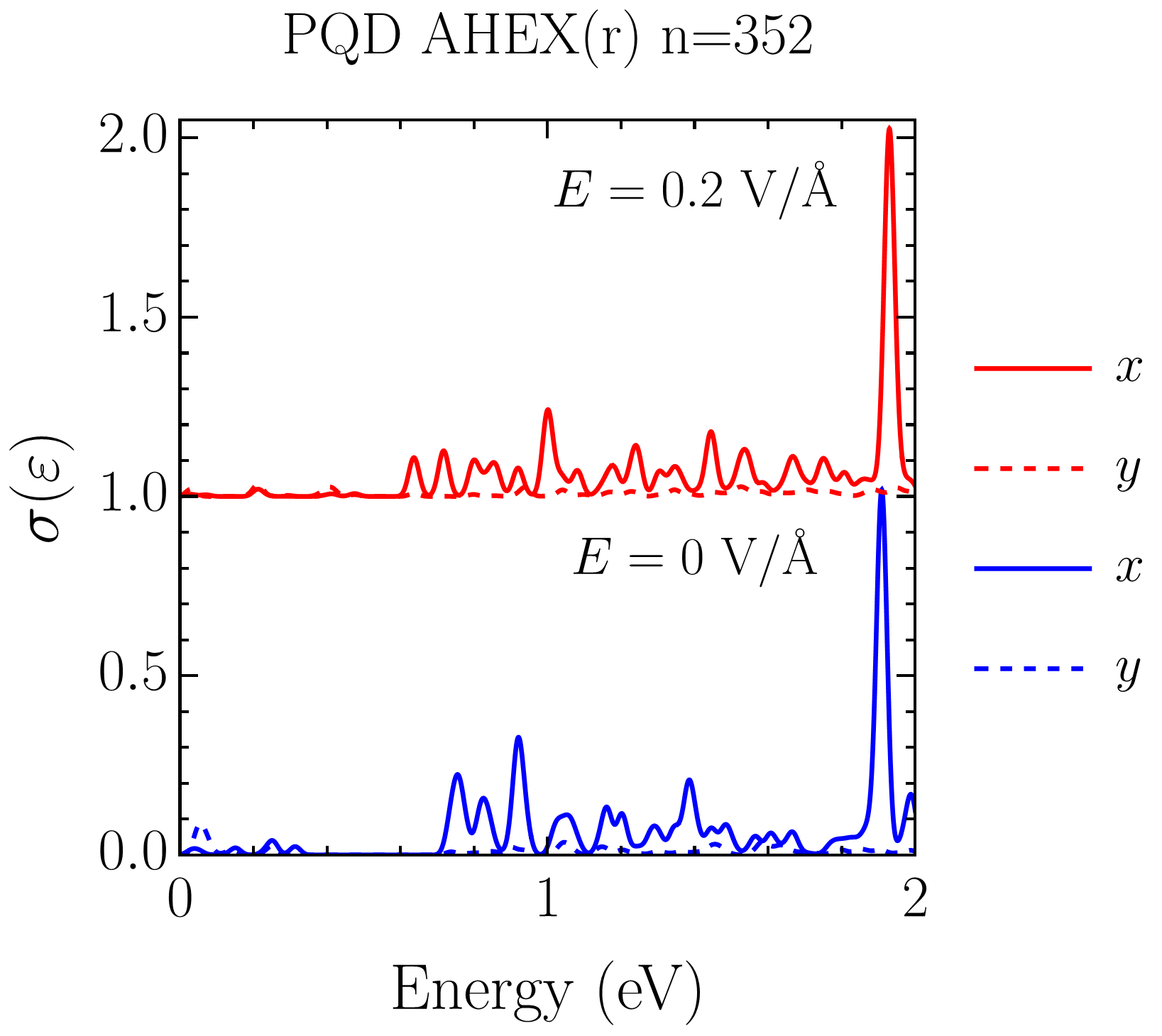}
    \caption{}
\end{subfigure}

\begin{subfigure}{.223\textwidth}
     \centering
     \includegraphics[width=\textwidth]{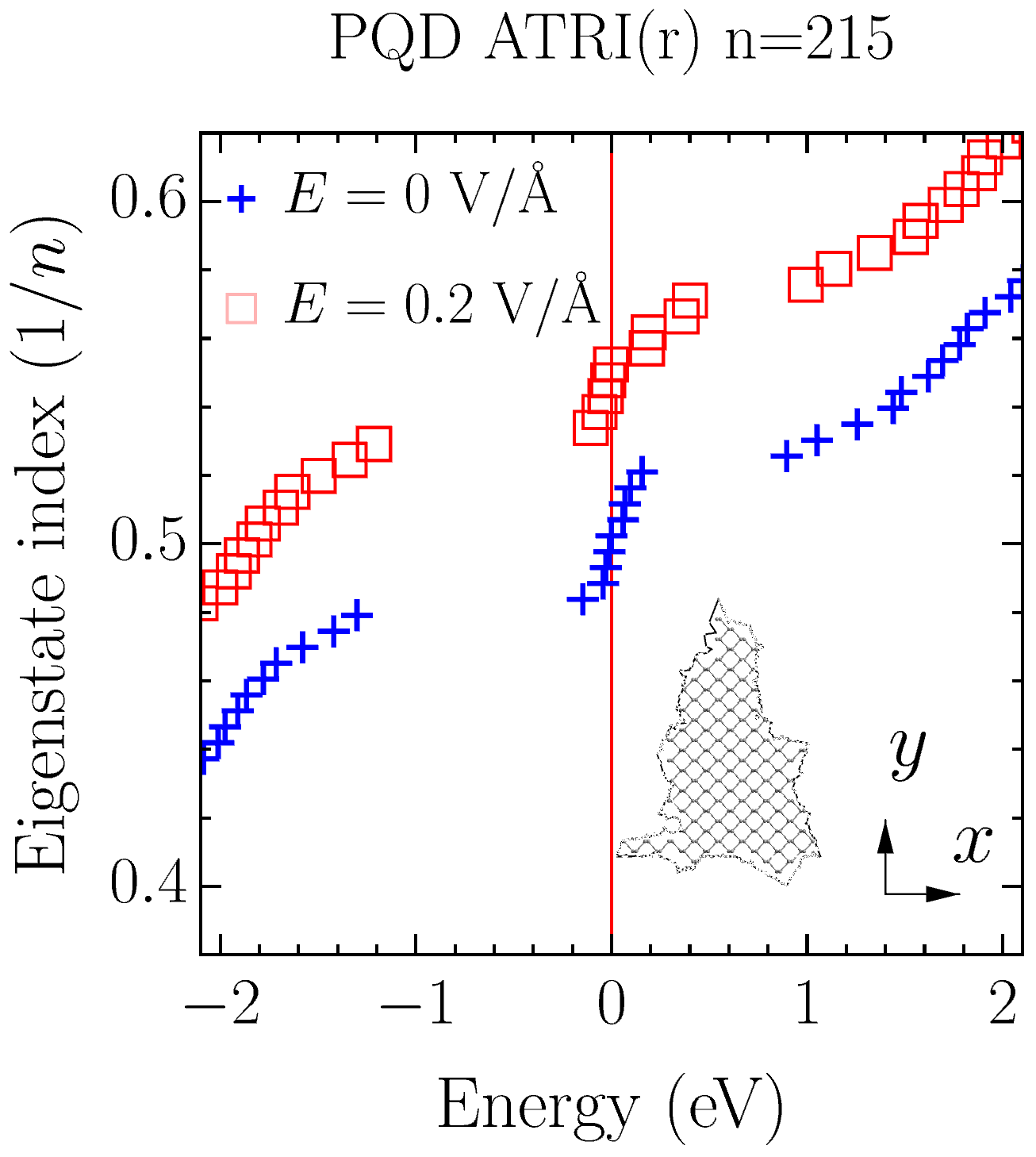}
     \caption{}
\end{subfigure}%
 \begin{subfigure}{.277\textwidth}
     \centering
     \includegraphics[width=\textwidth]{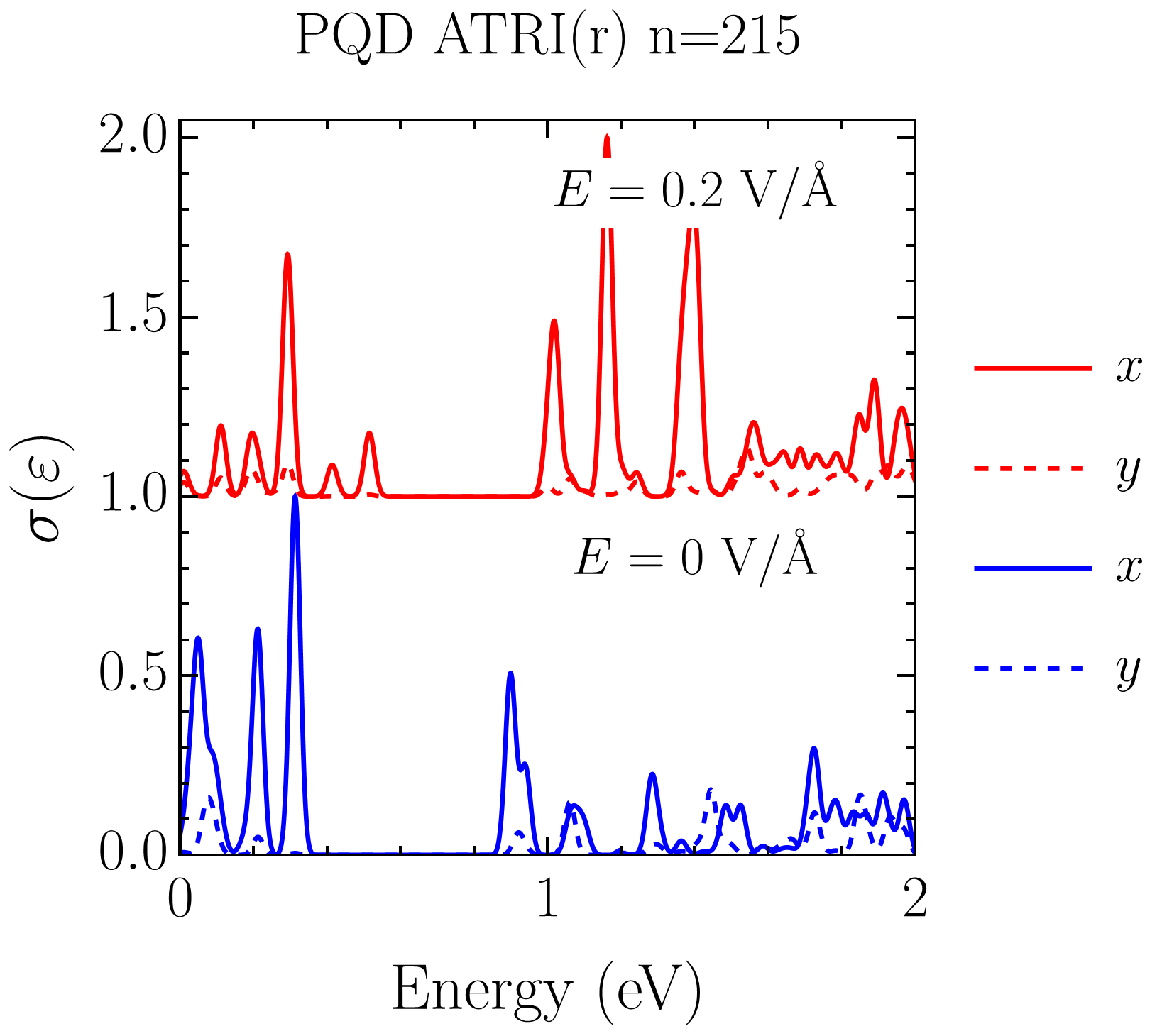}
     \caption{}
\end{subfigure}%
\caption{The effect of the electric field on the energy levels and optical absorption of disordered PQDs based on those with armchair edges. (a, c) The energy levels and (b, d) normalized absorption cross-sections of hexagonal and triangular clusters, respectively.}
\label{fig:EdgeRoughnessArmchair}
\end{figure}

\begin{figure}[htbp]
\centering
\begin{subfigure}{.223\textwidth}
    \centering
    \includegraphics[width=\textwidth]{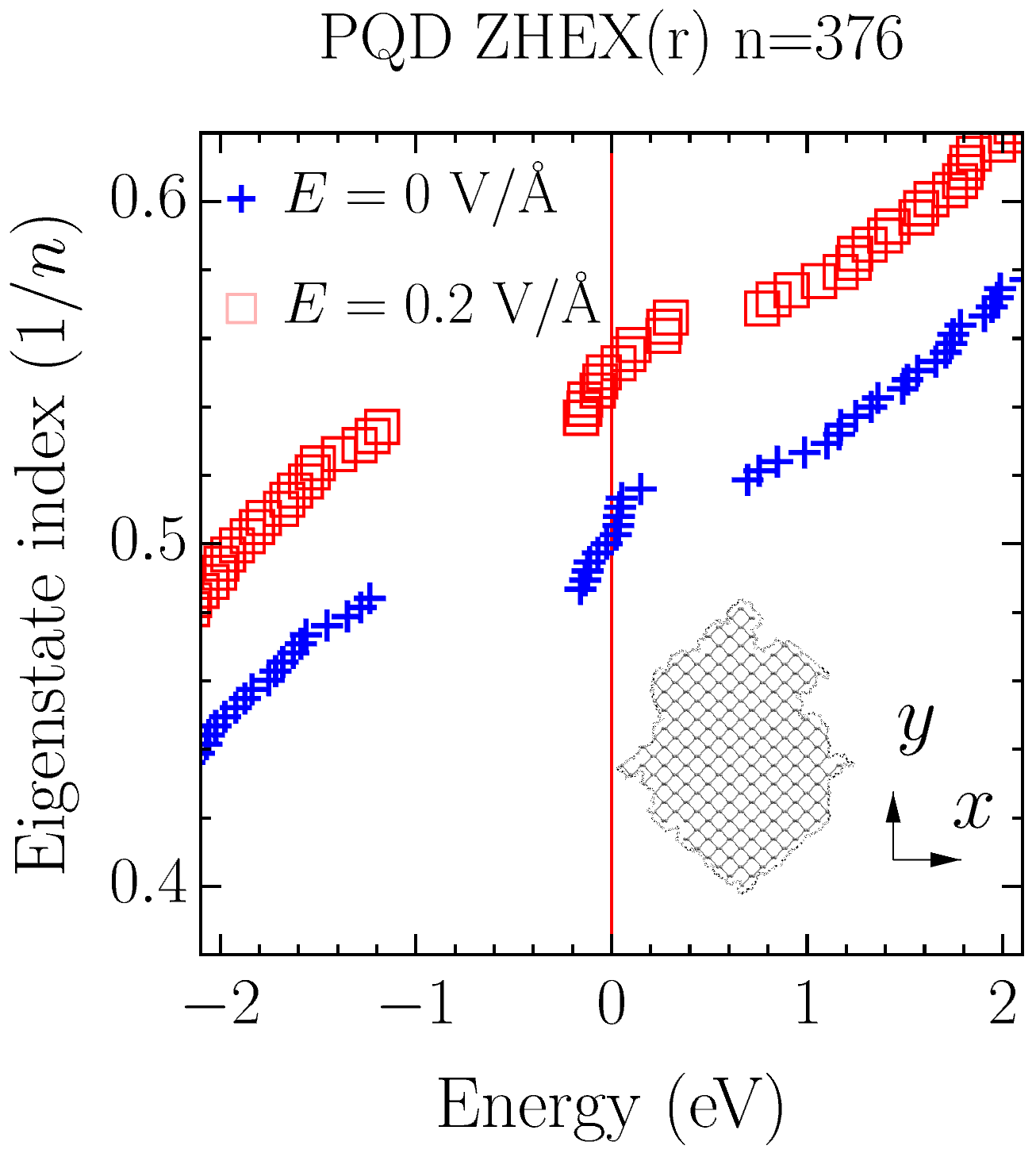}
    \caption{}
\end{subfigure}%
\begin{subfigure}{.277\textwidth}
    \centering
    \includegraphics[width=\textwidth]{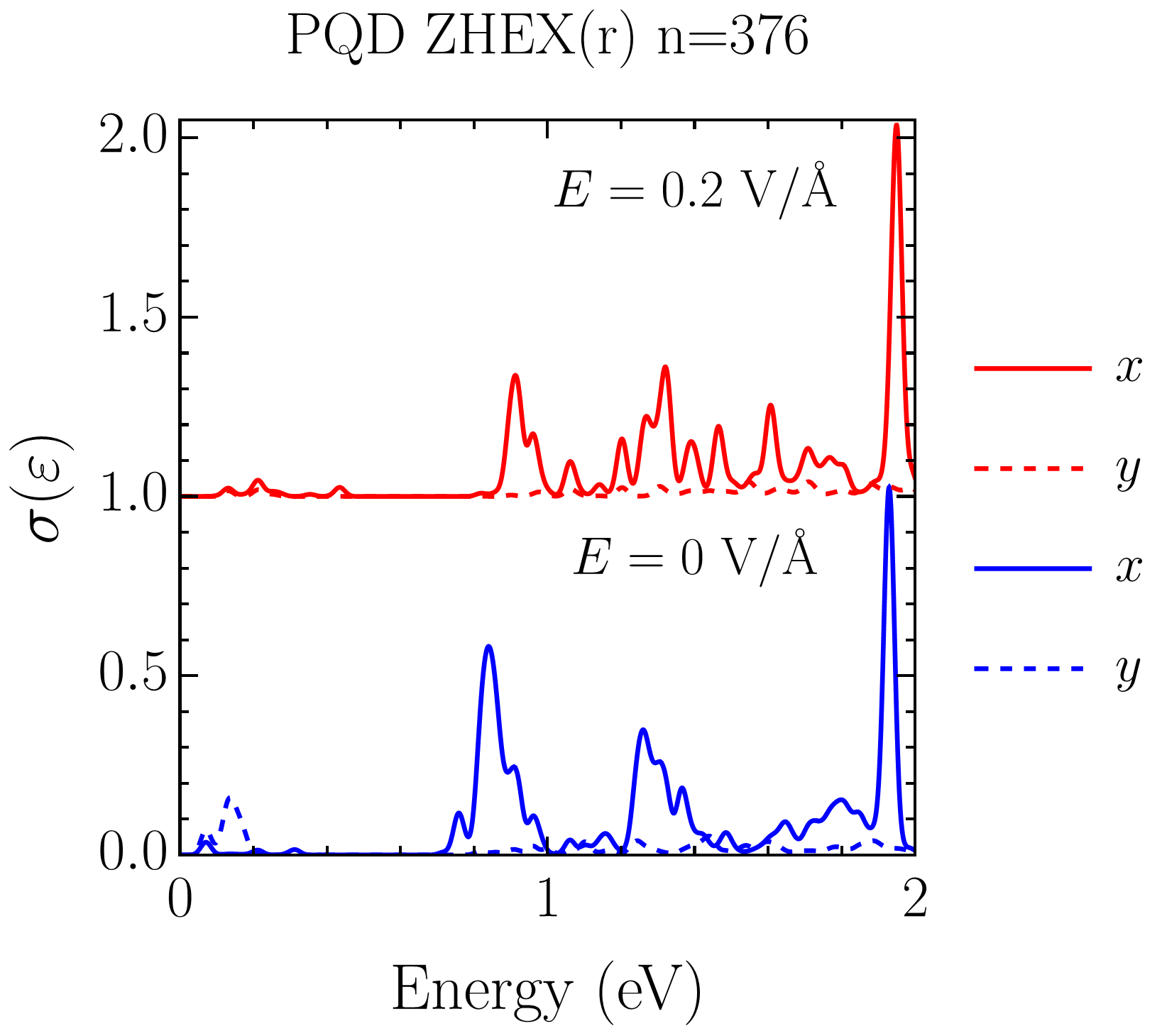}
    \caption{}
\end{subfigure}

\begin{subfigure}{.223\textwidth}
    \centering
    \includegraphics[width=\textwidth]{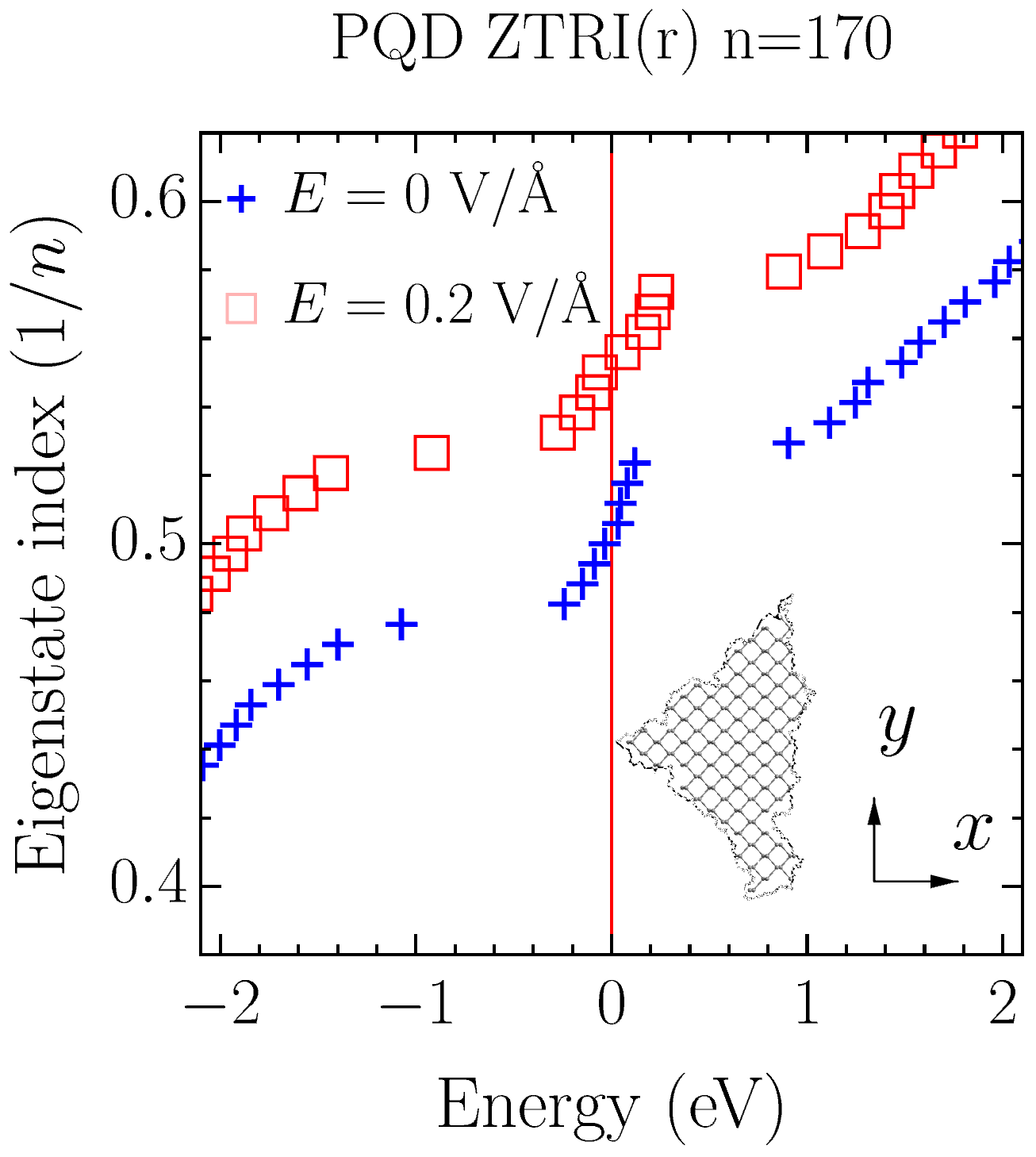}
    \caption{}
\end{subfigure}%
 \begin{subfigure}{.277\textwidth}
    \centering
    \includegraphics[width=\textwidth]{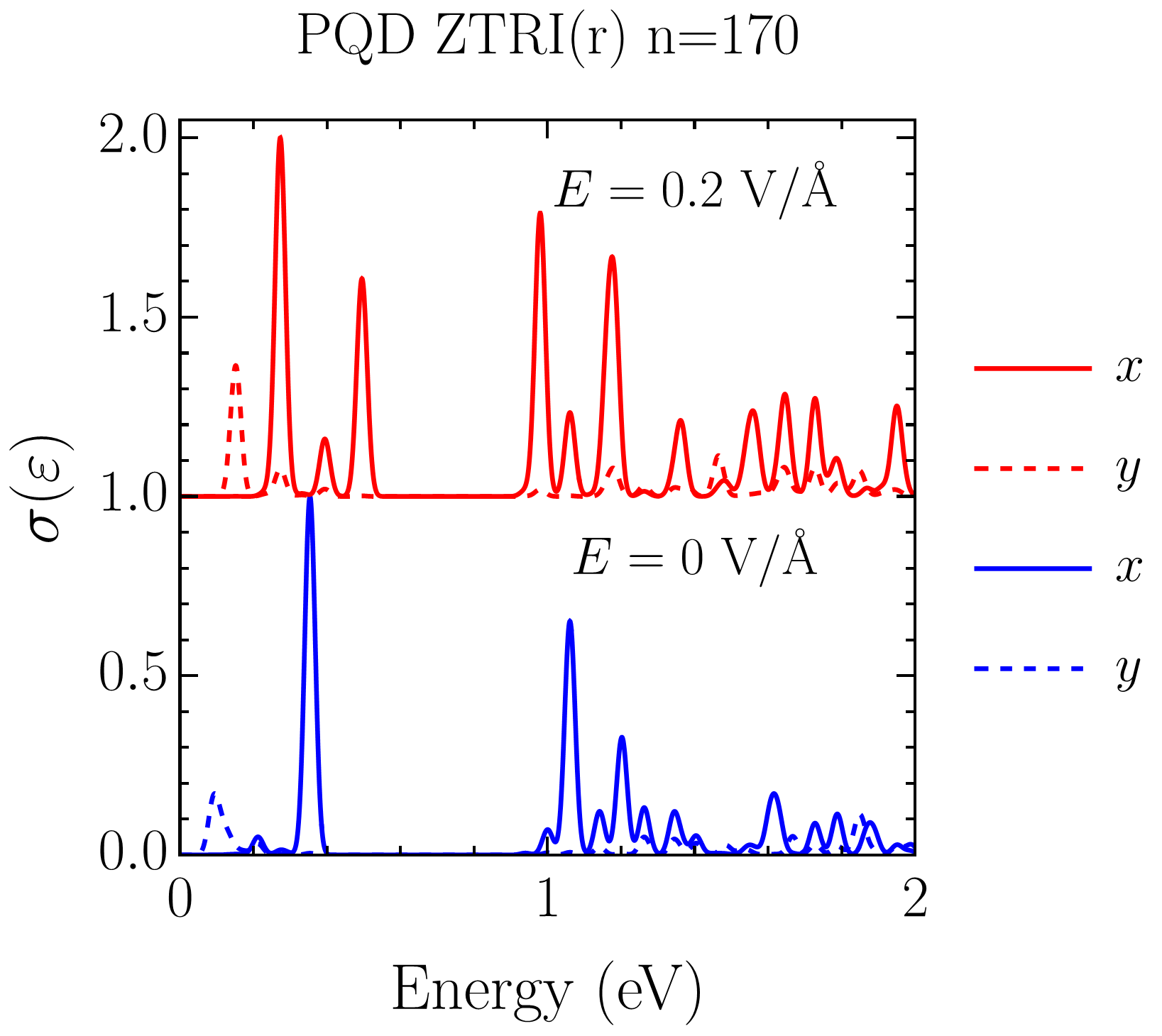}
    \caption{}
\end{subfigure}%
\caption{The same as Fig.~\ref{fig:EdgeRoughnessArmchair} but for disordered PQDs based on those with zigzag edges. (a, c) Energy levels and (b, d) absorption cross-section of hexagonal and triangular clusters, respectively.}
\label{fig:EdgeRoughnessZigzag}
\end{figure}

In Figure~\ref{fig:EdgeRoughnessArmchair} we present the energy levels and absorption spectra with and without a normal electric field for quantum dots with rough edges based on those with armchair edges -- AHEX(r) with $n=352$ and ATRI(r) with $n=215$. For clarity, in this picture, the energy levels for $E=0.2$~{V/\AA} are vertically shifted by $0.05$ with respect to those at $E=0$~{V/\AA}. A vertical shift of $1$ is used for corresponding normalized absorption spectra of both $x$- and $y$-polarization. The similar plots for irregular phosphorene dots based on QDs with zigzag edges are shown in Fig.~\ref{fig:EdgeRoughnessZigzag}.

In all cases the quasi-zero energy states within the bulk gap, i.e. between conduction and valence band states, survive but become more dispersed forming wider energy band around the Fermi level. The number of QZES in random structures is changed compared to the regular ones but it correlates with the number of unpaired phosphorous atoms (highlighted in Fig.~\ref{fig:PhosphoreneQuantumDotsClassification}) as discussed for regular QDs. The deviation from the rule was found in the cases when two atoms without a $t_2$ hopping pair were linked by $t_1$ hopping. We did not obtain dielectric structures, e.g. without QZES, in $10$ random seeds for each type of the irregular PQDs but we checked that the QZES disappear if all phosphorous atoms are paired by $t_2$ hopping. Thus, dielectric clusters can be, in principle, engineered (see Appendix). The effect of the electric field is further broadening of the zero energy band. Unlike the case of regular ATRI and AHEX PQDs the splitting is not that sharp for the corresponding QDs with irregular edges and the two groups of the QZES are less distinctive. One can also see from Figs.~\ref{fig:EdgeRoughnessArmchair} and~\ref{fig:EdgeRoughnessZigzag} that edge disorder can suppress QZES-associated transitions in the case of hexagonal structures, whereas transitions between the QZES or from QZES to HOEL and LUEL usually stay strong for triangular shapes of the dots.

The predicted properties of the individual clusters with edge disorder could be probed by such spectroscopy techniques as micro-photoluminescence at high frequencies~\cite{Kim1999,Nguyen2015a} and scanning near-field optical microscopy at infra-red frequencies~\cite{Johnson2002,Cvitkovic2006,Samson2006}.

\section{\label{sec:Conclusions} Conclusions}
In summary, we have investigated the electronic and optical properties of phosphorene quantum dots of triangular and hexagonal shapes, with regular and irregular edges and with armchair and zigzag crystallographic orientations. All studied types of PQDs are metallic due to a set of energy states formed around the Fermi level. These states are absent in graphene dots of similar shape and are attributed to the puckered structure of phosphorene. Similar states exist only in triangular graphene and silicene counterparts with zigzag edges~\cite{Abdelsalam2015b, Abdelsalam2016}. We found that for each type of phosphorene dot with regular edges the number of these peculiar states is related to the dot size indexed by the number of hexagonal elements at one edge, see Table~\ref{tab:QZESTable}.

As a more general rule, which also works for the structures with disordered edges, the number of quasi-zero-energy states is equal to the number of phosphorous atoms which do not have a counterpart atom in the opposite layer. The unpaired atoms connected by the $t_1$ hopping parameter do not contribute to the number of new states. Thus, producing dielectric phosphorene clusters should be a more technologically challenging problem compared to the metallic ones.

The absorption spectra due to the in-plane $x$- and $y$-polarizations of the incident light are very different in phosphorene QDs, whereas such two spectra have similar shapes in graphene dots. The $y$-polarization mostly contributes to the transitions within the new set of quasi-zero-energy states. These new states play a decisive role in the optical properties of PQDs, increasing the number of absorption peaks in the low-energy region ($<2$~eV) of phosphorene quantum dots compared to  graphene ones.
Applying an external electric field to the structure in the out-of-plane geometry greatly influences these absorption peaks by blue-shifting and splitting them, thereby
modifying the absorption gaps. Due to the quasi-zero-energy states's robustness against the edge disorder and their optical activity in the infrared range, the small clusters of phosphorene could be used as a filler material for producing composites for electromagnetic shielding. A strong linear dichroism makes small phosphorene quantum dots a promising material for infrared polarizers and tunable polarization-sensitive detectors. In particular, hexagonal dots with armchair edges demonstrate the most appealing behaviour having an extremely strong, well-isolated absorption peak tunable in a wide frequency range.

A natural extension of our calculations is to use the first principles methods~\cite{Matthes2016}. The many-body effects can also be taken into account, since they are known to redistribute energy levels shifting the positions of some absorption peaks in graphene-based clusters~\cite{Guclu2010,Guclu2011,Potasz2012,Sheng2012,Potasz2015,Guclu2016}. Accounting for the deeper $s$-orbitals should result in additional absorption peaks at high-energies. However, this should not affect the main conclusions of our work.

\begin{acknowledgments}
This work was supported by the EU FP7 ITN NOTEDEV (FP7-607521); EU H2020 RISE project CoExAN (H2020-644076); FP7 IRSES projects CANTOR (FP7-612285), QOCaN (FP7-316432), InterNoM (FP7-612624); Graphene Flagship (Grant No. 604391). The authors are very grateful to R. Keens for a careful reading of the manuscript and proposed corrections.
\end{acknowledgments}

\appendix*
\section{\label{App} A dielectric phosphorene quantum dot}
\begin{figure}[htbp]
\centering
\begin{subfigure}{.24\textwidth}
    \centering
    \includegraphics[width=\textwidth]{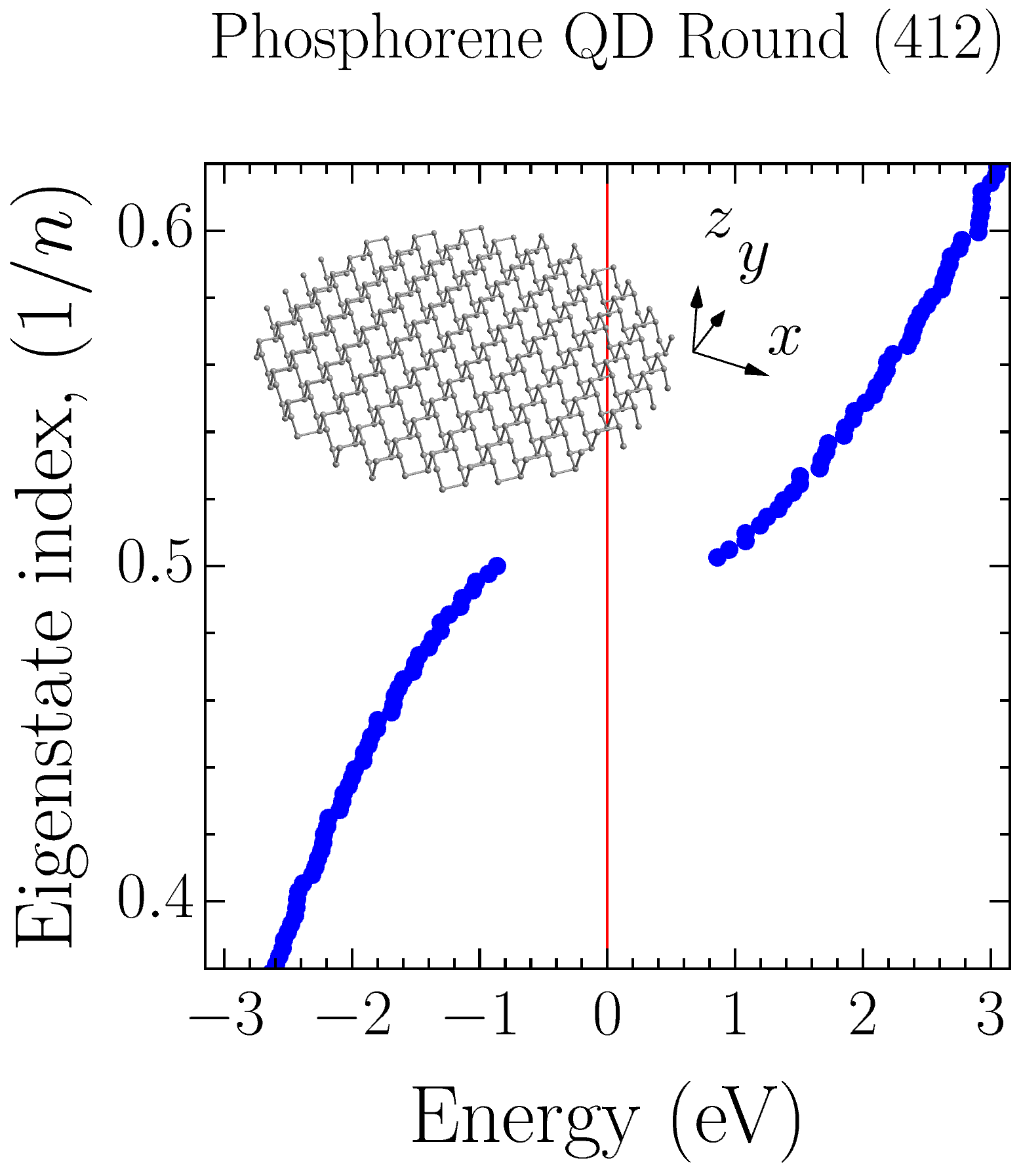}
    \caption{}
\end{subfigure}%
\begin{subfigure}{.24\textwidth}
    \centering
    \includegraphics[width=\textwidth]{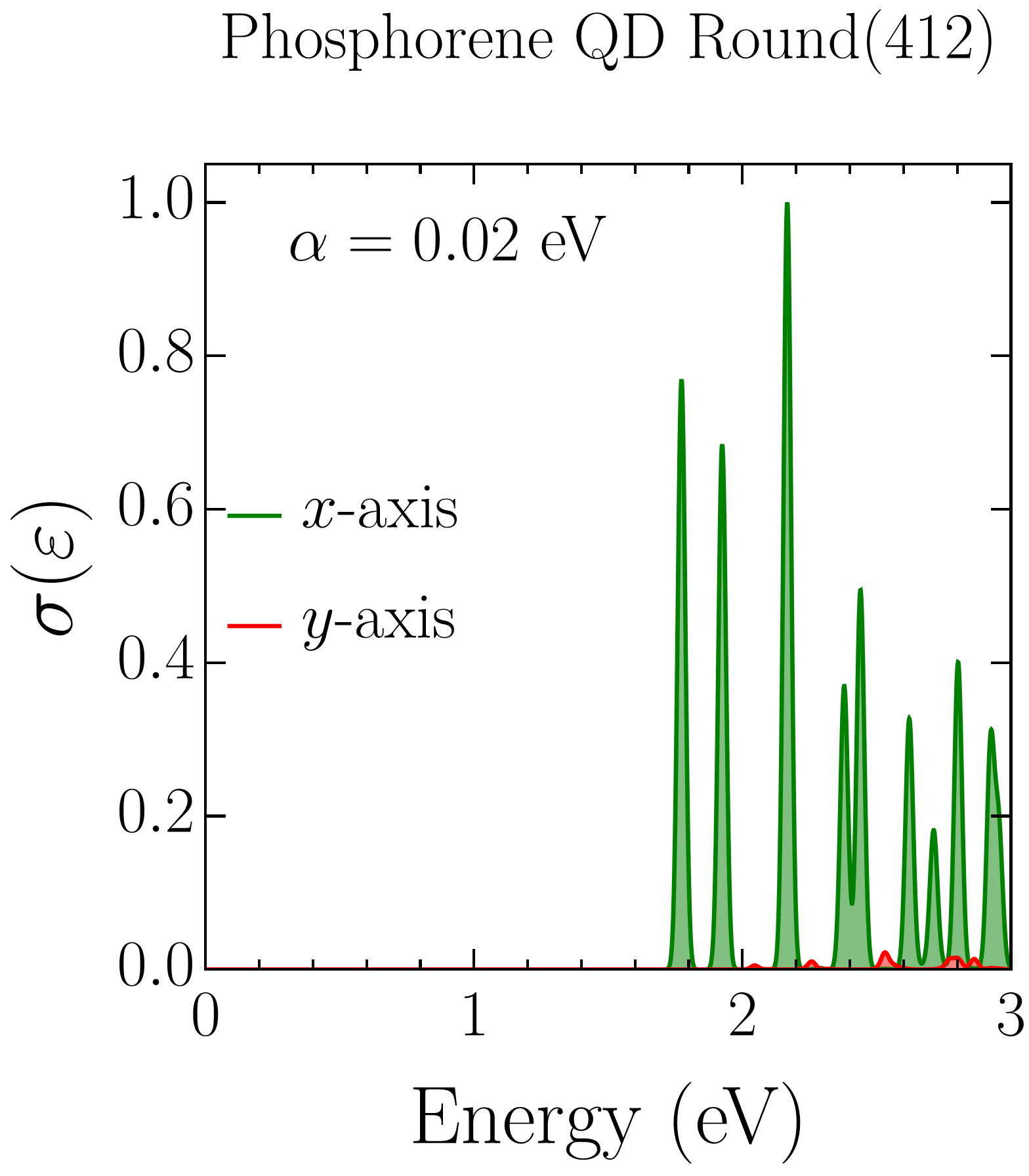}
    \caption{}
\end{subfigure}
\caption{The energy levels (a) and optical absorption cross-section (b) for a dielectric phosphorene quantum dot. The inset in (a) shows how the coordinate system is oriented with respect to the cluster.}
\label{fig:DielectricPQD}
\end{figure}

In this appendix we demonstrate that dielectric phosphorene quantum dots without quasi-zero energy states (QZES) are possible though they are to be much more rare compared to those with QZES. Figure~\ref{fig:DielectricPQD} shows the energy levels and absorption spectrum of a dielectric phosphorene cluster of round shape with $n=412$. Note that it is not the shape but rather the phosphorous atoms pairing with $t_2$ hopping that defines the absence of the QZES. The round phosphorene clusters with different size have QZES in their electronic spectra. The inset of Fig.~\ref{fig:DielectricPQD}(a) demonstrates that the above mentioned condition for dielectric cluster existence is fulfilled leading to the empty energy gap of about $2$~eV. This gap is also present in the absorption spectrum in Fig.~\ref{fig:DielectricPQD}(b). As one can see, in this case the spectrum is entirely defined by the $x$-polarized transitions between valence and conduction band states, and $y$-polarized absorption, which is strong for transitions involving QZES, is negligible. According to our calculations the application of an electric field normal to the structure plane up to $E=0.4$~V/{\AA}, does not noticeably change the presented energy levels and optical spectrum.

\end{document}